\documentclass[aps,prd,preprint,groupedaddress,nofootinbib,byrevtex]{revtex4-1}%\documentclass[pteplogo]{ptephy_v1}
\usepackage{amsmath,amssymb}
\usepackage[dvips]{graphicx}
\usepackage{mathrsfs}
\begin{document}
\title{ Finite size corrections to the Fermi's golden rule (2) Scattering probability .
Wavefunctions  of interacting many-body states at finite time interval:   wave packets  }
\author{Kenzo Ishikawa${}^1$} \author{Yutaka Tobita${}^2$}

\affiliation{${}^1$Department of Physics, Faculty of Science, Hokkaido
University, Sapporo 060-0810, Japan}  
 \affiliation{${}^2$Department of Mathematics and Physics, Faculty of Science, Hirosaki 
University, Hirosaki 036-8561, Japan }
\date{\today}
%\maketitle
%\begin{picture}(0,0)(0,0)
% \put(320,230){\makebox{arXiv:0000.0000 [hep-ph]}}
%\end{picture}
\begin{abstract}
Field theory is formulated with normalized one-particle states, and  scattering probability for the finite-time interval is studied   from  the probability principle of the quantum mechanics following  von Neumann.  Using normalized one-particle states,  we find  the many-body wavefunctions  without   encountering the divergence difficulty of plane waves.
  Owing to Using the wave packet representations, the wave functions of the interacting systems and the transition probability, which   become finite at the finite time 
 owing to their compact  natures and  space- time dependence, are obtained.  The   specific properties  at finite time interval  remain 
  at  $ t \rightarrow \pm \infty$,  and     appear in scattering  processes  at finite $t$ and  $ t \rightarrow \pm \infty$.  
A  scalar theory  and  Quantum Electrodynamics are studied  and the finite-size corrections to the Fermi's golden rule which are  not included in the standard S-matrix are obtained.  They  have  distinctive and intriguing properties, and   may become sizable,  which lead   new phenomena beyond those derived from the golden rule. 
  New perspectives of halos and other wave-like phenomena are presented.    
\end{abstract} 
%\subjectindex{xxx}
\maketitle
%arXiv:1106.4968[hep-ph]
%\newpage
% insert suggested PACS numbers in braces on next line
%\pacs{}
% insert suggested keywords - APS authors don't need to do this
%\keywords{}

\section{ Evolution of many-body states   } 

1. Solving  time-dependent Schroedinger equation of continuous  energy spectrum 

 2.  Initial wave function and its asymptotic behaviors  
 
 3. Green's function

Wave functions   satisfying a many-body      Schr\"{o}dinger  equation of  a Hamiltonian $H=H_0+H_\text{int}$,
 where  $H_0$ is a free part and 
    $H_\text{int}$ is an interaction part of fields,   are obtained with perturbative 
expansion with respects to  $H_\text{int}$ using  eigenstates of $H_0$ of eigenvalue $E_i$. 
These  states are plane waves, and their matrix elements are described by the Dirac delta function.  The Dirac delta function diverges but its integration
converges when that is combined  with smooth and well-behaved functions which decrease rapidly. Using them,   solutions  at macroscopic distances can  be computed uniquely normally.  Transition amplitude  is defined with these  wave functions.     Particularly those which are   defined at the infinite time interval, S-matrix, 
 under the interaction  $e^{-\epsilon |t|}H_\text{int}$, where $\epsilon$ is infinitesimal positive number and the limit $\epsilon \rightarrow 0$ is taken  before the limit of $t \rightarrow \pm \infty$, is used as the standard method.  Obviously this  interaction  switches off adiabatically (ASI) artificially. In this method, the physical quantities are convergent, and  the diverging norms of the plane waves cause no difficulty,
as are described in most literature and textbooks.   These have been applied in wide area 
 and lead successful results.     However, these are applicable in idealistic situations at $t \rightarrow \infty$ of ASI , but not in many 
 realistic cases.  There are many
  physical systems where this condition is not met. In fact, the time interval is not infinite in real systems, 
 and the interaction does not switch off rapidly but remains.    It is not clear if ASI is applicable  for the finite time interval.  
  If the standard one under ASI is applied,    inconsistency with experiments or  divergence in the calculations were found.   
    One of these  difficulties was noticed  for the transition in a finite time interval sometime
        ago by Stuckelberg. At a finite time,  the interaction remains. For the plane waves, the converging  factor is  absent, and  the integral
          becomes not  well defined. The probability also becomes divergent  
         at  a finite $t$. Obviously this difficulty is connected with divergence of the norm of the plane waves, and should  not appear for the wave packets, as the wave functions are normalized. With  these wave functions at the finite $t$,  the probability is finite, and  leads  new physical quantities, as by product.   The physical
          quantities diverged   in ASI ,but become converged in the proper treatment may  show  prominent effects for  extended
          wave-like states.    These have intriguing properties and          imply new phenomena.     
   
    {\bf Many-body wave function}
    
 The wave functions under the interaction $H_{int}$  reveal the physical states of separating each others and overlapping ones. The former is expressed by ASI but the  
   latter is not. These  states hold wavelike properties caused by  remaining interactions. That continue while  they  overlap and the effects  remain in the many-body 
   wave function even after they separate, since the Schr\"{o}dinger   equation is first order in time.      These  can be studied with  neither  ASI nor 
 the plane waves. Using a wave packet formalism, which uses a complete set of normalized functions,  the many-body wave functions are solved without
   encountering    the difficulty. 
    
   The wave functions  at  $t=T$ of $ T \rightarrow \infty$  under  the interaction $e^{-\epsilon |t|}H_{int}$ (ASI), where the limit $T \rightarrow \infty$ is taken first while  $ \epsilon$ is kept small, were  useful  to study fluctuations in 
   short distance region in  renormalization program
     \cite{Tomonaga,Schwinger,Feynman}.  These express the particle-like states in the asymptotic region $| t| \rightarrow \infty$.  
  
   In  the overlapping waves, the interaction energy $\langle \Psi(t)| H_\text{int} |\Psi(t) \rangle $  is finite. These   are expressed by  the states 
    of   the final energy  $E_f \neq E_i$, where $E_i$ is the initial energy and  are  included in solutions   of the 
   Schr\"{o}dinger  equation  $| \Psi(t) \rangle $ of $\langle \Psi(t)| H_\text{int} |\Psi(t) \rangle \neq 0$  at $t=T_1$.    Due to  this   expectation
  value, which we call interaction energy, the  states  satisfy $E_f \neq E_i$ and give unique contributions.  
    We call this state   a quasi-stationary  composite state (QCS).    These wave functions 
   at  $T$ are written explicitly with  the wave packet of a spatial size, $\sigma$, and    
     effects of  QCS  were found in \cite{Ishikawa-Tobita-PTEP,Ishikawa-Tobita-ANA,Ishikawa-Tajima-Tobita-PTEP}. 

 QCS  is  superposition of  states of continuous
 spectrum of the kinetic energy $E_f$ different from the initial value $E_i$, and  appears in various channels  of decaying one-particle state or many-body scattering states.  The wave packets are normalized states,  and  are  specified by not only  the momentum ${\vec P}$ but also  the position   ${\vec X}$ at an instant of time.
   Because the momentum and the position  are transformed  under the space-time transformation, the theory is invariant manifestly but the probability is slightly modified under the transformations.  
Unusual enhancement in a forward direction at high energy was also found. This is characterized by   a new characteristic length  $\frac{\hbar E}{m^2c^3} $, where $\hbar = \frac{h}{2\pi}$, $h$, $E$, and $m$ are the Planck constant, the energy, and the mass, which is longer than  the de Broglie wave length $\frac{h}{p}$, where $p$ is   the momentum, in many situations.  This is of  macroscopic length for the light particles, and can be longer than   the size of the detector.  Detailed account of these 
was given  in \cite{Ishikawa-Tobita-ANA}.

{\bf Transition probability}  
  
  The transition probability involving  QCS  is studied with the wave packets without ASI, and denoted as $S[T]$ and satisfies the necessary boundary conditions.   The standard S-matrix $S[\infty]$ does not give $T$-dependence from its definition and is useless.
   
       A transition of an eigenstate of $H_0$ of eigenvalue $E_i$  due to $H_\text{int}$  in ASI occurs   to     final states of the energy $E_f=E_i$ but QCS leads to
  those of $E_f \neq E_i$.  The former contribution   reveals of particle-like states in the asymptotic region
   $| t| \rightarrow \infty$.    The latter computed with the wave packets is finite, but divergent if   the plane waves  are used    due to these  waves 
   of   energy $E_f \neq E_i$, which includes $E_f \rightarrow \infty$ as   was  found by 
    Stueckelberg \cite{Stuckelberg}.   Towards its resolutions,  diffuse boundaries  \cite{Stuckelberg},  Schr\"{o}dinger  representation of the field theory \cite{Symanzik},  nonequilibrium dynamics \cite{Baacke} , and  analysis  from  different viewpoints \cite{Nomoto} have been studied.   An importance of the problem was emphasized   in \cite{Bogoliubov-Shirkov} and  \cite{Schweber}. However,   its connection with  the von Neumann's fundamental principle of the quantum mechanics (FQM ) has not been paid attention, and the connections  with  
  experimental observations  and the natural phenomena have been unclear. 
                Difficulties of computing transition probabilities   for overlapping waves   in quantum mechanics 
       were pointed  by( Peierles), Feyman ,( Sakurai, Greiner), Stuckelberg, Goldberger-Watson. The difficulty remains.  
       The present series of works compute  the rigorous 
       transition probability that includes QCS, by which natural phenomena are governed.

  Using  the wave function  $| i ,0\rangle$ at $t=0$ and  $|f, T \rangle $ at $t=T$, the transition probability  is  expressed from  
   FQM  as, $P(T)=|\langle f ,T | i,0  \rangle |^2$, for  normalized states.   
   For  plane waves or  stationary states,     a modification   
      introduced by Dirac \cite{Dirac-t,Dirac} is,   for $P(T) \ll1$, 
to use   the average rate $\Gamma=\frac{P(T)-P(T_1)}{T-T_1}$ between  a small $ T_1$ and a large $T$ is used normally.  
   The average transition probability of stationary states was expressed by  $\Gamma$,  known as the Fermi's golden rule for the final state of continuous spectrum,
   which is computed easily with the Dirac delta function. 
Solving  $P(T)$ at a large $T$ ,   
\begin{eqnarray}
P(T)=\Gamma  T+P^{(d)} , P^{(d)}=P(T_1)-\Gamma T_1. \label{probability1}
\end{eqnarray}
Unless   $P^{(d)}  \ll \Gamma T$, both of    $\Gamma$  and  $P^{(d)}$ are inevitable for physical phenomena. 
 In  the previous works,   $P(T)$  in various weak and electromagnetic decays  have been shown to resolve  puzzles,   \cite{Ishikawa-Tobita-PTEP,Ishikawa-Tobita-ANA,Ishikawa-Tajima-Tobita-PTEP}.    We compute Eq.$(\ref{probability1})$  using the wave packets   in a  scalar field theory and Quantum
  Electrodynamics (QED).
     The connections  of   the divergences  found by Stuckelberg and  those due to the  intermediate states are clarified.  The latter one is 
subtracted by the universal counter terms,  by  the renormalized  mass and  the renormalized charge \cite{Tomonaga,Schwinger,Feynman}, and the former one 
becomes finite due to the wave packets and   affects   the  transitions.

 The   $\Gamma T$ comes from the dynamics in the de Broglie wave length region, which represents the  transitions of the  states that  separate 
 quickly. The  de Broglie  wave length is $10^{-10}$ meters for the electron of $1$ KeV or $10^{-15}$ for $200$ MeV, and  is much shorter than the
  typical length $1$ meter of the experiments or natural phenomena, which can be approximated with   $T=\infty$.  This corresponds to       
 the  Fermi's golden rule \cite{ Dirac, Schiff-golden, Landau} or   its extension, an S-matrix $S[\infty]$, in quantum field theory     
\cite{LSZ,Low,Winter,Goldberger-Watson,Newton,Taylor,QFT}  under ASI.
 The states  are 
described by   the free Hamiltonian , and  S-matrix  satisfies  $[S[\infty],H_0]=0$. The states  of   $E_f = E_i$, where $E_i$ and $E_f$ are eigenvalues of $H_0$, 
contribute.
 The cross sections or decay rates defined by the ratios of the fluxes are 
  manifestly Lorentz covariant.   The amplitudes are expressed with  
 Green's functions defined with ASI. $\Gamma$,  cross sections, or  other corresponding
 rates computed from the ratios of the fluxes of the stationary states of the Schr\"{o}dinger  equation under  ASI reveal in fact the particle properties.

  Contrary to $\Gamma $, $P^{(d)}$  depends on  the spatial size of the initial and final waves $\sigma$, \cite{Ishikawa-Tobita-PTEP,Ishikawa-Tobita-ANA,Ishikawa-Tajima-Tobita-PTEP},  and is  enhanced for large $\sigma$.  They can satisfy 
      $P^{(d)} \geq  \Gamma T $ or $P^{(d)} \gg  \Gamma T $, and   lead     rapid transitions at small $T$, and macroscopic phenomena for light particles.    
        Importance of       transitions of   overlapping waves   were pointed 
out in \cite{Feynman2, G-W}.    
The  probability  in the laboratory frame,  in which the measurement is made, are computed and compared directly with the experimentally  observed values \cite{Gell-Mann, Ishikawa-Shimomura}.

Using perturbative expansions  with respect to a relevant coupling $g$   in a scalar theory and QED ,   $P^{(d)}$ and $\Gamma$ are expressed as,    
\begin{eqnarray}
& &P^{(d)}=g^2 C_0 +g^3C_1+ \cdots \\
& &\Gamma= g^2 D_0+g^3 D_1+\cdots \nonumber
\end{eqnarray}
where  $C_i$ and $D_i$ are numerical constants and   $g$ is the renormalized constant.  Although the constants $D_i$ are known long \cite{Jauch-Rohlich,Kinoshita},
  in these theories  $C_i$ are not. The lowest order term $g^2 C_0$  is a  unique
 consequence of the theory.  These are compared  with   
experiments.   Experiments  are hard  and  still unknown in many processes.  Nevertheless transition   processes  in nature follow the probability $\Gamma T+P^{(d)}$. 
  $P^{(d)}$  is similar to a correction  to the  scattering-into-cones theorem \cite{scattering-cones} in   a potential scattering, but is connected with 
  the fundamental physical quantities.   
Thomson scattering, pair
annihilation, and two photon scattering  in finite angles are known long, but  new corrections in the extreme forward direction are found in the present work.   The large effects  of intriguing properties in dilute systems are derived. These would be crucial for 
environmental problems.  In majority places the natural unit $\hbar=c=1$ is used, and   $t$ and $T$ are used for a time and a time-interval respectively. 
 
Plane wave limit   $\sigma \rightarrow \infty$ is unusual but gives an interesting insight to field theory.  $P^{(d)} \rightarrow \infty$ while  $\Gamma T$ remains the same. Accordingly, the total probability and the physical phenomena are governed   by $P^{(d)}$ and   the relative weight of $\Gamma T$   vanish.     Interestingly,
 this is very close  to the Haag's theorem, i.e.,
 an absence of the scattering in   a manifesly invariant formalism \cite{Haag}. 
 In the   standard method for physicists employing ASI \cite{Tomonaga,Schwinger,Feynman} and the plane waves,  the limit $T \rightarrow \infty$ is taken first  and $\epsilon \rightarrow 0$ is taken next. Finite $T$ is not allowed and the probability at finite $T$ is not computable as is seen in the  divergent probability by Stueckelberg.   FQM bridges the Stueckelberg
divergence for the plane wave of the relativistic field theory with the Haag theorem  

This paper is organized in the 
following manner. In Section 2, the scattering formalism beyond ASI in a scalar theory and unique properties of QCS in the
particle decays are
presented. In Section 3 and 4, 
these formalism   is applied to QED, and the probability $P(T)$ of the Thomson scattering is presented. In Section 5, unique features of QCS are summarized. In Section 6 and 7, implications and summary  are given.  Miscellaneous problems   of QCS and $P^{(d)}$ such as the connections with the principles of the statistical
 mechanics are presented in the Appendix.  
\newpage
%%%%%%%%%%%%%%%%%%%%%%%%%%%%%%%%%%%%%%%%%%%%%%%%%%%%%%%%%%%%%%%%%%%%%%%%%%      
 \section{ Scattering probability beyond the Fermi's golden rule} 
 %%%%%%%%%%%%%%%%%%%%%%%%%%%%%%%%%%%%%%%%%%%%%%%%%%%%%%%%%%%%%%%%%%%% 
 One dimensional potential scattering  for  initial state of finite spatial extension that satisfies an initial   condition  is suitable for clarifying 
   a distinctive property of the  transition probability in the quantum mechanics.   Namely,
    for a complete description of the scattering process   the cross section is not sufficient  and another  component, which corresponds to the correction to the Fermi's golden rule,  is necessary. 
 
 In classical dynamics, the  intensity of the particles or of the waves is a physical quantity,  which is measured directly,  hence  the cross section  defined by
  the intensity is necessary and sufficient for describing  scattering processes.   In quantum mechanical scattering, the transition probability  is a physical quantity measured directly in experiments. That     is a square of an inner product of the initial and final states  and   uniquely defined  for the normalizes states. 
  The quantum mechanical description    is not identical to  the classical one.        
%%%%%%%%%%%%%%%%%%%%%%%%%%%%%%%%%%%%%%%%%%%%%%%%%%%%%%
\subsection{Square well potential in one dimension}   
%%%%%%%%%%%%%%%%%%%%%%%%%%%%%%%%%%%%%%%%%%%%%%%%%%%%%%
 
   %%%%%%%%%%%%%%%%%%%%%%%%%%%%%%%%%%%%%%%%%%%%%%%%%%%%%%%%%%%%%%%%%%%%%%%%%%      
 \section{ Evolution of many-body states without adiabatic switching of interaction and quasi-stationary composite states }
 %%%%%%%%%%%%%%%%%%%%%%%%%%%%%%%%%%%%%%%%%%%%%%%%%%%%%%%%%%%%%%%%%%%%
 Many-body wave functions  are solved in a scalar field theory with the wave packets without encountering a divergence appeared for the plane waves. A new component  of     intriguing properties is found to appear.  In Eq.$(8)$ in a following section and  sub-section 2.3, $\hbar$ is written explicitly,  although the natural unit is used in the majority of  places.

A  free system of scalar fields $\varphi_1(x)$ and $\varphi_2(x)$   described 
by a Lagrangian 
 \begin{eqnarray}
 \label{scalar-lagrangian}  L_0={1 \over 2}\sum_l {\partial_{\mu} \varphi_l(x) }{\partial^{\mu}
   \varphi_l(x)}-{1 \over 2} \sum_l m_l^2 \varphi_l(x)^2,
\end{eqnarray}
 where  $m_2 > 2m_1$ is studied.  The theory is tranlationally invariant   and rotationally invariant and the energy-momentum tensor and the rotation tensors   
\begin{eqnarray}
& &T_{\mu \nu}= \sum_l {\partial_{\mu}} \varphi_l(x) {\partial_{\nu} \varphi_l(x) }
-g_{\mu  \nu} L ,\\
& & K^{\mu \nu\lambda}= ( x^{\nu} T^{\mu\lambda}-x^{\lambda} T^{\mu\nu})  \nonumber
\end{eqnarray}
are  conserved. The    energy-momentum vector and the rotation and boost operators 
\begin{eqnarray}
 P^{\nu}= \int d  {\vec x} T^{0 \nu}, K^{\nu\lambda}=\int d{\vec x}   K^{0 \nu\lambda}  \label{Poincare}
\end{eqnarray}
are constant and form Poincare algebra.%%%%%%%%%%%%%%%%%%%%%%%%%%%%%%%%%%%%%%%%%%%%%%%%%%%%%%%%%%%%%%%%%%%%%%%%%%%
\subsection{ Representation : plane waves vs wave packets}
 %%%%%%%%%%%%%%%%%%%%%%%%%%%%
 
%%%%%%%%%%%%%%%%%%%%%%%%%%%%%%%%%%%%%%%%%%%%%%%%%%%%%
\subsubsection{ Plane wave  } 
%%%%%%%%%%%%%%%%%%%%%%%%%%%%%%%%%%%%%%%%%%%%%%%%%%%%%

The  Hamiltonian $H=P^{0}$ describes the evolution of the state.   The filed is expanded in a set of momentum eigenstates of an energy 
$E_l({\vec p})=\sqrt{p^2+m_l^2}$,
\begin{eqnarray}
& &\varphi_l({\vec x},t)=\int d{\vec p} e^{i{\vec p}\cdot {\vec x}} \rho_l({\vec p}) a_l({\vec p},t)+h.c., \rho_l({\vec p})=\frac{1}{\sqrt{ (2 \pi)^3 2E_l({\vec p})}}, \\
& &[a_{l_1}({\vec p}_1,t_1),a_{l_2}^{\dagger}({\vec p}_2,t_2)]\delta(t_1-t_2)=\delta_{l_1l_2}\delta({\vec p}_1-{\vec p}_2)\delta(t_1-t_2),l_1,l_2=1,2, \nonumber 
\end{eqnarray}
where $h.c.$ stands for  a Hermitian conjugate. The states  defined by creation and annihilation operators  
\begin{eqnarray}
& &| 0\rangle, a_{l_1}^{\dagger}({\vec p},t ) |0 \rangle, a_{l_2}^{\dagger}({\vec p}_2,t )a_{l_1}^{\dagger}({\vec p}_1,t)|0 \rangle, \cdots, \\
& &a_{l_i}({\vec p},t) |0 \rangle =0 \nonumber
\end{eqnarray}
are eigenstates of the Hamiltonian,
\begin{eqnarray}
& &H_0 | {\vec p}_1,{\vec p}_2, \cdots \rangle =E_0| {\vec p}_1,{\vec p}_2, \cdots \rangle, E_0=\sum_i \sqrt{ {\vec p}_i^2+m_{l_i}^2},  \\
& &H_0= \sum_l \int {d {\vec p}}E_l(p) a_l^{\dagger}({\vec p},t)   a_l({\vec p},t).  \nonumber
\end{eqnarray}
They satisfy a Schr\"{o}dinger  equation,  
\begin{eqnarray}
& &i\hbar{\partial \over \partial t}|\Psi(t)\rangle=H_0|\Psi(t) \rangle,
\label{Schroedinger0}  \\
& &|\Psi(t) \rangle =e^{\frac{E_{\alpha}}{i \hbar}t} |\Psi_{\alpha}(0) \rangle. \nonumber 
\end{eqnarray}  
These   form  many-body states   satisfying   the completeness condition 
\begin{eqnarray}
& &I=|0 \rangle \langle 0 |+ \sum_l \int d {\vec p}   a_l^{\dagger}({\vec p},t)|0 \rangle \langle 0 | a_l({\vec p} , t) + \text {(many-body ~states)}, \\ 
& &|\Psi (T)\rangle =N_0(T) |0 \rangle + N_1({\vec p},T) | {\vec p}  \rangle+ \text{ many-body states},  \nonumber \\
& &N_0(T)= \langle 0| \Psi(T) \rangle, N_1({\vec p},T)=\langle {\vec p}| \Psi(T) \rangle, N_2({\vec p}_1,{\vec p}_2,T)=\langle {\vec p}_1,{\vec p}_2| \Psi(T) \rangle. \nonumber
\end{eqnarray} 
One particle states  are normalized with the Dirac's delta function,
\begin{eqnarray}
& &\langle {\vec p}_1, l_1| {\vec p}_2,l_2 \rangle=\delta_{l_1 l_2} \delta ({\vec p}_1-{\vec p}_2),\\
& &\langle {\vec p}, l_1| {\vec p} ,l_2 \rangle=\delta_{l_1 l_2} \delta (0) =\infty, \nonumber
\end{eqnarray}
and transformed by   Poincare transformation  $ U(\Lambda, d)$, where $\Lambda$ represents  a rotation and a boost and $d$ represents a four-dimensional 
translation,  and  constructed by the operators of Eq.$(\ref{Poincare})$ 
\begin{eqnarray}
U(\Lambda, d) |{\vec p} \rangle = e^{i p\cdot d} |{\Lambda} {\vec p}  \rangle.
\end{eqnarray}
Under a rotation of the coordinates ${ \tilde  {\vec p}}_i= {\overrightarrow{ R p}}_i.$, matrix elements of a scalar operator satisfy 
\begin{eqnarray}
%& &\langle {\vec p}_1,  {\vec p}_2,\cdots |O(x)|  {\vec p}_1',{\vec p}_2', \cdots  \rangle=e^{i(p'_T-p_T) \cdot x}  \langle {\vec p}_1,  {\vec p}_2,\cdots |O(0)|  {\vec p}_1',{\vec p}_2', \cdots  \rangle, \\
& &\langle {\vec p}_1,  {\vec p}_2,\cdots |O(0)|  {\vec p}_1',{\vec p}_2', \cdots  \rangle= \langle {\vec p}_1,  {\vec p}_2,\cdots |U^{-1}(\theta)O(0)U(\theta)|  {\vec p}_1',{\vec p}_2', \cdots  \rangle \label{scalar-matrix}
\nonumber\\
& &=\langle \tilde {\vec p}_1, \tilde {\vec p}_2,\tilde \cdots |O(0)|  \tilde {\vec p}_1',\tilde {\vec p}_2', \tilde \cdots  \rangle,  
\end{eqnarray}
and the same equality holds for Lorentz transformations. The matrix elements are invariant under the transformations of the momenta, and are  functions of the
 invariant combinations of the momenta.  
The norm is divergent but the rigorous treatments \cite{Schwartz} proved that the  normal physical quantities are finite in a space of distributions. Cross sections and decay rates are computed from the ratios of fluxes for the stationary states, despite of the diverging norm.  They are not identical to the probability computed from the normalized states. 
%%%%%%%%%%%%%%%%%%%%%%%%%%%%%%%%%%%%%%%%%%%%%%%%%%%%%
\subsubsection{ A wave packet  representation } 
%%%%%%%%%%%%%%%%%%%%%%%%%%%%%%%%%%%%%%%%%%%%%%%%%%%%%

A representation of a field operator  in terms of normalized base functions   
 \cite{Ishikawa-Shimomura}  is applied hereafter.  Dirac's brackets notation $\langle  A| B  \rangle $ is used for  one-particle space  and for many-body space. In the former, the states $|A \rangle  $ and $|B \rangle $ represent  of one particle states, and in the latter they represent of the many-body states.     In the following equations from Eq.$(\ref{wave-packet0-1})$  to    Eq.$(\ref{one-body})$, the brackets are those of one-particles.        

A normalized solution of the free wave   
equation  of a central momentum ${\vec P}$, 
position ${\vec X}$, and a time ${ T}_0$   in the momentum
space  is
\begin{eqnarray}
\langle t,{\vec p}\,|{\vec P},{\vec X},T_0,l
 \rangle=(\frac{ \sigma } {\pi})^{3/4}e^{ -iE_l({\vec p}\,) (t-T_0)-i{\vec p}\cdot{\vec
 X}-{\sigma \over 2}({\vec p}-{\vec P})^2}.
\label{wave-packet0-1} 
\end{eqnarray}
If the time $t-T_0$ is not very large, the solution in the configuration space is approximated well with    
  \begin{eqnarray}
& &w({\vec P},{\vec x},t;{\vec X})= Ne^{-{1 \over 2\sigma}({\vec x}-{\vec X}-{\vec v}_l (t-T_0))^2 }e^{-iE_l({\vec
 P})(t-T_0)+i{\vec P}\cdot({\vec x}-{\vec X})}, 
\label{coordidate-wave}  \\
& &N=(\pi \sigma)^{-3/4}, {({\vec 
 v}_l)}_i={\partial \over \partial p_i}E_l({\vec p})|_{{\vec p}={\vec
 P}},\nonumber 
 \end{eqnarray}
where $\sigma$ shows a size in the configuration space, and a case of $\sigma {\vec P}^2 \gg 1 $ is studied. 
 For the sake of simplicity, we use the Gaussian form in most parts
 of  this paper.   
 For a Fourier transformation,    $\langle {\vec x}|{\vec p}\, \rangle=(2 \pi)^{-3/2}e^{i{\vec p}\cdot{\vec x}}$ is used.   In this region, a spreading effect is negligible , and  the  center moves  with a velocity ${\vec v}_l$.  This paper concentrates this   region.   For brevity,  $\chi$ is used for    $({\vec P},{\vec X})$.  
 From Eq.$(\ref{wave-packet0-1})$  
 \begin{eqnarray}
& &\langle \chi_1,T_1,l| \chi_2,T_2,l \rangle=(\frac{\sigma}{\pi})^{3/2}e^{-\frac{\sigma}{4}({\vec P}_1-{\vec P}_2)^2} \int d{\vec p} e^{-i(E_l({\vec p})(T_1-T_2)-{\vec p}({\vec X}_1-{\vec X}_2))-{\sigma}({\vec p}-{\vec P}_0)^2}, \\
& &{\vec P}_0=\frac{{\vec P}_1+{\vec P}_2}{2}, \nonumber
\end{eqnarray} 
 and the detailed property of the matrix element was given in \cite{Ishikawa-Shimomura}.  The states satisfy   one-particle completeness relation,  
\begin{eqnarray}
 & &1=\int d \chi |{\chi} ,T_0 \rangle
 \langle \chi ,T_0|, ~d \chi =\frac{d{\vec X} d{\vec P}}{  (2\pi)^3}, \\
 & &|\psi \rangle=\int d \chi |{\chi} ,T_0 \rangle
 \langle \chi ,T_0|\psi \rangle, \langle \psi_1|\psi_2 \rangle=\int d \chi \langle \psi_1 |\chi ,T_0\rangle  \langle \chi,T_0|\psi_2 \rangle. \nonumber
\end{eqnarray}  
Creation and annihilation operators  of the wave-packet states defined  as   
\begin{eqnarray}
A_l^{\dagger}(\chi,t,T_0) =\int d{\vec p} a_l^{\dagger}({\vec p},t) \langle t,{\vec
 p}|\chi,T_0 \rangle , A_l(\chi,t,T_0)=\int d {\vec p} \langle \chi,T_0|{\vec p},t
 \rangle a_l({\vec p},t), \label{one-body}
 \end{eqnarray}
form  many-body states   
\begin{eqnarray}
& &| 0\rangle, A_l^{\dagger}(\chi,t,T_0) |0 \rangle, A_{l_1}^{\dagger}(\chi_1,t,T_0)A_{l_2}^{\dagger}(\chi_2,t,T_0)|0 \rangle, \cdots, \label{wave-packets}\\
& &a_l({\vec p},t) |0 \rangle =0, \nonumber
\end{eqnarray}
satisfying  the completeness condition 
\begin{eqnarray}
I=|0 \rangle \langle 0 |+ \sum_l \int d \chi  A_l^{\dagger}(\chi,t,T_0)|0 \rangle \langle 0 | A_l(\chi, t,T_0)  +\text {(many-body ~states)}. \label{complete-set} 
\end{eqnarray}  

{\bf Lemma 1}

 (Expansion of a state) 

From Eq.$(\ref{complete-set} )$, a many-body state is expanded with unique coefficients  
\begin{eqnarray}
& & \label{expansion} |\Psi (T)\rangle =N_0(T) |0 \rangle + N_1(\chi,T) | \chi \rangle+ \text{ many-body states} , \\
& &N_0(T)= \langle 0 |\Psi (T)\rangle, N_1(T)=\langle 0 | A_l(\chi, t,T_0) |\Psi (T)\rangle , \text{many-body-components} ,\nonumber\\
& & \langle \Psi_1(T)|\Psi_2(T) \rangle=(N_0^1(T))^{*}N_0^2(T) + \int d\chi (N_1^1(\chi, T))^{*}N_1^2(\chi,T) + \text{ many-body states},  \nonumber
\end{eqnarray}  
and $  \langle \Psi(T)|\Psi(T) \rangle =0$ only when $ |\Psi(T) \rangle=0$.

The field operator is expanded  as 
\begin{eqnarray}
& &\varphi_l(x)=   \int d \chi  ( C_l({\vec x};\chi )A_l(\chi ,t,T_0) + C_l^{*}({\vec x};\chi )A_l^{\dagger}(\chi,t,T_0)) ,         \label{wave-operator} \\
& &C_l({\vec x};\chi )=\int \frac{ d{\vec p}}{\sqrt{ 2E_l({\vec p})}}} \langle {\vec x}| {\vec p \rangle \langle {\vec p}|\chi,T_0,l \rangle,  \nonumber
\end{eqnarray}
and is substituted to the free Hamiltonian, 
\begin{eqnarray}
& &H_0= \sum_l \int {d {\vec p}}E_l({\vec p}) a_l^{\dagger}({\vec p},t)   a_l({\vec p},t)  \nonumber   \\
& &= \sum_l \int d \chi_1 d \chi_2 d{\vec p} A_l^{\dagger} (\chi_1,t,T_0) \langle \chi_1,T_0|{\vec p} \rangle E_l({\vec p}) \langle {\vec p}|\chi_2,T_0 \rangle A_l(\chi_2,t,T_0).                     
%& &H_{int}=-\frac{g}{2} \int d{\vec x} (C_{l_1}({\vec x};\chi_1) A_{l_1}(\chi_1,t)+C_{l_1}({\vec x},\chi_1)^{*}A_{l_1}^{\dagger}(\chi_1,t) )(C_{l_2}({\vec x};\chi_2)A_{l_2}(\chi_2,t)\nonumber \\
%& &+C_{l_2}({\vec x},\chi_2)^{*}A_{l_2}^{\dagger}(\chi_2,t)) (C_{l_3}({\vec x},\chi_3) A_{l_3}(\chi_3,t)+C_{l_3}({\vec x},\chi_3)^{*}A_{l_3}^{\dagger}(\chi_3,t)).
\end{eqnarray}
The  states Eq.$(\ref{wave-packets})$ are normalized solutions of the Schr\"{o}dinger  equation,Eq.$(\ref{Schroedinger0})$ and satisfies  
\begin{eqnarray}
\langle 0| A_l( {\vec P}_1,{\vec X}_1,  t,T)A_l^{\dagger}( {\vec P}_2,{\vec X}_2, t,T)|0 \rangle=\langle {\vec P}_1,{\vec X}_1,T|{\vec P}_2,{\vec X}_2,T \rangle, 
\end{eqnarray}
where ${\vec P},{\vec X}$ are used instead of $\chi$, and the matrix element of the scalar operator satisfies
\begin{eqnarray}
\langle  {\vec P}_1,{\vec X}_1;  {\vec P}_2,{\vec X}_2;  \cdots |O(0)|  {\vec P}_1',{\vec X}_1'; {\vec P}_2',{\vec X}_2'; \cdots   \rangle=   \langle  \tilde {\vec P}_1,\tilde{\vec X}_1;  \tilde{\vec P}_2,\tilde{\vec X}_2;  \tilde\cdots |O(0)|  \tilde{\vec P}_1',\tilde{\vec X}_1'; \tilde\cdots   \rangle,     \label{scalar-matrix2}
\end{eqnarray}
where $\tilde {\vec X}$ and $,\tilde {\vec P}$ are $ {\overrightarrow{R  X}}$ and $ {\overrightarrow{ R  P}}$.  Not only momenta but also positions in the configuration space are transformed. For a matrix element that is factorized in the form,
\begin{eqnarray}
\langle  {\vec P}_1,{\vec X}_1;  {\vec P}_2,{\vec X}_2;  \cdot |O(0)|  {\vec P}_1',{\vec X}_1'; \cdot  \rangle=  \langle  {\vec P}_1,  {\vec P}_2;  \cdot |O(0)|  {\vec P}_1',{\vec P}_2'; \cdot  \rangle  \langle  {\vec P}_1,{\vec X}_1;  {\vec P}_2,{\vec X}_2;  \cdot|  {\vec P}_1',{\vec X}_1'; \cdot  \rangle \label{short-range}
\end{eqnarray}  
the sum over the positions is made easily. 
  By a Poincare transformation,   $ U(\Lambda, d) $, ${\vec P},{\vec X}, T_0 $  is transformed to ${\vec P}',{\vec X}', T_0' $.  
 The invariance of the theory is reduced to complicated relations of the matrix elements Eq.$(\ref{scalar-matrix2})$ generally.  For the short-range correlations, Eq.$(\ref{short-range})$, the momentum dependent term is equivalent to Eq.$(  \ref{scalar-matrix})$, and     the probability integrated over the positions  are expressed by $ |\langle  {\vec P}_1,  {\vec P}_2;  \cdot |O(0)|  {\vec P}_1',{\vec P}_2'; \cdot  \rangle |^2$, which becomes manifestly invariant.
A experimental test of the transformation property of the short-range term is straightforward, but that of the long-range term is difficult,   because  the transformed position ${\vec X}'$ depends on ${\vec P}$ and $T_0$. 
A detection at this coordinate is actually impossible.     
   Various aspects of the manifest Lorentz invariance in the finite-size correction to the golden rule 
  has been presented  in \cite{Ishikawa-Tobita-ANA} and will be studied in a latter section.         
  
  Thus    a complete set of normalized states are constructed and will be used for the computation of the probability in the laboratory frame, and for the transitions in nature.  
  %%%%%%%%%%%%%%%%%%%%%%%%%%%%%%%%%%%%%%%%%%%
  \subsection{ Many-body Schr\"{o}dinger  equation}
  %%%%%%%%%%%%%%%%%%%%%%%%%%%%%%%%%%%%%%%%%%%%%
The interacting scalar fields is described by the Langrangian in terms of bare fields, 
  \begin{eqnarray}
& &L=L_0+L_{int}\\
& &L_\text{int}= -\frac{g}{2!} ( \varphi_2(x)\varphi_1(x)^2 +h.c.), \label{interaction} 
\end{eqnarray}
in the classical level. Short-distance fluctuations  of intermediate states in the perturbative expansions induce the divergences. They are eliminated by the counter terms in the renormalized theory \cite{Tomonaga,Schwinger,Feynman}. In the present theory, divergence appears only in the mass terms, and the Lagrangian 
in terms of the renormalized fields is
\begin{eqnarray}
& &L={1 \over 2}\sum_l {\partial_{\mu} \varphi_l^r(x) }{\partial^{\mu}
   \varphi_l^r(x)}-{1 \over 2} \sum_l (m_l^2 -\delta m_l^2) \varphi_l^r(x)^2-\frac{g}{2!} ( \varphi_2^r(x)\varphi_1^r(x)^2 +h.c.) . \label{counter-term} 
\end{eqnarray} 
 In the interaction representation \cite{Tomonaga,Schwinger},  a wavefunction $ |\Psi(t)\rangle$ satisfies a Schr\"{o}dinger  equation,  
\begin{eqnarray}
& &i\hbar{\partial \over \partial t}|\Psi_i(t)\rangle=H_\text{int}(t)|\Psi_i(t) \rangle, \\
& &|\Psi(t) \rangle= e^{i H_0 t \over \hbar} |\Psi_i(t) \rangle  ,H_\text{int}(t)=e^{i H_0 t \over \hbar}H_\text{int} e^{-i H_0 t \over \hbar}, \label{Schroedinger2}
\end{eqnarray} 
Solutions are obtained with  perturbative expansions with respect to $H_{int}$, and  are  expressed  with the operator $U(t,t_0)$ as

\begin{eqnarray}
& & |\Psi(t) \rangle =U(t,t_0) |\Psi(t_0) \rangle, \label{solution} \\
& &U(t,t_0)=\int_{t_0}^{t} dt_1 {\mathcal T}e^{-iH_\text{int}t_1/\hbar }=1+\int_{t_0}^{t} {dt_1
  \over \hbar} (-i H_\text{int}(t_1)) \nonumber \\
&  &  + \int_{t_0}^{t} {dt_1 \over \hbar}
  \int_{t_0}^{t_1} {dt_2 \over \hbar} ( (-i)^2 H_\text{int}(t_1)H_\text{int}(t_2))+\cdots, \nonumber  \\
& &=1+\sum_{n=1}\frac{1}{n!} \int_{t_0}^{t} {dt_1}{dt_2}\cdots {dt_n} {\mathcal T}(H_\text{int}(t_1)\cdots H_\text{int}(t_n))
\end{eqnarray}
where $\mathcal T$ stands for the time-ordered product.
%%%%%%%%%%%%%%%%%%%%%%%%%%%%%%%%%%%%%%%
 \subsubsection{Plane waves: Stueckelberg divergences }
 %%%%%%%%%%%%%%%%%%%%%%%%%%%%%%%%%%%%%%%
 The solution Eq.$(\ref{solution})$    is expressed with plane waves. That   in  the first  order of the coupling constant, for  an initial  state $|\Psi_1(0) \rangle =|{\vec p}_1 \rangle$,  $H_0 |\Psi_1(0)\rangle =E_1({\vec p}_1) |\Psi_1(0) \rangle ,E_1=\sqrt{p_1^2 +m^2}$,   is
\begin{eqnarray}
& & |\Psi_1(t) \rangle =N_1(|\Psi_1(0) \rangle + \int dn D(\Delta E,t)| n \rangle  \langle n|(-i H_\text{int}(0) |\Psi_1(0) \rangle), \label{first-order}  \\
& &D(\Delta E,t)=e^{-i\frac{\Delta E t}{2\hbar}} \frac{2 \sin \frac{\Delta E t}{2\hbar}}{\Delta E}, \Delta E= E_0-E_n,\nonumber 
\end{eqnarray}  
where  $N_1$ is a normalization constant, the state $|n \rangle$ denotes a  state of the energy $E_n$ of the volume element $dn$.  The inner product of the state $|\Psi_1(t) \rangle $ with a state  $|\Psi_2(t) \rangle $ of momentum ${\vec p}_2$   is given by
\begin{eqnarray}
& &\langle \Psi_1(t)| \Psi_2(t) \rangle=\delta({\vec p}_1-{\vec p_2})|N_1|^2(1+\int dn |D(\Delta E,t)|^2  | \langle n|(-i H_\text{int}(0) |\Psi(0) \rangle|^2  \label{plane-wave-})  \nonumber  \\
& &=\delta({\vec p}_1-{\vec p_2})|N_1|^2(1+\int \prod_i \frac{d^3 {\vec k}_i}{ 2E(k_i) (2 \pi)^3} (k_i)^q \frac{(2 \sin \frac{\Delta E t}{2\hbar})^2 }{ ( \Delta E)^2} (2 \pi)^3 \delta ({\vec p}_1-\sum_i {\vec k}_i))    
\end{eqnarray}
where in the integral ${\vec p}_1={\vec p}_2$ was substituted and $q=0$ now. The integral Eq.$( \ref{plane-wave-})$ behaves differently depending upon   a spectrum of  $ \Delta E=E_n-E_0 $. We write s a minimum value  as $E_g$, i.e.,
 $\Delta E \geq E_g$. For  positive  definite $E_g$, this  shows  an energy gap between the initial and final states. There is no state of satisfying $\Delta E=0$.   
  For negative  $E_g$,   there is no  energy gap, and there are   states of  $\Delta E=0$.
    In the integrand  in Eq.$( \ref{plane-wave-})$,
\begin{eqnarray}
(\frac{2 \sin \frac{\Delta E t}{2\hbar}}{\Delta E})^2 \approx  t \delta(\Delta E)
 \end{eqnarray} 
due  to a sharp peak at    $\Delta E \approx 0$, and is proportional to $t$. The proportional constant is given by the integral over the phase space and converges.
Outside this region,  this is not proportional to $t$.  This function does not vanish  there  and at
   $\Delta E \rightarrow \infty$, $(  2 \sin \frac{\Delta E t}{2\hbar})^2 \approx 2$. The integrand  is independent of $t$, and  converges for $q=0$, but 
 diverges for $q \geq 1$.  The  norm is proportional to   $\delta({\vec p}_1-{\vec p_2}) $, and the proportional constant  diverges. This  leads the divergence found by Stueckelberg  of the transition probability at the finite $t$ diverges   for the plane waves. This behaves differently for the wave packets.

%%%%%%%%%%%%%%%%%%%%%%%%%%%%%%%%%%%%%%%%%%%%%%%%%%%%%%%%%%%%%%%%%%%%%%%%%%%%
\subsubsection{  Wave packets: Stationary  states }
%%%%%%%%%%%%%%%%%%%%%%%%%%%%%%%%%%%%%%%%%%%%%%%%%%%%%%%%%%%%%%%%%%%%%%%%%%%
The wavefunctions are obtained with the wave packet representation, in which the  interaction Hamiltonian  is expressed with the operators $  A_{l_1}(\chi_1,t)$ and  $A_{l_1}^{\dagger}(\chi_1,t) $,
\begin{eqnarray}
& &H_{int}=-\frac{g}{2} \int d{\vec x} (C_{l_1}({\vec x};\chi_1) A_{l_1}(\chi_1,t)+C_{l_1}({\vec x};\chi_1)^{*}A_{l_1}^{\dagger}(\chi_1,t) )(C_{l_2}({\vec x};\chi_2)A_{l_2}(\chi_2,t)\nonumber \\
& &+C_{l_2}({\vec x};\chi_2)^{*}A_{l_2}^{\dagger}(\chi_2,t)) (C_{l_3}({\vec x};\chi_3) A_{l_3}(\chi_3,t)+C_{l_3}({\vec x};\chi_3)^{*}A_{l_3}^{\dagger}(\chi_3,t)),
\end{eqnarray}
using  the coefficients $ C_{l_1}({\vec x};\chi_1)   $ given in Eq.$(\ref{wave-operator})$. Substituting the expressions   Eq.$(\ref{complete-set})$  to  Eq.$(\ref{solution})$,  the coefficients of the solutions  are determined. They become simple forms for stationary states.

%%%%%%%%%%%%%%%%%%%%%%%%%%%%%%%%%%%%%%%%%%%%%%%%%%%%%%%%%%%%    
{\bf  Vacuum}
%%%%%%%%%%%%%%%%%%%%%%%%%%%%%%%%%%%%%%%%%%%%%%%%%

The ground state  $|0 \rangle$ is expressed   in the first order of  $g$ as,   
\begin{eqnarray}
& &|\Psi_0 (T) \rangle =N_0(T) |0 \rangle + \int  d^3\chi  N_3(\chi_1,\chi_2,\chi_3,T) A^{\dagger}_i(\chi_1,t) A^{\dagger}_j ({\chi}_2,t) A^{\dagger}_k ({\chi}_3,t)|0 \rangle ;i,j,k=1,2 \label{vacuum-amplitude},\nonumber \\      
& &N_0(T)=1+O(g^2), d^3 \chi=d\chi_1d\chi_2 d\chi_3 ,      
\end{eqnarray}
where three-particle-state is disconnected part of the energy gap $E_g=2m_1+m_2$ shown in Fig. 1.   $N_3(\chi_1,\chi_2,\chi_3,T)$ is the Gaussian integral over space-time coordinate, and found  in Appendix A and  B as      
\begin{eqnarray}
& &|N_0(T)|^2=1, \int d^3 \chi  |N_3(\chi_1,\chi_2,\chi_3,T)|^2=0,  \label{qcs-vacuum} \\
& &\langle \Psi_0(T)|H_{int}| \Psi_0 (T)\rangle =0.  \nonumber 
\end{eqnarray}
The  ground state is stationary and unique.  From Appendix A and B, the following lemma holds.
%%%%%%Fig%%%%%%%%%%%%%%%%%%%%%%%%%%%%%%%%
\begin{figure}[t]
\centering{ \includegraphics[scale=.8,angle=0]{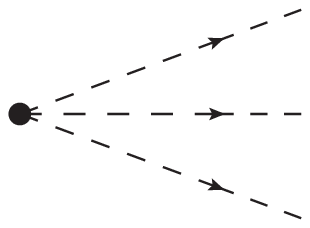}
\caption{Diagram of the vacuum amplitude that three particles are produced. }
}\label{fig:vacuum}
\end{figure}
%%%%%%%%%%%%%%%%%%%%%%%%%%%%%%%%%%%%%%%%%%%

{\bf Lemma 2}

The integral over the space-time coordinate of the Gaussian functions for the states of the energy gap $E_g$ in the bulk region is proportional to $e^{-\frac{\sigma_t E_g^2}{ 2}}$, where $\sigma_t$ 
is the square of the size in the temporal direction of the overlapping region of the waves.  The space-time positions which contribute to the boundary term are restricted to microscopic region and the probability is negligible.   

%%%%%%%%%%%%%%%%%%%%%%%%%%%%%%%%%%%%%%%%%%%%%%%%%%%%
{\bf Stable one-particle state: lightest  one}
%%%%%%%%%%%%%%%%%%%%%%%%%%%%%%%%%%%%%%%%%%%%%%%%%%%%

  One-particle state of lighter scalar $ A^{\dagger}_1(\chi) |0 \rangle$ is expressed in the order $g$ by  
\begin{eqnarray}
& &|\Psi_1^L(T) \rangle =N_1^L(T)  A^{\dagger}_1(\chi) |0 \rangle +\int d^2 \chi   N_2^L(\chi_1,\chi_2,T) A^{\dagger}_1(\chi_1) A^{\dagger}_2(\chi_2)|0 \rangle \nonumber\\
& &+ \int  d^4\chi  N_4^L(\chi_1,\chi_2,\chi_3,\chi_4, T)  A^{\dagger}_1(\chi_1) A^{\dagger}_i(\chi_2) A^{\dagger}_j(\chi_3) A^{\dagger}_k(\chi_4) |0 \rangle ,   \\
& &|N_1^L(T)|^2=1, \int d^2 \chi |N_2^L(\chi_1,\chi_2 ,T)|^2=0,  \int d^4 \chi |N_4^L(\chi_1,\chi_2\chi_3,\chi_4,T)|^2=0; i,j,k=1,2,  \nonumber
\end{eqnarray}
In the above equation, $N_2^L(\chi_l,T)$ and $N_4^L(\chi_l,T)$ are  the Gaussian integrals  in the bulk of  the finite energy gaps of $ \sigma_t \neq 0$. From Lemma,  they vanish.  
   The four particle state in disconnected part has also a large energy gap and vanishes  from the same reasons.  The contribution from the boundary also negligible as in the vacuum. 
 The lightest particle is stationary and unique.

For the vacuum state and the stable one-particle state at $t=0$, the first-order corrections do not modify the wavefunctions.  
%%%%%%%%%%%%%%%%%%%%%%%%%%%%%%%%%%%%%%%%%%%%%%%%%%%%%%%%%%%%%%%%%%%%%%%%%%%%
\subsubsection{  Quasi-stationary  composite states }
%%%%%%%%%%%%%%%%%%%%%%%%%%%%%%%%%%%%%%%%%%%%%%%%%%%%%%%%%%%%%%%%%%%%%%%%%%%
%%%%%%%%%%%%%%%%%%%%%%%%%%%%%%%%%%%%%%%%%%%%%%%%%%%%%%%%%%%%%%
%\subsubsection{Exceptional vertex part }
%%%%%%%%%%%%%%%%%%%%%%%%%%%%%%%%%%%%%%%%%%%%%%%%%%%%%%%%%%%%

 Heavier scalar decays to two lighter scalars, and is not stationary.  
The many-body wavefunction of the decaying state is a superposition of the parent and daughters in a finite time.  In a space-time region where these waves fully overlap, which is called a bulk,  that is a symmetric  superposition , whereas in a   boundary regions
that is asymmetrical.  That    behaves differently in two cases.   

%%%%%%%%%%%%%%%%%%%%%%%%%%%%%%%%%%%%%%%%%%%%%%%%%%%%
 {\bf Metastable one-particle state: heavier  one}
 %%%%%%%%%%%%%%%%%%%%%%%%%%%%%%%%%%%%%%%%%%%%%%%%%%%
 
 We assume that  the heavier field is almost stable and described by a field operator.  
A solution  Eq.$(\ref{solution})$  starting from the heavier state of the momentum and position $({\vec P},{\vec X})$ at $t=0$ is 
\begin{eqnarray}
& &|\Psi_1^H(T) \rangle = N_1^H (T) A^{\dagger}_2(\chi)|0 \rangle   + |\Psi_2(T) \rangle   \label{wave-function1},\\
& &|\Psi_2(T) \rangle=\int d^2\chi ( N_2^H( \chi,T) A^{\dagger}_1(\chi_1) A^{\dagger}_1(\chi_2) + \cdots ) |0 \rangle,  \nonumber 
\end{eqnarray}
where the disconnected part written by $''\cdots''$ vanishes due to the energy gap and is dropped.  $N_1^H(T)$ and $N_2^H(\chi,T)$ are  finite and were studied in \cite{Ishikawa-Tobita-PTEP,Ishikawa-Tobita-ANA} and  are briefly summarized  hereafter, using  the variables $({\vec P}, {\vec X})$  instead of $\chi$, and the matrix element  $ F_{\beta,\varphi_2} (\omega)=\langle \beta|H_{int}(0)| \varphi_2 \rangle$, $\omega=E_{\varphi_2}-E_{\beta}$.   The initial state is an approximate eigenstate of $H_0$.
%\newpage

The coefficient $N_2^H( \chi,T)$ is the integral over the space-time coordinate 
\begin{eqnarray}
 N_2^H(\chi,T)=\mathcal{M}=\int_{T_{2}}^{T_{1}} dt\int d^3x  \langle {\vec k}_{1},{\vec k}_{1'};{\vec X}_{1},{\vec X}_{1'}
 |H_{int}(x)| {\vec p}_{2} , {\vec X}_{2}   \rangle,
\end{eqnarray}
which gets contributions from the bulk and the boundary from Appendix A, B, and C.

{\bf Bulk contribution}

The integral  over $t$ from $-\sqrt { 2\sigma_t} <t-T_0< \sqrt{2 \sigma_t}$ is given in   Eq.$(\ref{Gaussian-integral1})$ of Appendix A as  
\begin{eqnarray}
N_2^H(\chi,T;bulk)=N_{bulk}({\vec X})e^{-\frac{1}{2 \sigma_s}(\delta {\vec p})^2-\frac{1}{2 \sigma_t}(\delta \omega)^2},
\end{eqnarray}  
where $N_{bulk}({\vec X})$ is independent of the center position in time.  From the exponential behavior, the wave function  decreases rapidly  with $\delta \omega$.  This component  shows the   conservation of  kinetic energy, which is  characteristic of point particles.   
  
{\bf Boundary contribution }

 The integral   over $t$ from the boundary is given in Eq.$(\ref{Gaussian-integral2})$ of Appendix A as  
\begin{eqnarray}
N_2^H(\chi,T;boundary)=N_{boundary}({\vec X})e^{-\frac{1}{2\sigma_s}( \delta {\vec p})^2}\frac{\sqrt{2\sigma_t}}{1+ i \sqrt{ 2\sigma_t }(\delta \omega)},
\end{eqnarray}  
 where $N_{boundary}({\vec X})$ is defined at a  boundary in time. This follows  a  power  behavior instead of the exponential one, and  the wavefunction decreases slowly with $\omega$. This  shows the non-conservation of kinetic energy, which is  characteristic of waves.    
The wavefunction is a sum, 
\begin{eqnarray}
|\Psi_2(T_1)\rangle =|\Psi_{(2,p)}(T_1)\rangle +|\Psi_{(2,w)}(T_1) \rangle, \label{p_w_wave} 
\end{eqnarray}
where    $|\Psi_{(2,p)}(T_1)\rangle$   is derived from   $\omega \approx 0$, which is computed in a standard method from  $|D(\omega, T_1)|^2 \approx 2\pi T_1\delta(\omega )$, and   $|\Psi_{(2,w)}(T_1)\rangle$ is 
 from outside of $\omega$ of   $|D(\omega, T_1)|^2 \approx constant  $. Their norms   are  expressed by
 \begin{eqnarray}
& & \langle \Psi_{(2,p)}(T_1)|\Psi_{(2,p)}(T_1) \rangle=\Gamma T_1, \Gamma= \int  d\beta \delta(\omega)|F_{\varphi_2,\beta}(0)|^2 2\pi \tilde \rho(0),\label{weight1}  \\
& &\langle \Psi_{(2,w)} (T_1)|\Psi_{(2,w)} (T_1)\rangle =P^{(d)}=\int_{|\omega| \geq \frac{1}{\sqrt 2 \sigma_t} } d {\vec p}_1 d{\vec p}_2  |F_{\varphi_2,\beta}(\omega)|^2  |D(\omega,T_1)|^2. \nonumber
\label{weight2}
\end{eqnarray}
where $\tilde \rho(0)$ is the density of the states at $\omega =0$.

The norm of the parent and  the expectation value of the interaction Hamiltonian of the state $| \Psi_2(T) \rangle $, which we call interaction energy for brevity,   up to  $g^2$ are, 
\begin{eqnarray}
& &|N_1^H(T)|^2=1 -  T \Gamma -P^{(d)} \nonumber\\   
& &E_{int}(T)= \langle \Psi_2(T)|H_{int}|  \Psi_2(T)  \rangle=\int d {\vec p}_1 d{\vec p}_2  \omega |F_{\varphi_2,\beta}(\omega)|^2  |D(\omega,T)|^2. \label{interaction-energy0}
\end{eqnarray}
$|\Psi_{(2,p)}(T) \rangle$ denotes  the state of $|\omega| \approx 0 $, and  $|\Psi_{(2,w)}(T) \rangle$ denotes the state  of $|\omega| \neq 0 ( > \frac{1}{\sqrt{ 2 \sigma_t}}) $, 
and satisfy
\begin{eqnarray}  
& & \frac{\langle \Psi_{(2,p)} (T)|H_{int}| \Psi_{(2,p)}(T) \rangle}{\langle  \Psi_{(2,p)} (T)| \Psi_{(2,p)}(T) \rangle} \leq  \frac{1}{\sqrt{ 2\sigma_t}} , \\
& & \frac{\langle \Psi_{(2,w)} (T)|H_{int}| \Psi_{(2,w)}(T) \rangle}{ \langle \Psi_{(2,w)} (T)| \Psi_{(2,w)}(T) \rangle} > \frac{1}{\sqrt{ 2 \sigma_t}} . \nonumber 
\end{eqnarray}  
The former 
 is included in ASI  \cite{Kayser,
 Giunti,Nussinov,Kiers,Stodolsky,Lipkin}. Now, 
 $|\Psi_{(2,w)}(T) \rangle$  is   a correlated  state similar to a stationary 
 state, and gives various physical effects  
 \cite{Ishikawa-Tobita-PTEP,Ishikawa-Tobita-ANA,
 Ishikawa-Tajima-Tobita-PTEP,G-W,Asahara}. 
 
 $|\Psi_{(2,w)}(T)\rangle$ is a quasi-stationary composite states (QCS) of features; (1) does not vanish at macroscopic $T$, (2) the continuous  spectrum of kinetic energy, (3) $T$-independent norm at large $T$, Eq. $(\ref{weight2})$, (4) finite interaction energy Eq.$(\ref{interaction-energy0})$.  QCS is accompanied by  the 
decaying particle states 
in the normal cases.  
In an extremely  small $T$, the integral Eq.$(\ref{weight2})$ is not exactly constant, but varies steeply to reach the constant
 \cite{Ishikawa-Tobita-PTEP, Ishikawa-Tobita-ANA}. The final states expressed by QCS appear rapidly, and  its probability $P^{(d)}$  remains at later times.     

In an interaction Eq. $(\ref{total-deivative})$, the coupling at $\omega =0$ vanishes 
$F_{\varphi_{\mu,\beta}}(0)=0$,
and the state  is expressed as 
\begin{eqnarray}
& |\Psi(T) \rangle =N_1^H(T) |\varphi_{\mu} \rangle +    |\Psi_{(2,w)}(T) \rangle,\label{t_derivative_wf}|N_1^H(T)|^2=1-P^{(d)}, \\
& P^{(d)}=\frac{\sqrt{\pi \sigma_1}}{4\pi} g^2  \frac{1}{E_{V} }\int \frac{d^3 p_1}{E_1} \text {log}({ y }) \theta(m_2^2-2P_{V}\cdot p_1) \neq 0, \label{t_derivative_P}
\end{eqnarray}
 where $P_{V}$ is the energy-momentum of the initial state, and $y$ is given in Eq.$(\ref{p^d-distribution})$ \cite{Ishikawa-Tajima-Tobita-PTEP}. 
The state is composed of QCS and the parent, and  varies with time. The  normal
component of the decaying particles $|\Psi_{(2,p)} \rangle $ is not included.

%%%%%%%%%%%%%%%%%%%%%%%%%%%%%%%%%%%%%%%%%%%%%%%%%%%%%%%%%%%
 \subsection{ Quasi Stationary Composite States(QCS): Correlations }
 %%%%%%%%%%%%%%%%%%%%%%%%%%%%%%%%%%%%%%%%%%%%%%%%%%%%%%%%%%
QCS has unique properties which do not arise in the normal transition under ASI.

 %%%%%%%%%%%%%%%%%%%%%%%%%%%%%%%%%%%%%%%%%%%%  
\subsubsection{Kinetic energy vs interaction energy  }
%%%%%%%%%%%%%%%%%%%%%%%%%%%%%%%%%%%%%%%%%%%%
 
From  Eqs.$(\ref{wave-function1})$ and $(\ref{interaction-energy0})$,  the expectation 
values of the  total kinetic energy   at $T_1$ is
\begin{eqnarray}
& & E_\text{kin}(T_1)  =\langle \Psi_1^H (T_1)|H_0| \Psi_1^H(T_1) \rangle =E_2(\vec{ P}) -E_{int} (T_1) \label{energy-expectation}, \\
& &  E_\text{int}(T_1)  =  2\int d{\vec p}_1 \left\{\frac{E_\text{total}}{2}-E({\vec
  p}_1)\right\}\int d{\vec p}_2 |D(\omega,T_1)|^2  |F_{\varphi_2,\beta}(\omega)|^2, \label{interaction-energy}
\end{eqnarray}
where $E_{total}$ is the total energy and agrees with $E_2(\vec{ P}) $. 
In the right-hand side of Eq.$(\ref{interaction-energy})$,    a symmetric nature of the two particle wave function was used.   
Thus the interaction energy depends on the single particle spectrum.
 The $|\Psi_{(2,w)}(T) \rangle$  has    interaction energy and a continuous  kinetic energy which  differs  from that of  an isolated particle.

%%%%%%%%%%%%%%%%%%%%%%%%%%%%%%%%%%%%%%%%%%%%%%%%%%%%%%%%%%%%%%%  
%   \subsubsection{Space-time symmetry of QCS  }
%%%%%%%%%%%%%%%%%%%%%%%%%%%%%%%%%%%%%%%%%%%%%%%%%%%%%%%%%%%%%%%%
\subsubsection{  Space-time symmetry   }

QCS  is a superposition
of a continuous mass spectrum  $P_{\mu}P^{\mu}=(E_1+E_2)^2-({\vec p}_1+{\vec p}_2)^2$, which is that of the visible energy and momentum as is shown later and is not a constant.    This 
differs from a bound state which belongs to an  irreducible representation of a definite mass. 

%%%%%%%%%%%%%%%%%%%%%%%%%%%%%%%%%%%%%%%%%%%%%%%%%%%%%%%%%%
\subsubsection{Order parameter : interaction energy  }
%%%%%%%%%%%%%%%%%%%%%%%%%%%%%%%%%%%%%%%%%%%%%%%%%%%%%%%%
QCS is the state that  has an expectation value of the interaction
Hamiltonian 
\begin{eqnarray}
O_{qcs}(\pm \frac{T}{2})= \lim_{t \rightarrow \pm \frac{T}{2}}\langle \Psi (t)| H_\text{int}(t)| \Psi (t)\rangle. \label{qcs-order}
\end{eqnarray} 
Accordingly,  $O_{qcs}(\pm \frac{T}{2})$  is an order parameter for QCS. 
A state is QCS if  $O_{qcs} (\pm \frac{T}{2})\neq 0$, and is not if $O_{qcs}(\pm \frac{T}{2})=0$.
$|\Psi_{(2,w)}(x) \rangle $ in Eq. $( \ref{p_w_wave})$ and $|\Psi_{(2,w)}(x) \rangle $ in 
Eq. $(\ref{t_derivative_wf})$ have $O_{qcs} (T)\neq 0$ and are  QCS, but
$|\Psi_{(2,p)}(x)\rangle $  has $O_{qcs}(T) = 0$, and  is not. QCS in scattering
processes are also identified by this order parameter. If all the states
of Eq.$(\ref{Schroedinger2})$ at $t \rightarrow \infty$ have the value $O_{qcs}(\infty)=0$, QCS 
does not appear, and  states are composed of particle-like states.  
  
In ASI, 
\begin{eqnarray}
O_{qcs}(\pm \infty)&=&\lim_{t \rightarrow \pm \infty}\langle \Psi(t) |e^{-\epsilon
 |t|}H_\text{int}| \Psi(t) \rangle =0. 
\end{eqnarray}
Accordingly all the asymptotic states are  particle-like states that
 satisfy  the kinetic-energy
conservation.    If ASI is not required,  there may be states of $O_{qcs} (\pm \infty)=0$ and those of  $O_{qcs} (\pm \infty) \neq 0$.    The amplitude and probability   in the space of 
$O_{qcs}(\pm \infty)=0$, and  of $O_{qcs} (\infty)\neq 0$ describe the physical processes. 
   Contributions from the latter are obtained without ambiguity  in the tree level of the perturbative expansion, where  the scattering is studied without assuming   ASI.  Accordingly, the method of Tomonaga-Schwinger-Feynman \cite{Tomonaga, Schwinger, Feynman} developed for  the higher
 order corrections in QED under  ASI  is not applied, but a naive   Schr\"odinger equation is applied.   Using it,   the wave functions  for the interaction Hamiltonian $H_\text{int}$ instead of $e^{-\epsilon|t|}H_\text{int} $ are obtained and the transition probabilities  are computed.   
 
%%%%%%%%%%%%%%%%%%%%%%%%%%%%%%%%%%%%%%%%%%%%%%%%%%%%%%%%

%%%%%%%%%%%%%%%%%%%%%%%%%%%%%%%%%%%%%%%%%%%%%%%%%%%%%%%%%%%%%%%
\subsubsection{Norm  }
%%%%%%%%%%%%%%%%%%%%%%%%%%%%%%%%%%%%%%%%%%%%%%%%%%%%%%%%%%%%
The norm of the initial state and those of the final states are modified at  later times, and affect the transition probability.  Higher order terms in $g$ relevant 
to the probability are found.  The  operator in  Eq.$(\ref{solution})$ is written    as,
\begin{eqnarray}
& &U(t,t_0)=1+iK(t,t_0),\ K(t,t_0)=\int_{t_0}^{t} dt_1 \tilde K(t_1),\label{unitarity1}\\ 
& &i(K(t,t_0)-K^{\dagger}(t,t_0))+K(t,t_0) K^{\dagger}(t,t_0)=0, \nonumber
\end{eqnarray}
where $\tilde K(t_1)=H_\text{int}(t_1)+\int_{t_0}^{t_1} dt_2H_\text{int}(t_1)H_\text{int}(t_2)+\dots $, is the integrand of $K(t,t_0)$.
Substituting a one-particle  state $|\alpha \rangle$ and a complete set $|n \rangle$ of approximate eigenstates of $H_0$ of the energies $E_{\alpha}$ and $E_n$,
\begin{eqnarray}
& &\langle n| K(t,t_0)| \alpha \rangle=D(\omega,\delta t) \langle n| \bar  K(t_0)| \alpha \rangle, \ \omega=E_{\alpha}-E_n,\ \delta t=t-t_0,\\
& & \langle \alpha |K(t,t_0) K^{\dagger}(t,t_0) |\alpha \rangle = |D(\omega,\delta t)|^2 \langle
| \alpha| \bar K(t_0)|n \rangle|^2 ,          \nonumber
\end{eqnarray}
where the phase factors of the states are extracted to define $D(\omega, \delta t)$, and the rest is denoted as  $\bar K(t_0)$,   
\begin{eqnarray}
\langle \alpha| K(t,t_0)-K^{\dagger}(t,t_0)| \alpha \rangle=-i|D(\omega,\delta t )|^2 \langle
| \alpha| \bar K(t_0) | n \rangle|^2.
\end{eqnarray}

%%%%%%%%%%%%%%%%%%%%%%%%%%%%%%%%%%
\subsection{ Transition probability }
%%%%%%%%%%%%%%%%%%%%%%%%%%%%%%%%%%%%
\subsubsection{Wave packets of $H_0$}
 We assume that   $|\Psi_\alpha (0)\rangle $ is  a normalized state defined by  an approximate  eigenstate of $H_0$ of the central energy $E_0$, and  $|\Psi_\beta (T)\rangle $ at $t=T$ be  normalized 
and an approximate  eigenstate of the central energy  $E_1$,  then a product of their matrix element, 
\begin{eqnarray}
P_{\beta,\alpha}(T)= |\langle \Psi_{\beta} (T)| \Psi_{\alpha} (0) \rangle|^2=  |  | D(\Delta E,T)| \langle \beta  |{H_{int}}|  \Psi_{\alpha}(0) \rangle|^2 \label{norm2},
\end{eqnarray}
where $\Delta E=E_1-E_0$ satisfies 
\begin{eqnarray}
0 \leq P_{\beta,\alpha} (T) \leq 1, \sum_{\beta}P_{\beta,\alpha} (T) =1.
    \end{eqnarray}
  By    substituting the relation for the large $T$
 \begin{eqnarray}
 \frac{|D( \Delta E,T)|^2-|D(\Delta E,T_1)|^2  }{T-T_1}= 2\pi \delta (\Delta E),
 \end{eqnarray}
  the summation over $\beta$ of these  around  a center $\bar \beta$ ,
 $ P_{\bar \beta \alpha}(T)=\sum_{\beta;\bar \beta} P_{\beta,\alpha}$  is expressed from Eq.$(\ref{first-order})$ for a large $T$ as  
\begin{eqnarray}
& &P_{\bar \beta \alpha}(T)=P_{\bar \beta \alpha}(T_1)+(T-T_1)  \Gamma_{\bar \beta \alpha}\\
& & ~~\Gamma_{\bar \beta \alpha}= \int_{\bar \beta} d {\beta}  |\langle\beta  |{H_{int}}| \Psi_{\alpha}( 0) \rangle|^2  2 \pi \delta( \Delta E).   \nonumber
\end{eqnarray}
  The coefficient  $ \Gamma_{\bar \beta \alpha}$ is derived from the wavefunctions of $\Delta E=0$.  $P_{\bar \beta \alpha}(0)=0$ but $\frac{ d^2 P_{\bar \beta \alpha}(T_1)}{d T_1^2}$ for the plane waves  corresponds to the integral Eq.$(\ref{plane-wave-})$ of $q=2$, and diverges   at $T_1=0$  due to the Stueckelberg  divergence.   Now
$P_{\bar \beta \alpha}(T_1)-\Gamma_{\bar \beta \alpha} T_1$ with the normalized states     is convergent  \cite{Ishikawa-Tobita-PTEP}. Next we study  the
 many-body wavefunctions explicitly.

        $P^{(d)}$ is the contribution form the boundary of the interval  $\sqrt {2 \sigma_t }$, and its convergence  is proved in Appendix B.   Hence, $T_1=\sqrt{ 2\sigma_t}$.  For a case that the wave packet of the final state $\sigma_1$ and that of the initial state $\sigma_2$, and   $ { \sigma_1 }  < \sigma_2$,  $P^{(d)}$ is computed easily at a high energy region with  the correlation function obtained by  
 interchanging the order of the integrations over the space-time positions and momenta, in which   the light-cone singularity derived from 
  $|{\vec p}_2| \rightarrow \infty $  gives  the most important
  contribution \cite{Ishikawa-Tobita-PTEP}, 
 \begin{eqnarray}
P^{(d)}= \frac{8g^2}{E_2({\vec P})}
 \sigma_1\int \frac{d^3 p_1}{(2\pi)^3 E({\vec p}_1)} T_1\tilde g(y) \theta(m_2^2-m_1^2-2P \cdot p_1), y=\frac{m_1^2 T_1}{2E_1({\vec p}_1)},\label{p^d-distribution}
\end{eqnarray}
where 
 the function $\tilde g(y ) $ is derived from the light-cone singularity and proportional to $\frac{1}{y}$ at  large $y$ region and is $\pi$  at small $y$.  $T_1 \tilde g(y)$ is independent of $T_1$ in the former region and  is proportional to $T_1$ in the latter region. This   component shows a wider energy distribution which 
 shifts  to   lower energy region  than $\Gamma$  \cite{Ishikawa-Tobita-PTEP,Ishikawa-Tobita-ANA,von neumann}.

The norm of the parent and  the expectation value of the interaction Hamiltonian of the state $| \Psi_2(T) \rangle $, which we call interaction energy for brevity,   up to  $g^2$ are, 
\begin{eqnarray}
& &|N_1^H(T)|^2=1 -  T \Gamma -P^{(d)} \nonumber\\   
& &E_{int}(T)= \langle \Psi_2(T)|H_{int}|  \Psi_2(T)  \rangle=\int d {\vec p}_1 d{\vec p}_2  \omega |F_{\varphi_2,\beta}(\omega)|^2  |D(\omega,T)|^2. \label{interaction-energy0}
\end{eqnarray}
$|\Psi_{(2,p)}(T) \rangle$ denotes  the state of $|\omega| \approx 0 $, and  $|\Psi_{(2,w)}(T) \rangle$ denotes the state  of $|\omega| \neq 0 ( > \frac{1}{\sqrt{ 2 \sigma_t}}) $, 
and satisfy
\begin{eqnarray}  
& & \frac{\langle \Psi_{(2,p)} (T)|H_{int}| \Psi_{(2,p)}(T) \rangle}{\langle  \Psi_{(2,p)} (T)| \Psi_{(2,p)}(T) \rangle} \leq  \frac{1}{\sqrt{ 2\sigma_t}} , \\
& & \frac{\langle \Psi_{(2,w)} (T)|H_{int}| \Psi_{(2,w)}(T) \rangle}{ \langle \Psi_{(2,w)} (T)| \Psi_{(2,w)}(T) \rangle} > \frac{1}{\sqrt{ 2 \sigma_t}} . \nonumber 
\end{eqnarray}  
The former 
 is included in ASI  \cite{Kayser,
 Giunti,Nussinov,Kiers,Stodolsky,Lipkin}. Now, 
 $|\Psi_{(2,w)}(T) \rangle$  is   a correlated  state similar to a stationary 
 state, and gives various physical effects  
 \cite{Ishikawa-Tobita-PTEP,Ishikawa-Tobita-ANA,
 Ishikawa-Tajima-Tobita-PTEP,G-W,Asahara}. 
 
 $|\Psi_{(2,w)}(T)\rangle$ is a quasi-stationary composite states (QCS) of features; (1) does not vanish at macroscopic $T$, (2) the continuous  spectrum of kinetic energy, (3) $T$-independent norm at large $T$, Eq. $(\ref{weight2})$, (4) finite interaction energy Eq.$(\ref{interaction-energy0})$.  QCS is accompanied by  the 
decaying particle states 
in the normal cases.  
In an extremely  small $T$, the integral Eq.$(\ref{weight2})$ is not exactly constant, but varies steeply to reach the constant
 \cite{Ishikawa-Tobita-PTEP, Ishikawa-Tobita-ANA}. The final states expressed by QCS appear rapidly, and  its probability $P^{(d)}$  remains at later times.     

In an interaction Eq. $(\ref{total-deivative})$, the coupling at $\omega =0$ vanishes 
$F_{\varphi_{\mu,\beta}}(0)=0$,
and the state  is expressed as 
\begin{eqnarray}
& |\Psi(T) \rangle =N_1^H(T) |\varphi_{\mu} \rangle +    |\Psi_{(2,w)}(T) \rangle,\label{t_derivative_wf}|N_1^H(T)|^2=1-P^{(d)}, \\
& P^{(d)}=\frac{\sqrt{\pi \sigma_1}}{4\pi} g^2  \frac{1}{E_{V} }\int \frac{d^3 p_1}{E_1} \text {log}({ y }) \theta(m_2^2-2P_{V}\cdot p_1) \neq 0, \label{t_derivative_P}
\end{eqnarray}
 where $P_{V}$ is the energy-momentum of the initial state, and $y$ is given in Eq.$(\ref{p^d-distribution})$ \cite{Ishikawa-Tajima-Tobita-PTEP}. 
The state is composed of QCS and the parent, and  varies with time. The  normal
component of the decaying particles $|\Psi_{(2,p)} \rangle $ is not included.

{\bf Single particle spectrum}

A single-particle distribution of the particle $\varphi_1$ from $P^{(d)}$ for large $y$ is 
  \begin{eqnarray}
   \frac{ d P^{(d)}({\vec p}_1,T_1)}{d {\vec p}_1}=\frac{8g^2}{E_2({\vec P})}
 \sigma_1  \frac{1}{(2\pi)^3 }  \frac{1}{m_1^2} \theta(m_2^2-m_1^2-2P \cdot p_1),\label{p^d-distribution2}
  \end{eqnarray}
from   Eq. $(\ref{p^d-distribution})$ .  
   The average kinetic energy is  about a 
half of the kinetic-energy-conserving value. Accordingly the
interaction 
energy is positive, and the total
energy $E_{total}$ is larger than an
energy  estimated  from the single-particle distribution by
about a factor two. A kind of Virial theorem holds. 
\subsubsection{Wave packets of $H$: Uniqueness of the scattering amplitude}
%%%%%%%%%%%%%%%%%%%%%%%%%%%%%%%%%%%%%%%%%%%%%%%
% \subsection{Uniqueness of the scattering amplitude: Eigenstates of $H$}
%%%%%%%%%%%%%%%%%%%%%%%%%%%%%%%%%%%%%%%%%%%%%%
We show that  the scattering amplitude for  wave packets  defined from the eigenstates  of $H$,
\begin{eqnarray}
& &H|E({\vec k}) ,{\vec k} \rangle = E({\vec k}) | E({\vec k}), {\vec k} \rangle,\\
& &H|E({\vec p}_1+{\vec p}_2) ,{\vec p}_1,{\vec p}_2 \rangle = E({\vec p}_1,{\vec p}_2) | E({\vec p}_1+{\vec p}_2), {\vec p}_1,{\vec p}_2 \rangle, \nonumber
\end{eqnarray}
is equivalent to that  defined from the eigenstates of $H_0$.   The eigenstates     in the perturbative expansions with respect to  $H_{int}$  are   
\begin{eqnarray}
& &|E({\vec k}) ,{\vec k} \rangle =  |  {\vec k} \rangle+ \mathcal{P}\int d{\vec p}_1 d{\vec p}_2 \frac{1}{E({\vec k})-E({\vec p}_1)-E({\vec p}_2)} |{\vec p}_1 ,{\vec p}_2 \rangle \langle {\vec p}_1,{\vec p}_2|H_{int}| {\vec k}  \rangle, \label{eigen-state} \\
& &|E({\vec p}_1+{\vec p}_2),{\vec p}_1,{\vec p}_2  \rangle =  |  {\vec p}_1,{\vec p}_2 \rangle+ \mathcal{P}\int d{\vec k}  \frac{1}{-E({\vec k})+E({\vec p}_1)+E({\vec p}_2)} |{\vec k}  \rangle \langle {\vec k} |H_{int}| {\vec p}_1, {\vec p}_2  \rangle, \nonumber
\end{eqnarray}
up to the first order , where $\mathcal{P}$ stands for the principle value, and $|E({\vec k}),{\vec k} \rangle $ is the one particle state of the heavier scalar and $|E({\vec p}_1+{\vec p}_2),{\vec p}_1,{\vec p}_2 \rangle $ is the two particle state of the lighter scalar.
These are orthogonal each other 
\begin{eqnarray}
\langle E({\vec k}),{\vec k}|E({\vec p}_1+{\vec p}_2),{\vec p}_1,{\vec p}_2 \rangle =0,
\end{eqnarray}
but are not normalized.  Wave packets constructed from the states of Eq.$(\ref{eigen-state})$ as 
\begin{eqnarray}
& &| \Psi_{\alpha} \rangle =\int d{\vec k} w({\vec P},{\vec X},{\vec k}) |E({\vec k}), {\vec k} \rangle,  \\
%=\int d{\vec k} w({\vec P},{\vec X},{\vec k}) |{\vec k} \rangle \label{wave-packet_H_2}\\
%& &+\int d{\vec k} w({\vec P},{\vec X},{\vec k}) \mathcal{P}\int d{\vec p}_1 d{\vec p}_2 \frac{1}{E({\vec k})-E({\vec p}_1)-E({\vec p}_2)} |{\vec p}_1 ,{\vec p}_2 \rangle \langle {\vec p}_1,{\vec p}_2|H_{int}| {\vec k}  \rangle \nonumber \\
& &| \Psi_{\beta} \rangle =\int d{\vec p}_1 d {\vec p}_2 w({\vec P}_1,{\vec X}_1,{\vec p}_1) w({\vec P}_2,{\vec X}_2,{\vec p}_2) |E({\vec p}_1+{\vec p}_2), {\vec p}_1,{\vec p}_2 \rangle, %\langle \Psi_{\alpha}| \Psi_{\beta} \rangle =0 
 \nonumber
%=\int d{\vec k} w({\vec P},{\vec X},{\vec k}) |{\vec k} \rangle \label{wave-packet_H_3} \nonumber\\
%& &+\int d{\vec k} w({\vec P},{\vec X},{\vec k}) \mathcal{P}\int d{\vec k}   \frac{1}{-E({\vec k})+E({\vec p}_1)+E({\vec p}_2)} |{\vec k}  \rangle \langle {\vec k} |H_{int}| {\vec p}_1,{\vec p}_2  \rangle  \nonumber
\end{eqnarray}
are normalized and satisfy $\langle \Psi_{\alpha}| \Psi_{\beta} \rangle =0$.  
 As the  state is evolved as   
\begin{eqnarray}
|\Psi_{\alpha} (T) \rangle =U(T,0) |\Psi_{\alpha}(0) \rangle, 
\end{eqnarray} 
the transition amplitude for these states is given by 
\begin{eqnarray}
\langle \Psi_{\beta}|U(T,0) |\Psi _{\alpha} \rangle=  \langle {\chi}_1 {\chi}_2| \int_0^T \frac{dt}{i \hbar} H_{int}(t)| {\chi}_0  \rangle,
\end{eqnarray} 
where Eq.$(\ref{Schroedinger2})$ was substituted. This  is in agreement with that of the wave packets defined from the eigenstates of   $H_0$. Accordingly, the transition 
amplitude at the finite time interval  is  universal.  

%%%%%%%%%%%%%%%%%%%%%%%%%%%%%%%%%%%%%%%%%%%%%%%%%%%%%%%%%%%%%%%%%%%%%
\subsection{Large T behavior}
\subsubsection{  $\Gamma T$: T-dependent norm}
%%%%%%%%%%%%%%%%%%%%%%%%%%%%%%%%%%%%%%%%%%%%%%%%%%%%%%%%%%%%%%%%%%%%%
From  Eq.$(\ref{unitarity1})$,
\begin{eqnarray}
& &2 Im \langle \alpha| K(t,t_0) | \alpha \rangle= 2\pi\delta (0)\Gamma
 =\delta t \Gamma,  \\
& &\Gamma=2\pi \int d \beta \delta(\omega) | \langle \beta |\bar K(t_0) |\alpha
 \rangle |^2, \nonumber
\end{eqnarray}
where $\beta$ represents a final state.
From the unitarity relation  at an arbitrary $\delta t$,
\begin{eqnarray}
|\langle \alpha |U(t,t_0) |\alpha \rangle|^2=1-\int_{\beta \neq \alpha} d \beta |\langle \beta
 |U(t,t_0)|\alpha \rangle|^2.
\end{eqnarray}
For a  small $|K(t,t_0)|$,
 \begin{eqnarray}
|\langle \alpha |U(t,t_0) |\alpha \rangle|^2=1-\int_{\beta \neq \alpha} d \beta |\langle \beta
 |U(t,t_0)|\alpha \rangle|^2=1-\delta t \Gamma. \label{small P}
\end{eqnarray}
For  $1-\delta t \Gamma <0$,  Eq.$(\ref{small P})$ is invalid, and higher order terms are added. For $|\langle \alpha |U(t,t_0)|\alpha
\rangle | \gg
|\langle \beta|U( t,t_0)| \alpha \rangle|$ for $\beta \neq \alpha$, ignoring the off-diagonal
matrix element for a positive small $t'-t$, 
\begin{eqnarray}
 |\langle \alpha|U(t',t_0)|\alpha \rangle |^2 = |\langle \alpha|U(t',t)
  U(t,t_0)|\alpha \rangle |^2 = |\langle \alpha|U(t,t_0)|\alpha \rangle
  |^2(1-(t'-t) \Gamma).
\end{eqnarray}
 Integrating over the time difference  for a large $T$, it follows that
\begin{eqnarray}
& &|\langle \alpha |U(t+T,t) |\alpha \rangle|^2=  e^{-T \Gamma} \label{exponential1}
,\\
& &\sum_{\beta \neq \alpha}|\langle \beta |U(t+T,t) |\alpha \rangle|^2= 1-e^{-T \Gamma}. \label{exponential2} \nonumber
\end{eqnarray}
This is equivalent to Weisskopf- Wigner formula \cite{Weisskopf-Wigner}. 
%%%%%%%%%%%%%%%%%%%%%%%%%%%%%%%%%%%%%%%%%%%%%%%%%%%%%%%%%%%%%%%%%%%%%%%
\subsubsection{ $P^{(d)}$:T-independent  norm}
 %%%%%%%%%%%%%%%%%%%%%%%%%%%%%%%%%%%%%%%%%%%%%%%%%%%%%%%%%%%%%%%%%%%%%
For the interaction Eq. $(\ref{total-deivative})$,  
\begin{eqnarray}
 | \langle \beta |H_{int} |\alpha \rangle |_{\omega=0}=0.
\end{eqnarray}
The transition occurs to the states of  $\omega \neq 0$, and the probability varies   
 rapidly in small  $t$ from Eq.$( \ref{t_derivative_P})$.
For a small $|K(t,t_0)|$, a variation of the initial state is negligible, and 
\begin{eqnarray}
& &\int_{\omega \neq 0} d \beta |\langle \beta |U(t,t_0)| \alpha \rangle|^2
 =P^{(d)}, \\
& & |\langle \alpha |U(t,t_0)|\alpha \rangle |^2=1-P^{(d)}. \nonumber
\end{eqnarray}
For $1-P^{(d)} <0$, which occurs at large $K(t,t_0)$, the higher order effect of the $t$-independent $P^{(d)}$    is added. That becomes  to $1-P^{(d)}+(-P^{(d)})^2+\cdots$,  and
\begin{eqnarray}
& &\int_{\omega \neq 0} d \beta |\langle \beta |U(t,t_0)| \alpha \rangle|^2
 =\frac{P^{(d)}}{1+P^{(d)}}, \label{power-law}\\
& & |\langle \alpha |U(t,t_0)|\alpha \rangle |^2=\frac{1}{1+P^{(d)}}. \nonumber
\end{eqnarray}
The  initial state  $|\alpha \rangle $ and the final state $|\beta \rangle$ co-exist at later times.  This feature that the initial norm decreases with $P^{(d)}$ substantially 
 arose also 
  in the the decay $0^{-}
 \rightarrow l+\nu$ of violating  the helicity suppression \cite{Ishikawa-Tobita-PTEP,Ishikawa-Tobita-ANA} and
 in the decay $1^{+} \rightarrow 2\gamma$  of violating  
the Landau-Yang's theorem 
\cite{Ishikawa-Tajima-Tobita-PTEP}.     $\Gamma$ is suppressed, but $P^{(d)}$  is not suppressed.
The rapid change in  an extremely  small  $T$ region is a feature of $P^{(d)}$.
%%%%%%%%%%%%%%%%%%%%%%%%%%%%%%%%%%%%%%%%%%%%%%%%%%%%%%%%%%%%%%%%%%%%%%%
\subsubsection{ $\Gamma$ and $P^{(d)}$}
%%%%%%%%%%%%%%%%%%%%%%%%%%%%%%%%%%%%%%%%%%%%%%%%%%%%%%%%%%%%%%%%%%%%%%%
  For  general case of $P^{(d)} \gg \Gamma T$,  
\begin{eqnarray}
& &|\langle \alpha| U(T+t,t) |\alpha \rangle|^2 =\frac{ 1}{ 1+P^{(d)}} e^{-\Gamma T}, \label{two-components}\\
& &\sum_{\beta,\omega \approx 0}|\langle  \beta^{(n)}| U(T+t,t) |\alpha \rangle|^2 =\frac{ 1}{ 1+P^{(d)}}
 (1-e^{-\Gamma T}), \nonumber \\
& &\sum_{\beta,\omega \neq 0}|\langle  \beta^{(d)} | U(T+t,t) |\alpha \rangle|^2 =\frac{ P^{(d)}}{ 1+P^{(d)}}
 \nonumber, 
\end{eqnarray}
where $|\beta^{(n)} \rangle $ and   $|\beta^{(d)} \rangle $ are the kinetic-energy-conserving 
and    non-conserving states.   
At a large $\Gamma T$,  the initial state disappears, and the final states include the states of
conserving kinetic energy and those of non-conserving it.  The
former has the kinetic energy $E_{\alpha}$ and the latter is assumed 
$E_{\beta} \approx \frac{E_{\alpha}}{r_{ke}}$, \cite{Ishikawa-Tobita-PTEP}, where $r_{ke}>1$ . In that case, the ratio of the visible energy
over the total energy is
\begin{eqnarray}
r_\text{visible}= \frac{E_{visible}}{E_{total}}=\frac{1}{1+P^{(d)}} ( \frac{P^{(d)}}{r_{ke}}+1), 
\end{eqnarray}
which becomes the unity at $P^{(d)}=0$, and agrees with $\frac{1}{r_{ke}}$  at $P^{(d)} =\infty$.

%%%%%%%%%%%%%%%%%%%%%%%%%%%%%%%%%%%%%%%%%%%%%%%%%%%
\subsection{Higher order effects :renormalization}
%%%%%%%%%%%%%%%%%%%%%%%%%%%%%%%%%%%%%%%%%%%%%%%%%
 In higher order corrections with respect to the coupling constant, the integration  are made over more than two space-time positions. A boundary term of the integral of  the second variable, which  is proportional to $\frac{1}{ \delta \omega}$, appears at  large momenta of the intermediate states. The phase factor $\Phi$ which depends on the energy- momentum  and the spatial position of the intermediate states remains in the integrand, and the integrals over these variables  for  the loop
  diagrams diverge.  The divergence due to the intermediate states  for the wave packets is equivalent to  that  for the plane waves  from the completeness.  
     These divergences  caused by the fluctuations of the infinite momentum arise 
in the extremely  short-distance region, and are subtracted by the local counter terms in the Lagrangian Eq.$(\ref{counter-term} )$.   
   Thus  the ultraviolet divergences   due to the intermediate states are subtracted and both of   $\Gamma$ and $P^{(d)}$ are expressed  by the renormalized quantities.

%%%%%%%%%%%%%%%%%%%%%%%%%%%%%%%%%%%%%%%%%%%%%%%%%%%%%%%%%%%%%%%%%%%%%%
%%%%%%%%%%%%%%%%%%%%%%%%%%%%%%%%%%%%%%%%%%%%%%%%%%%%%%%%%%%%%%%%%%%%%% 
\section{Quantum Electro-Dynamis (QED)    }
%%%%%%%%%%%%%%%%%%%%%%%%%%%%%%%%%%%%%%%%%%%%%%%%%%%%%%%%%%%%%%%%%

The wave functions of QED at a finite time interval are studied with the wave packets. The vacuum and one particle states are equivalent to those in ASI, but   a new component of 
  overlapping waves of finite interaction energy appears in two particle states.

The  QED  
Lagrangian    
\begin{eqnarray}
L=\bar \psi(x)(p_{\mu}\gamma^{\mu}-m_e)\psi(x)- e \bar \psi(x) \gamma^{\mu} \psi(x)
 A_{\mu}(x)-{1 \over 4}F_{\mu \nu}F^{\mu \nu},
\end{eqnarray}
is written with the renormalized fields,
\begin{eqnarray}
\psi(x)=Z_2^{1/2} \psi_r(x), \bar \psi(x)= Z_2^{1/2} \bar \psi_r(x), A_{\mu}(x)=Z_3^{1/2} A_{\mu}(x)_r
\end{eqnarray}
and decomposed into the free parts and the interaction parts
\begin{eqnarray}
& &L=  \bar \psi^r(x)(p_{\mu}\gamma^{\mu}-m_e^r)\psi^r(x)-{1 \over 4} F_{\mu \nu}^r {F^{\mu \nu}}^r +Z_2 \delta m \bar \psi^r(x)  \psi^r(x)\nonumber\\
 & &+(Z_2-1) \bar \psi^r(x)(p_{\mu}\gamma^{\mu}-m_e^r)\psi^r(x)- eZ_2 Z_3^{1/2}\bar \psi^r(x) \gamma^{\mu} \psi^r(x)
 A_{\mu}^r(x)-{1 \over 4}(Z_3-1) F_{\mu \nu}^r {F^{\mu \nu}}^r. \nonumber \\
 & & \label{renormalized-QED}
\end{eqnarray}
In the above equations, $m_e^r$ and $eZ_2Z_3^{1/2}$ are the physical  mass and charge of the electron and   $Z_2-1$, $Z_3-1$, and $\delta m$ are order $e^2$ and higher orders.
The  wave functions in  a tree level, i.e., the lowest order in $\hbar$,  is   studied     in this section by substituting the  free part  $ H_0$ and the interaction part $  H_{int}$  to the solution Eq.$(\ref{solution} )$. Connections with  the renormalization  procedure will be given   in  a latter section.
%%%%%%%%%%%%%%%%%%%%%%%%%%%%%%%%%%%%%%%%%%%%
%%%%%%%%%%%%%%%%%%%%%%%%%%%%%%%%%%%%%%%%%%%
%\subsection{Wave packet representation }
%%%%%%%%%%%%%%%%%%%%%%%%%%%%%%%%%%%%%%%%%%%%%

Hereafter $\psi(x)$ and $A_{\mu}(x)$ express the renormalized fields.  These  fields  in the interaction representation are expanded  with the wave packets, 
\begin{eqnarray}
& &\psi(x)= \sum_l \int d\chi (\mathcal U({\vec x};\chi,l)B(\chi,l)+\mathcal V^{\dagger}({\vec x};\chi,l) D^{\dagger}(\chi,l)), \label{QED-fields}\\
& &A_{\mu}(x)=\sum_s \int d\chi (\mathcal{E}_{\mu}({\vec x};\chi,s) A(\chi,s)+\mathcal{E}_{\mu}^{*}({\vec x};\chi,s)A^{\dagger}(\chi,s)), \nonumber 
\end{eqnarray}
with spinors and polarization vectors  
\begin{eqnarray}
& &\mathcal U({\vec x};\chi,l)=\int d{\vec k}  \rho_e({\vec k})e^{{i {\vec k}\cdot ({\vec x}-{\vec X}_e) -i E_e({\vec k})(t-T_0)} -\frac{\sigma_e}{2}({\vec k}-{\vec P}_e )^2}  u({\vec k},l) , \\
& &  \mathcal V^{\dagger}({\vec x};\chi,l)=\int d{\vec k}   \rho_e({\vec k}) e^{{i {\vec k} \cdot ({\vec x}-{\vec X}_{\bar e}) -i E_e({\vec k})(t-T_0)} -\frac{\sigma_e}{2}({\vec k}-{\vec P_{\bar e}} )^2}  v^{\dagger}({\vec k},l),  \nonumber \\
& &\mathcal{E}_{\mu}({\vec x};\chi,s)= \int d{\vec k} \rho_{\gamma}({\vec k})e^{{i {\vec k} \cdot ({\vec x}-{\vec X}_{\gamma}) -i E_{\gamma}({\vec k})(t-T_0)} -\frac{\sigma_{\gamma}}{2}({\vec k}-{\vec P}_{\gamma} )^2}  \epsilon_{\mu}({\vec k},s), \nonumber
\end{eqnarray}
where $ \rho_{\gamma}({\vec
 k})=\left(\frac{1}{2E_{\gamma}(\vec{k}\,)(2\pi)^3}\right)^{\frac{1}{2}}$ and $\rho_{e}({\vec k})=\left(\frac{m_e}{E_e({\vec k}) (2\pi)^3}\right)^{\frac{1}{2}}$.  ${\sigma_{\gamma}}$ is the range in space covered by 
 the nucleus or atomic  wave function that the photon interacts with,  and  $\sigma_e$ is that for the electron,
which  are  estimated later.  
%%%%%%%%%%%%%%%%%%%%%%%%%%%%%%%%%%%%%%%%%%%%%%%%%%%%%%%
  \subsection{Vacuum and one-body states }
%%%%%%%%%%%%%%%%%%%%%%%%%%%%%%%%%%%%%%%%%%%%%%%%%%%%%%%
   The Schr\"{o}dinger  equation Eq.$(\ref{Schroedinger2})$  for QED is expressed by 
\begin{eqnarray}
H_{int}=e \int d{\vec x} \bar \psi(x) \gamma_{\mu} \psi(x) A^{\mu}(x), 
\end{eqnarray}
in which Eq.$(\ref{QED-fields})$ is substituted, and is solved with   the unitary operator $U(t,t_0)$ of Eq.$(\ref{solution})$.
%%%%%%%%%%%%%%%%%%%%%%%%%%%%%%%%%%%%%%%%%%%%
\subsubsection{Stationary states: vacuum and one particle states}
%%%%%%%%%%%%%%%%%%%%%%%%%%%%%%%%%%%%%%%%%%%%%
%\subsubsection{vacuum}
The solution satisfying the Schroedinger equation and the initial condition  $|0 \rangle $, which
 satisfies $  B(\chi,l)|0 \rangle =D(\chi,l)|0 \rangle =A(\chi,l)|0 \rangle =0$ at $t=0$ is, 
\begin{eqnarray}
| \Psi_0(T) \rangle =(N_0(T) +\int d^3 \chi N_3(\chi,T)A^{\dagger}(\chi_1,s_1) B^{\dagger}(\chi_2,s_2) D^{\dagger}(\chi_3,s_3)) | 0 \rangle, 
\end{eqnarray}
where the second term shows the state of   the energy gap $E_g=2m_e$. As is shown in Appendix B,  the coefficients are given by
\begin{eqnarray}
N_3(\chi,T)=0, N_0(T)=1.   
\end{eqnarray}
 $  \Gamma=0, P^{(d)}=0$, and the  vacuum is stable.   
%%%%%%%%%%%%%%%%%%%%%%%%%%%%%%%%%%%%%%%%%%%%%
%\subsubsection{One particle states}
%%%%%%%%%%%%%%%%%%%%%%%%%%%%%%%%%%%%%%%%%%%%

An electron and  a positron have the mass $m_e$, and a photon has no  mass.   Due to $U(1)$ symmetry, the electric charge is conserved, and an  electron and  a positron do not decay. A  massless photon  does not decay to a pair of massive electron and positron.  Thus one particle states are stable and  stationary.
%%%%%%%%%%%%%%%%%%%%%%%%%%%%%%%%%%%%%%%%%%%%%%
\subsection{Electron and photon : two particle state}
 %%%%%%%%%%%%%%%%%%%%%%%%%%%%%%%%%%%%%%%%%%%%%
 For an initial state of an electron and a photon of an energy $E _{e_1}+E_{{\gamma}_1}$, final states  of $\Delta E=0$ exist and $\Gamma \neq 0$. If these waves  overlap long,  they have a finite interaction energy  as in the decay, Eq.$(\ref{interaction-energy0})$. The wave component appears and gives $P^{(d)}$ in the scattering probability.  
For a two-particle state of the  
electron and photon at $t=0$
\begin{align}
|\Psi(0) \rangle = A^{\dagger}(\zeta_1,s_{\gamma_1})B^{\dagger}(\chi_2,s_{e_1} )|0 \rangle, \zeta_1= ({\vec P}_{\gamma_1}, {\vec
 X}_{\gamma_1}),   \chi_2=({\vec P}_{e_2}, {\vec X}_{e_2}),
\end{align}
the matrix element of the operator $U(T,0)$ in Eq.$(\ref{solution})$ in the second order in $e$,  is expressed by, 
  \begin{eqnarray}
& &  U(T,0)=\frac{1}{(i\hbar)^2} \int_0^T d t_1 \int_0^{t_1} d t_2 F,  \label{second-order} \\
& &F=\langle e (\chi_2),\gamma (\zeta_2) | H_{int}(t_1) H_{int}(t_2) | e(\chi_1), \gamma (\zeta_1)\rangle \nonumber \\
 & &=\langle  e (\chi_2),\gamma (\zeta_2)|H_{int}(t_1)| e(\chi_3) \rangle  \langle e(\chi_3)|H_{int}(t_2) | e(\chi_1), \gamma (\zeta_1)\rangle \nonumber\\
 & & + \langle e (\chi_2),\gamma (\zeta_2) | H_{int}(t_1) |e(\chi_3),  \gamma(\zeta_1),  \gamma(\zeta_2)  \rangle \langle \gamma(\zeta_1), \gamma(\zeta_2), e(\chi_3)     |H_{int}(t_2) | e(\chi_1), \gamma (\zeta_1)\rangle. \nonumber
   \end{eqnarray} 
The electron state $|e(\chi_3)\rangle $ is an intermediate state and summed over.
 In the case of plane waves,  time-dependent exponential factors  in 
the   $t_1$ and $ t_1-t_2$ are,
\begin{eqnarray}
& &\int_0^T dt_1 e^{i{\delta E_{total} t_1}}  \int_0^T d(t_1-t_2) e^{i \delta E_{rel}^{(\prime)} (t_1-t_2)}, \delta E_{total}=  E_{\gamma_1}+E_{e_1}- E_{\gamma_2}-E_{e_2}, \\
& & \delta E_{rel}=   E_{\gamma_1}+E_{e_1}- E_{e_3} ,    \delta E_{rel}'=   E_{e_1}-E_{e_3}- E_{\gamma_2},    \nonumber 
\end{eqnarray}  
   where $E_{rel}$  and $\delta E_{rel}'$ correspond to  the first and the second terms in Eq.$(\ref{second-order})$ of Fig.2  and Fig.3. %%%%%%Fig%%%%%%%%%%%%%%%%%%%%%%%%%%%%%%%%
%%%%%%%%%%%%%%%%%%%%%%%%%%%%%%%%%%
\begin{figure}[t]
\centering{ \includegraphics[scale=.4,angle=0]{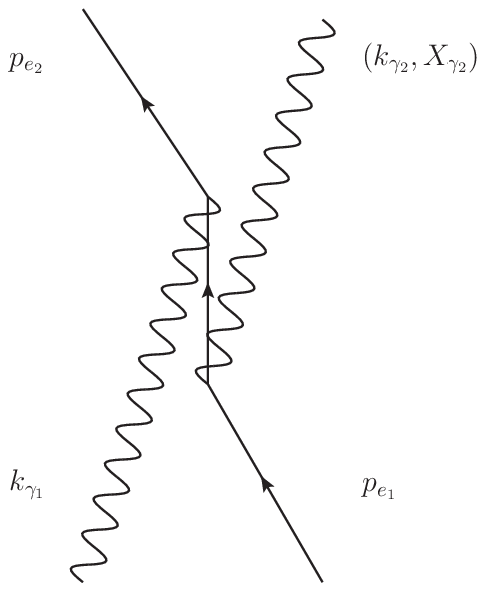}
\caption{One of the diagrams of the Thomson scattering. Solid line shows the electron and wavy line shows the photon.}
}\label{fig:thomson1}
\end{figure}
%%%%%%%%%%%%%%%%%%%%%%%%%%%%%%%%%%%%%%%%%%% 
%%%%%%Fig%%%%%%%%%%%%%%%%%%%%%%%%%%%%%%%%
\begin{figure}[t]
\centering{ \includegraphics[scale=.4,angle=0]{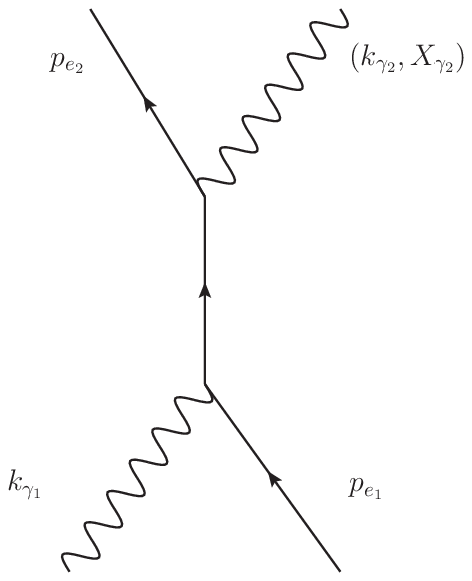}
\caption{One of diagrams of the Thomson scattering. Solid line shows the electron and wavy line shows the photon.}
}\label{fig:thomson}
\end{figure}
%%%%%%%%%%%%%%%%%%%%%%%%%%%%%%%%%%%%%%%%%%% 
The probability from the boundary  in time  converges as in the scalar decay of  Appendix B from the energy denominators $\frac{1}{\delta E_{rel} \delta E_{rel}' } $.
     Neither $\delta E_{rel}=0$ nor $\delta E_{rel}'=0$ has  solution for $E_{e_3}$, and the intermediate state lives a short period.  
     Because the deviations of these integrals  at a finite $T$  from those of  $T=\infty$ are of order   $e^{-\frac{T^2}{\sigma_t}}, \sigma_t \neq 0$, they  can be 
      evaluated  with  $T=\infty$ or with ASI.  Accordingly, the effective interaction for an electron moving in the time-like direction  is expressed by 
      \begin{eqnarray}
      S_{int}=-i \frac{1}{2!}  e^2 \int d^4x d^4y \bar \psi(x) \gamma_{\mu}S_F(x-y)\gamma_{\nu}\psi(y) A_o^{\mu}(x) A_i^{\nu}(y), 
      \end{eqnarray}
where $S_F(x-y)$ is the Feynman propagator, and $A_o^{\mu}(x)$ expresses  the out-going photon and $A_i^{\nu}(y)$ expresses  the in-coming photon. Adding the term where the two photons are interchanged, and substituting $\int d^4(x-y) S_F(x-y)=\frac{i}{m_e}$, the effective interaction  in the long-distance region is 
 \begin{eqnarray}
      S_{int}=\frac{e^2}{2m_e}  \int d^4x \bar \psi(x) g_{\mu \nu} \psi(x) A^{\mu}(x) A^{\nu}(x). 
      \end{eqnarray}
 Thus     the integral over time in the amplitude 
 \begin{eqnarray}
 & &U(T,0)=-i \frac{e^2}{2 m_e } \int_0^T dt_1 d {\vec x}_1  \langle  e (\chi_2)| \bar \psi(x)  \psi(x)  | e(\chi_1) \rangle \langle \gamma (\zeta_2)| A_{\mu}(x)  A^{\mu}(x)  |  \gamma (\zeta_1)\rangle,  
 \end{eqnarray} 
 is almost equivalent to that of the scalar theory, and  composed of  the bulk and boundary terms as in Appendix A.  $\sigma_s$, $\sigma_t$, ${\vec x}_0(t)$, and $T_0$ 
are expressed with the parameters of the one particle states as in Appendix A. 
%%%%%%%%%%%%%%%%%%%%%%%%%%%%%%%%%%%%%%%%%%%%%%%%
\subsubsection{ Integration over  time in  bulk  }
%%%%%%%%%%%%%%%%%%%%%%%%%%%%%%%%%%%%%%%%%%%%%%%%%
The integration over the time $t_1$  in the bulk $T_i +\sqrt {2 \sigma_t} \leq T_0^r \leq T_f -\sqrt{ 2 \sigma_t} $ is made with the Gaussian integral
 and   is proportional to $e^{- \frac{ {\delta \omega} ^2}{2 \sigma_t}}$.
Thus  the amplitude  decreases rapidly with $\delta \omega $ and is equivalent to the golden rule term.

%%%%%%%%%%%%%%%%%%%%%%%%%%%%%%%%%%%%%%%%%%%%%%%
\subsubsection{ Integration over  time at   boundary  }
 %%%%%%%%%%%%%%%%%%%%%%%%%%%%%%%%%%%%%%%%%%%%%%%
The integration in the boundary   $T_0^r \leq T_i +\sqrt{ 2 \sigma_t} $  and  $T_f -\sqrt {2 \sigma_t} \leq T_0^r $  is inversely proportional to $\delta \omega $ and behaves differently from the bulk.
 
 %%%%%%%%%%%%%%%%%%%%%%%%%%%%%%%%%%%%%%%%%%%%%%%%%%%%%%%%
  \section{ Thomson scattering}
%%%%%%%%%%%%%%%%%%%%%%%%%%%%%%%%%%%%%%%%%%%%%%%%%%%%%%%%%%  
%\subsection{Stationary states}
The new component of the wave functions leads  a new component to the transition probability    of the Thomson scattering.  
The low energy theorem, \cite{gell-mann-goldberger} holds, nevertheless.    
Various cases  which  are classified  by  a number of small or large wave packets,  $n_s$ and  $n_l$ denoted  as $(n_s,n_l)$, appear in nature. The
 formula for   $(4,0)$ case is presented first, and others are next.  In ground laboratory,  wave packets are small and $(4,0)$ is  applied normally, but 
 in nature others are realized often. Scattering for the time interval $T$  of $(1,3)$ case shows distinctive properties of observable and is  studied in details. 
%%%%%%%%%%%%%%%%%%%%%%%%%%%%%%%%%%%%%%%%%%%%%%%%%%%%%%%%%%%%%%%%%
   \subsection{Four  small wave-packet; (4,0) case}
%%%%%%%%%%%%%%%%%%%%%%%%%%%%%%%%%%%%%%%%%%%%%%%%%%%%%%%%%%%%%%%%%%%%%%%%%%
For the  wave packets of   photons and electrons,  
the transition amplitude for the time interval $T$
is given by
\begin{eqnarray}
& & \mathcal{M}=- i\frac{e^2 }{2m_e} \int_0^T dt \int d^3x \langle {\vec p}_{e_2},s_{e_2}|
 J(x)|{\vec p}_{e_1} ,{s_e}_1\rangle \langle {\vec k}_{\gamma_2},{\vec
 X}_{\gamma_2},s_{{\gamma}_2},T_{\gamma_2}|A_{\mu}(x) A^{\mu}(x)| {\vec k}_{\gamma_1}, s_{{\gamma}_1}\rangle, \label{amplitude-thomson} \\
& &\langle {\vec k}_{\gamma},{\vec X}_{\gamma},s_{\gamma},T_{\gamma}|A^{\mu}(x)|0\rangle  =N_{\gamma} \int
 d{\vec k}_2 \rho_{\gamma}({\vec k}_2)  e^{-{\sigma_{\gamma}\over 2}({\vec k}_2-{\vec
 k}_{\gamma_2})^2+i(E({\vec k}_2)(t-T_{\gamma_2})-{\vec k}_2\cdot({\vec
 x}-{\vec X}_{\gamma_2}))} \epsilon^{\mu}({\vec k}_{\gamma_2},s_{\gamma_2}), \label{couplings1} \nonumber\\
& &\langle {\vec k_{\gamma_1}},s_{\gamma_1}|A^{\mu}(x)|0\rangle = \rho_{\gamma}({{\vec k}_{\gamma_1}}) 
\epsilon^{\mu}({\vec k}_{\gamma_1},s_{\gamma_1}) e^{{i(E({{\vec k}_{\gamma_1}})t-({{\vec k}_{\gamma_1}}\cdot{\vec
 x} ))}},  \nonumber \\
& &\langle {\vec p}_{e_2},s_{e_2}|J(x)|{\vec p}_{e_1},s_{e_1} \rangle=( 2\pi)^{\frac{3}{2}}
 \rho_{e}({\vec p}_{e_1}) \rho_{e}({\vec p}_{e_2})\bar u_{e} ({\vec p}_{e_1},s_{e_1})  u({\vec p}_{e_2},s_{e_2})
e^{{-i((E_{e_1}-E_{e_2})t-({\vec{p}_{e_1}}-{\vec p}_{e_2})\cdot{\vec{x}})}}. \nonumber
\end{eqnarray}

 The  initial state is normalized,  and  the numerical constant in the last equation has  a coefficient 
$(2\pi)^{\frac{3}{2}}$,  and
\begin{eqnarray}
k_{\mu} \epsilon^{\mu}(k)=0, N_{\gamma}=\left(\frac{\sigma_{\gamma}}{\pi}\right)^{\frac{3}{4}}. 
\end{eqnarray}
 Let    
\begin{eqnarray}
& &\lambda=(t-T_{\gamma})^2-({\vec x}-{\vec X}_{\gamma})^2,\\
& &\xi(x)=\frac{1}{2\sigma_{\gamma}}(({\vec x}-{\vec{X}}_{\gamma})_L-{\vec v}_{\gamma}(t-T_{\gamma}))^2
+\frac{1}{2\sigma_{T}}({\vec x}-{\vec{X}}_{\gamma})_T^2,  \nonumber 
\end{eqnarray}
where $\sigma_{\gamma}^j$ for the longitudinal $L$ and the transverse $T$ directions are given by
\begin{eqnarray}
& &\sigma_{\gamma}^{L}=   \sigma_{\gamma} ,         \\ 
& &\sigma_{\gamma}^{T}= \sigma_{\gamma}- {i  \over E_{\gamma}}(t-T_{\gamma}). \nonumber
\end{eqnarray}

Substituting the integral     over ${\vec k}_2$ of  Eq.$(\ref{amplitude-thomson})$  \cite{Ishikawa-Tajima-Tobita-PTEP},   
$\mathcal M$ is written as   
\begin{eqnarray}
& & \mathcal{M}=- i\frac {e^2 }{2m_e} N \int_0^T dt \int d^3 x e^{-i(p_{e_1}-p_{e_2}+k_{\gamma_1})\cdot x}
 e^{i(E_{\gamma_2}(t-T_{\gamma_2}) -{\vec k}_{\gamma_2}\cdot ({\vec
 x}-{\vec X}_{\gamma_2}))-\xi(x) }  \mathcal{T}, \label{Th-amplitude}\\
& &  \mathcal{T}=\bar u(p_{e_2}) u(p_{e_1})\epsilon^{\mu}({\vec
 k}_{\gamma_1}) \epsilon_{\mu} ({\vec k}_{\gamma_2}+\delta k(x)), \nonumber \\
& &N= ({(2\pi)^3 \over \sigma_{\gamma} \sigma_T^2})^{1/2}
 N_{\gamma}(2\pi)^{3/2} \rho_{\gamma}({\vec k}_{\gamma_1}) \rho_{\gamma}({\vec
 k}_{\gamma_2}+\delta {\vec k}) {1 \over (2\pi)^{3/2}}({m_e^2 \over E_{e_1}
 E_{e_2}})^{1/2}, \nonumber
\end{eqnarray}
where  $\delta k(x)^\mu$ represents corrections due to the expansion of the wave packet and in the leading order  in 
$\frac{1}{\sigma_{\gamma}}$, 
\begin{eqnarray}
\delta k(x)^\mu=0, \sigma_T=\sigma_{\gamma}.
\end{eqnarray}
In this case,  
\begin{eqnarray}
& & \mathcal{M}=- i\frac {e^2 }{2m_e} N    \mathcal{T} I, \label{Th-amplitudes}\\
& &  I= \int_0^T dt \int  d^3 x e^{-i(p_{e_1}-p_{e_2}+k_{\gamma_1})\cdot x}
 e^{i(E_{\gamma_2}(t-T_{\gamma_2}) -{\vec k}_{\gamma_2}\cdot ({\vec
 x}-{\vec X}_{\gamma_2}))-\xi(x) },
 \end{eqnarray}
where $I$ agrees with $I_{bulk}$ of  Eq.$(\ref{Gaussian-integral1})$  or  $I_{boundary}$ of Eq.$(\ref{Gaussian-integral2})$ depending on the central time $T_0^r$. 

The probability is expressed by 
\begin{eqnarray}
P(T)={1 \over V} \int\frac{d{{\vec X}_{{\gamma}_2}}} {(2\pi)^3} d{\vec k}_{{\gamma}_2}
 d{\vec p}_{e_2} \sum_{spin}|\mathcal{M}|^2,  \label{probability-tomson}
\end{eqnarray}
where the normalization volume  $V$ for the initial state expressed by the wave packet size,  \cite{Ishikawa-Tobita-ANA}, satisfies 
\begin{eqnarray}
\frac{1}{V} \int \frac{d{{\vec X}_{{\gamma}_2}}}{(2\pi)^3}=1. \label{normalization}
\end{eqnarray}

The transition probability is finite  and the integrations over the space-time position $(t,{\vec x})$ and of the momentum of the final state can be interchanged.  Thus 
the transition probability is reduced to 
\begin{eqnarray}
& &\int d{\vec p}_{e_2}\sum |\mathcal{M}|^2=(e^2 )^2 N^2 \int d^4 x_1 d^4 x_2 \int d{\vec
 p}_{e_2} 
 e^{-i(p_{e_1}-p_{e_2}+k_{\gamma_1})\cdot (x_1-x_2)}   \nonumber \label{probability-tomson2} \\
& &\times e^{i (E_{\gamma_2}(t_1-t_2) -{\vec k}_{\gamma 2} \cdot
 ({\vec x}_1-{\vec x}_2))-\xi(x_1)-\xi(x_2)}  \sum |\mathcal{T}|^2. 
\end{eqnarray}
The  sum over the final spin and the average over the initial spin  
\begin{eqnarray}
& & \frac{1}{4} \sum_{spin} |\mathcal{T}|^2=2, 
\end{eqnarray} 
is substituted. The integral  over times  in Eq.$(\ref{probability-tomson2})$ consists of  the short-range term derived from the region of $|t_1-t_2| \approx 0$ and the long-range one 
derived from   large $|t_1-t_2|$. The former  gives $\Gamma T$
and the latter gives  $P^{(d)}$. 

%%%%%%%%%%%%%%%%%%%%%%%%%%%%%%%%%%%%%%%%%%%%%%%%%%%%%%%%%%%%%%%%%
   \subsection{QCS scattering; (1,3) case}
%%%%%%%%%%%%%%%%%%%%%%%%%%%%%%%%%%%%%%%%%%%%%%%%%%%%%%%%%%%%%%%%%%%%%%%%%%

The transition amplitude for the time interval $T$
is given by
\begin{eqnarray}
& & \mathcal{M}=- i\frac{e^2 }{2m_e} \int_0^T dt \int d^3x \langle {\vec p}_{e_2},s_{e_2}|
 J(x)|{\vec p}_{e_1} ,{s_e}_1\rangle \langle {\vec k}_{\gamma_2},{\vec
 X}_{\gamma_2},s_{{\gamma}_2},T_{\gamma_2}|A_{\mu}(x) A^{\mu}(x)| {\vec k}_{\gamma_1}, s_{{\gamma}_1}\rangle, \label{amplitude-thomson} \\
& &\langle {\vec k}_{\gamma},{\vec X}_{\gamma},s_{\gamma},T_{\gamma}|A^{\mu}(x)|0\rangle  =N_{\gamma} \int
 d{\vec k}_2 \rho_{\gamma}({\vec k}_2)  e^{-{\sigma_{\gamma}\over 2}({\vec k}_2-{\vec
 k}_{\gamma_2})^2+i(E({\vec k}_2)(t-T_{\gamma_2})-{\vec k}_2\cdot({\vec
 x}-{\vec X}_{\gamma_2}))} \epsilon^{\mu}({\vec k}_{\gamma_2},s_{\gamma_2}), \label{couplings1} \nonumber\\
& &\langle {\vec k_{\gamma_1}},s_{\gamma_1}|A^{\mu}(x)|0\rangle = \rho_{\gamma}({{\vec k}_{\gamma_1}}) 
\epsilon^{\mu}({\vec k}_{\gamma_1},s_{\gamma_1}) e^{{i(E({{\vec k}_{\gamma_1}})t-({{\vec k}_{\gamma_1}}\cdot{\vec
 x} ))}},  \nonumber \\
& &\langle {\vec p}_{e_2},s_{e_2}|J(x)|{\vec p}_{e_1},s_{e_1} \rangle=( 2\pi)^{\frac{3}{2}}
 \rho_{e}({\vec p}_{e_1}) \rho_{e}({\vec p}_{e_2})\bar u_{e} ({\vec p}_{e_1},s_{e_1})  u({\vec p}_{e_2},s_{e_2})
e^{{-i((E_{e_1}-E_{e_2})t-({\vec{p}_{e_1}}-{\vec p}_{e_2})\cdot{\vec{x}})}}. \nonumber
\end{eqnarray}

 The  initial state is normalized,  and  the numerical constant in the last equation has  a coefficient 
$(2\pi)^{\frac{3}{2}}$,  and
\begin{eqnarray}
k_{\mu} \epsilon^{\mu}(k)=0, N_{\gamma}=\left(\frac{\sigma_{\gamma}}{\pi}\right)^{\frac{3}{4}}. 
\end{eqnarray}
 Let    
\begin{eqnarray}
& &\lambda=(t-T_{\gamma})^2-({\vec x}-{\vec X}_{\gamma})^2,\\
& &\xi(x)=\frac{1}{2\sigma_{\gamma}}(({\vec x}-{\vec{X}}_{\gamma})_L-{\vec v}_{\gamma}(t-T_{\gamma}))^2
+\frac{1}{2\sigma_{T}}({\vec x}-{\vec{X}}_{\gamma})_T^2,  \nonumber 
\end{eqnarray}
where $\sigma_{\gamma}^j$ for the longitudinal $L$ and the transverse $T$ directions are given by
\begin{eqnarray}
& &\sigma_{\gamma}^{L}=   \sigma_{\gamma} ,         \\ 
& &\sigma_{\gamma}^{T}= \sigma_{\gamma}- {i  \over E_{\gamma}}(t-T_{\gamma}). \nonumber
\end{eqnarray}

Substituting the integral     over ${\vec k}_2$ of  Eq.$(\ref{amplitude-thomson})$  \cite{Ishikawa-Tajima-Tobita-PTEP},   
$\mathcal M$ is written as   
\begin{eqnarray}
& & \mathcal{M}=- i\frac {e^2 }{2m_e} N \int_0^T dt \int d^3 x e^{-i(p_{e_1}-p_{e_2}+k_{\gamma_1})\cdot x}
 e^{i(E_{\gamma_2}(t-T_{\gamma_2}) -{\vec k}_{\gamma_2}\cdot ({\vec
 x}-{\vec X}_{\gamma_2}))-\xi(x) }  \mathcal{T}, \label{Th-amplitude}\\
& &  \mathcal{T}=\bar u(p_{e_2}) u(p_{e_1})\epsilon^{\mu}({\vec
 k}_{\gamma_1}) \epsilon_{\mu} ({\vec k}_{\gamma_2}+\delta k(x)), \nonumber \\
& &N= ({(2\pi)^3 \over \sigma_{\gamma} \sigma_T^2})^{1/2}
 N_{\gamma}(2\pi)^{3/2} \rho_{\gamma}({\vec k}_{\gamma_1}) \rho_{\gamma}({\vec
 k}_{\gamma_2}+\delta {\vec k}) {1 \over (2\pi)^{3/2}}({m_e^2 \over E_{e_1}
 E_{e_2}})^{1/2}, \nonumber
\end{eqnarray}
where  $\delta k(x)^\mu$ represents corrections due to the expansion of the wave packet and in the leading order  in 
$\frac{1}{\sigma_{\gamma}}$, 
\begin{eqnarray}
\delta k(x)^\mu=0, \sigma_T=\sigma_{\gamma}.
\end{eqnarray}
In this case,  
\begin{eqnarray}
& & \mathcal{M}=- i\frac {e^2 }{2m_e} N    \mathcal{T} I, \label{Th-amplitudes}\\
& &  I= \int_0^T dt \int  d^3 x e^{-i(p_{e_1}-p_{e_2}+k_{\gamma_1})\cdot x}
 e^{i(E_{\gamma_2}(t-T_{\gamma_2}) -{\vec k}_{\gamma_2}\cdot ({\vec
 x}-{\vec X}_{\gamma_2}))-\xi(x) },
 \end{eqnarray}
where $I$ agrees with $I_{bulk}$ of  Eq.$(\ref{Gaussian-integral1})$  or  $I_{boundary}$ of Eq.$(\ref{Gaussian-integral2})$ depending on the central time $T_0^r$. 
The transition probability is finite  and the integrations over the space-time position $(t,{\vec x})$ and of the momentum of the final state can be interchanged.  Thus 
the transition probability is reduced to 
\begin{eqnarray}
& &\int d{\vec p}_{e_2}\sum |\mathcal{M}|^2=(e^2 )^2 N^2 \int d^4 x_1 d^4 x_2 \int d{\vec
 p}_{e_2} 
 e^{-i(p_{e_1}-p_{e_2}+k_{\gamma_1})\cdot (x_1-x_2)}   \nonumber \label{probability-tomson2} \\
& &\times e^{i (E_{\gamma_2}(t_1-t_2) -{\vec k}_{\gamma 2} \cdot
 ({\vec x}_1-{\vec x}_2))-\xi(x_1)-\xi(x_2)}  \sum |\mathcal{T}|^2. 
\end{eqnarray}
The  sum over the final spin and the average over the initial spin  
\begin{eqnarray}
& & \frac{1}{4} \sum_{spin} |\mathcal{T}|^2=2, 
\end{eqnarray} 
is substituted. The integral  over times  in Eq.$(\ref{probability-tomson2})$ consists of  the short-range term derived from the region of $|t_1-t_2| \approx 0$ and the long-range one 
derived from   large $|t_1-t_2|$. The former  gives $\Gamma T$
and the latter gives  $P^{(d)}$.

The  cross section in the low energy
limit agrees with the 
classical Thomson cross section
\begin{eqnarray}
\sigma_\text{Thomson}={8 \pi \over 3} r_e^2,\ r_e= {\alpha \over m_e}.
\end{eqnarray}

%%%%%%%%%%%%%%%%%%%%%%%%%%%%%%%%%%%%%%%%%%%%%%%%%%%%%%%%%%%%%%%%%%%%%%%%%%%%%%%%%%%%%%%%%%
\subsection{  $P^{(d)}_{\gamma}$ in the  (1,3)  case } 
%%%%%%%%%%%%%%%%%%%%%%%%%%%%%%%%%%%%%%%%%%%%%%%%%%%%%%%%%%%%%%%%%%%%%%%%%%%%%%%%%%%%%%%%%%%%

 In $(1,3)$ case, $\sigma_t =\infty$, and $T_1=T$ is substituted. From Eqs.$(\ref{probability-tomson})$, $(\ref{normalization})$, and $( \ref{probability-tomson2})$,   
  $\Gamma$ is derived from the term proportional to  $(2\pi)^4
 \delta^{(4)}(p_{e_1}+k_{\gamma_1}-p_{e_2}-k_{\gamma_2})$,  whereas 
   $P^{(d)}_{\gamma} $ is derived from the states of  $E_f \neq E_i$. This is   computed  easily in the configuration space with the light-cone
    singularity \cite{Wilson-OPE} of Eq.$(\ref{probability-tomson2})$   as in
  decays \cite{Ishikawa-Tobita-PTEP, Ishikawa-Tobita-ANA}.      After tedious calculations, it is found that  
the differential probabilities with respect to ${\vec k}_{\gamma_2}$,   which are integrated over the electron's momentum,  corresponding to  $\Gamma_{Thom} T$ and  $P^{(d)}_{Thom}$  are
\begin{eqnarray}
& & \frac{d \Gamma_{Thom }}{d{\vec k}_{\gamma_2}} T =T\frac{1}{2 E_{e_1}E_{\gamma_1} E_{\gamma_2} E_{e_2} \pi^2}   \delta(E_{e_1}+E_{\gamma_1}-E_{e_2}-E_{\gamma_2})  (e^2)^2 2 , \label{13probability}\\
& &\frac{d P^{(d)}_{Thom,\gamma}}{d {\vec k}_{\gamma_2}}=
 \sigma_{\gamma}\frac{1}{E_{e_1} E_{\gamma_1}  E_{\gamma_2}(2\pi)^3}  e^4 2 T\tilde
 g(\omega_{\gamma_2}T)   \theta(s-m_e^2-2k_{\gamma_2}(p_{e_1}+k_{\gamma_1}) ),  \label{13probability;P^d}
\end{eqnarray}
where  $E_{e_2}=\sqrt{{({\vec k_{\gamma_1}+{\vec p_{e_1}}-{\vec k}_{\gamma_2}})^2+m_e^2}}$, $s=(p_{e_1}+k_{\gamma_1})^2$,  and $\omega_{\gamma_2}=\frac{m_{\gamma}^2}{2E_{\gamma_2}}$. $m_{\gamma}$ is zero in the vacuum and is the photon's effective mass determined by the plasma frequency in medium at high energy.  Matter effect is described  by a refractive index, \cite{Ishikawa-Tajima-Tobita-PTEP} in extreme low energy.   We study the situation that the photon has an  effective mass in this paper.  The  function $\tilde g(\omega_{\gamma_2} T)$ and   the modified phase space  in  $P^{(d)}$ are 
almost equivalent to those in  the decays    Ref.\cite{Ishikawa-Tobita-PTEP, Ishikawa-Tobita-ANA}.  The average energy of the photon in $P^{(d)}$ is lower 
than that of  the golden rule.  
 
In a system of a finite electron density $n_e$, the total probability for one photon to make a transition  is
 \begin{eqnarray}
& &P(total)= n_e ( \Gamma_{Thom} T + P^{(d)}_{Thom,\gamma}).
 \end{eqnarray}
 The transition probability has  $P^{(d)}$ in addition to $\Gamma T$,  which is equivalent   to  the decay probability. If $P(total)$ is not small, the reduction of the 
 photons in the initial states are included, and the total probability behaves as Eq.$( \ref{two-components})$.    
%%%%%%%%%%%%%%%%%%%%%%%%%%%%%%%%%%%%%%%%%%%%%%%%%%%%%%%%%%%%%%%%%%%%%%%%%
\subsubsection{ Electron at rest: photon distribution} 
%%%%%%%%%%%%%%%%%%%%%%%%%%%%%%%%%%%%%%%%%%%%%%%%%%%%%%%%%%%%%%%%%%%%%%%%
The photon distribution 
in the rest system of the electron, $p_{e_1}=(m_e,{\vec 0})$ and the photon of momentum $k_{\gamma_1}=(E_{\gamma_1},0,0,k_{\gamma_1})$ is given
by the
phase space 
\begin{eqnarray}
m_e(E_{\gamma_1}-E_{\gamma_2})-E_{\gamma_1}E_{\gamma_2}(1-\cos \theta) \geq 0, \label{angle-energy}
\end{eqnarray}
where $\theta$ is 
the angle between the final photon and the initial photon.  The integration is made over the region  
\begin{eqnarray}
k_{\gamma_2} \leq k_{\gamma_2}^{max}= \frac{m_e E_{\gamma_1}}{E_{\gamma_1}(1-\cos
 \theta )+m_e}. \label{maximum-energy}
\end{eqnarray}

At a small $\omega_{\gamma_2} T$, $\tilde g(\omega_{\gamma_2} T)=\pi$, the differential
probability and the total
probability are proportional to $T$,
\begin{eqnarray}
& &\frac{d P^{(d)}}{d \cos \theta }=T(\frac {e^2}{2m_e})^2
 \sigma_{\gamma}\frac{1}{ 4\pi }\frac{2k_{\gamma_1}}{  m_e } [\frac{(2m_e +E_{\gamma_1}) ({k_{\gamma_2}^{max}})^2}{2} -\frac{ ({k_{\gamma_2}^{max}})^3} {3}]  ,\label{electron at rest 1} \\
%\begin{eqnarray}
& & \frac{P^{(d)}}{T} =  \Gamma_{eff} \label{electronat rest 2} =   (\frac {e^2}{2m_e})^2
 \sigma_{\gamma} \frac{1}{4 \pi }  2 E_{\gamma_1}^3  [ \frac{
 1}{3} +\frac{3 }{2} \xi 
 +\frac{1}{6}\xi^2 ], ~~\xi =\frac{m_e}{m_e+2E_{\gamma_1}}. 
\end{eqnarray}
Depending on the size $\sigma_{\gamma} $, $\Gamma_{eff} $ varies, 
\begin{eqnarray}
& & \sigma_{\gamma}  E_{\gamma}^2 \gg 1 ;~~ \Gamma_{eff} \gg \Gamma_{Thom} \label{P^d vs Gamma}\\
& & \sigma_{\gamma}  E_{\gamma}^2 \approx  1 ;~~ \Gamma_{eff} \approx  \Gamma_{Thom} \nonumber \\
& & \sigma_{\gamma}  E_{\gamma}^2 \ll 1 ;~~ \Gamma_{eff} \ll \Gamma_{Thom}.  \nonumber
\end{eqnarray}
At a large $\sigma_{\gamma}$, $\Gamma_{eff}$ is larger than $\Gamma_{Thom}$. In the low energy limit, $E_{\gamma} \rightarrow 0$, the probability agrees with the classical Thomson cross section. 
  
At a large $\omega_{\gamma_2} T$, $T \tilde g(\omega_{\gamma_2} T)=\frac{4
E_{\gamma_2}}{m_{\gamma}^2}$, the differential and the total probability are 
\begin{eqnarray}
& &\frac{d P^{(d)}}{d \cos \theta }=(\frac {e^2}{2m_e})^2
 \sigma_{\gamma}\frac{1}{ \pi^2 }\frac{2 k_{\gamma_1}}{  m_{\gamma}^2} [(2m_e +E_{\gamma_1}) \frac{({k_{\gamma_2}^{max})}^3}{3} -\frac{({ k_{\gamma_2}^{max})}^4} {4}]  ,\label{electron at rest 3 } \\
& &P^{(d)}=(\frac {e^2}{2 m_e})^2
 \sigma_{\gamma}\frac{1}{6\pi^2}\frac{1}{m_{\gamma}^2}m_e
 E_{\gamma,1}^3[
 E_{\gamma_1}+4m_e-\frac{2m_e+E_{\gamma_1}}{2}\xi^2+E_{\gamma_1} \xi^3].\label{electron at rest 4}
\end{eqnarray}

Next we study the energy dependence in the region $E_{\gamma_1} \approx m_e$.  From Eq.$(\ref{angle-energy})$ the angle is integrated in the region,  
\begin{eqnarray}
1-\cos \theta \leq \frac{m_e(E_{\gamma_1}-E_{\gamma_2})}{E_{\gamma_1}E_{\gamma_2}}.
\end{eqnarray}
Using a constant $C=(\frac {e^2}{2m_e})^2
 \sigma_{\gamma}\frac{1}{ 4\pi }\frac{2m_e^2 }{k_{\gamma_1}^2}$,  the energy spectrum is expressed as 
\begin{eqnarray}
\frac{d P^{(d)}}{d E_{\gamma_2}}
=
\begin{cases}
  C T \tilde g(0)(E_{\gamma_1}-E_{\gamma_2}) \\
 C  \frac{ 2 }{m_{\gamma}^2}  E_{\gamma_2}(E_{\gamma_1}-E_{\gamma_2}), \\
\end{cases} 
\end{eqnarray}
and the  average fraction of the kinetic  energy $r_{ke}=\frac{\langle E_{\gamma_2}\rangle }{E_{\gamma_1}} $ is 
\begin{eqnarray}
r_{ke}=\frac{E_{\gamma_2}}{E_{\gamma_1}}=  \label{energy-fraction2}
\begin{cases}
\frac{1}{2} ~~\text{for} ~P^{(d)}  \text{ in the small } \omega_{\gamma_2} T \\
 \frac{1}{3} ~~\text{ for} ~ P^{(d)}  \text{ in  the large}  \omega_{\gamma_2} T.
\end{cases}
\end{eqnarray}

%%%%%%%%%%%%%%%%%%%%%%%%%%%%%%%%%%%%%%%%%%%%%%%%%%%%%%%%%%%
\subsubsection{ High energy electron : energy conversion from the electron to the photon }
%%%%%%%%%%%%%%%%%%%%%%%%%%%%%%%%%%%%%%%%%%%%%%%%%%%%%%%%%%%
In a high energy region, the effective interaction
between the electron and the photon is that of the   Compton
scattering. Nevertheless it is useful to know the behavior and the
magnitude of those of the  Thomson scattering. 

For the parallel initial states, 
\begin{eqnarray}
p_{e_1}=(E_{e_1},0,0,p_{e_1}), p_{\gamma_1}=(E_{\gamma_1},0,0,
E_{\gamma_1}),  E_{e_1} \gg E_{\gamma_1}, 
\end{eqnarray}
the momentum satisfies 
\begin{eqnarray}
& &k_{\gamma_2} \leq k_{\gamma_2}^{max} = \frac{k_{\gamma_1} (E_{e_1}-p_{e_1})}{(p_{e_1}+k_{\gamma_1})(1-\cos \theta)+(E_{e_1}-p_{e_1})} , 
\end{eqnarray}
and the probability is given, for $\omega_{\gamma_2} T \ll 1$,  by
\begin{align}
\frac {dP^{(d)}}{d k_{\gamma_2} d \cos \theta}&=A (2\pi) T\tilde g(0)  k_{\gamma_2} (2 m_e^2+(E_{e_1}-p_{e_1})k_{\gamma_1}-k_{\gamma_2}(E_{e_1}-p_{e_1} \cos \theta )) \theta(B), 
\\
  A&= \sigma_{\gamma} \frac{2 k_{\gamma_1}}{E_{e_1} (2\pi)^3}(\frac{e^2 }{2 m_e})^2 ,                                    \nonumber \\
  B& =  (k_{\gamma_1}-k_{\gamma_2})(E_{e_1}-p_{e_1})-(p_{e_1}+k_{\gamma_1})k_{\gamma_2}(1-\cos \theta ),           \nonumber 
\end{align}
and for $\omega_{\gamma_2} T >1 $, by
\begin{align}
\frac {dP^{(d)}}{d k_{\gamma_2} d \cos \theta}&=&A (2\pi) \frac{4}{m_{\gamma}^2}    k_{\gamma_2}^2 (2 m_e^2+(E_{e_1}-p_{e_1})k_{\gamma_1}-k_{\gamma_2}(E_{e_1}-p_{e_1} \cos \theta )) \theta(B) .
\end{align}

For the anti-parallel case, 
\begin{eqnarray}
p_{e_1}=(E_{e_1},0,0,p_{e_1}), p_{\gamma_1}=(E_{\gamma_1},0,0,-
E_{\gamma_1}),  E_{e_1} \gg E_{\gamma_1}, 
\end{eqnarray}
 the momentum satisfies 
\begin{eqnarray}
& &k_{\gamma_2} \leq k_{\gamma_2}^{max}= \frac{k_{\gamma_1} (E_{e_1}+p_{e_1})}{-(p_{e_1}+k_{\gamma_1})(1-\cos \theta)+(E_{p_1}+p_{e_1})} , 
\end{eqnarray}
and the probability is given, for $\omega_{\gamma_2} T \ll 1$,  by
\begin{align}
\frac {dP^{(d)}}{d k_{\gamma_2} d \cos \theta}&=A (2\pi) T\tilde g(0)  k_{\gamma_2} (2 m_e^2+(E_{e_1}+p_{e_1})k_{\gamma_1}-k_{\gamma_2}(E_{e_1}-p_{e_1} \cos \theta )) \theta(B) ,
\\
  B&=  k_{\gamma_1}(E_{e_1}+p_{e_1})- k_{\gamma_2}(E_{e_1}+k_{\gamma_1}-(p_{e_1}-k_{\gamma_1}) \cos \theta ) ,          \nonumber 
\end{align}
and for $\omega_{\gamma_2} T >1 $, by
\begin{align}
\frac {dP^{(d)}}{d k_{\gamma_2} d \cos \theta}&=A (2\pi) \frac{4}{m_{\gamma}^2}    k_{\gamma_2}^2 (m_e^2+(E_{e_1}-p_{e_1})k_{\gamma_1}-k_{\gamma_2}(E_{e_1}-p_{e_1} \cos \theta )) \theta(B). 
\end{align}
$A$ is proportional to $\sigma_{\gamma}$, which is determined not only by the microscopic  quantity but also by the macroscopic quantities.
The probability is enhanced in a large $\sigma_{\gamma}$.   
%%%%%%%%%%%%%%%%%%%%%%%%%%%%%%%%%%%%%%%%%%%%%%%%%%%%%%%%%%%%%%%%%%%%%%%
\subsection{ $P^{(d)}_e$ for the electron in the (1,3) case}
%%%%%%%%%%%%%%%%%%%%%%%%%%%%%%%%%%%%%%%%%%%%%%%%%%%%%%%%%%%%%%%%%%%%%%%
For the events that the electron is detected or interacts with other microscopic objects, the wave function of the
electron in the final state determined by the reaction process is used. That in a solid state is normally atomic size, but   is large in a dilute gas.
 $\Gamma_{Thom}$ for the electron is equivalent to  that for the photon, but  $P^{(d)}_e$  is derived  from 
the light-cone singularity of the photon.  $\omega_e=\frac{m_{e}^2}{ 2
E_{e}}$ is much larger than $\omega_{\gamma}$ 
Ref.\cite{Ishikawa-Tobita-PTEP, Ishikawa-Tobita-ANA}. At the small $\omega_e T$,
$\Gamma_{eff}$ is the same as that of the photon, if $\sigma_e=\sigma_{\gamma}$.  At a large $\omega_e T$,
\begin{align}
P^{(d)}_{Thom,e}=
 (\frac{e^2}{2 m_e})^2\frac{4\sigma_e}{E_{e_1}} {E_{\gamma_1}}\int
 \frac{d{\vec
 k_{e_2}}}{2E_{e_2}(2\pi)^3}  \label{electron-diffraction}
(p_{e_1} \cdot p_{e_2} +m_e^2) T\tilde
 g(\omega_{e_2}T) \theta(s-m_e^2-2p_{e_2}\cdot (p_{e_1}+k_{\gamma_1})
 ) .
\end{align}
Eq. $(\ref{electron-diffraction})$ is almost identical to
Eq. $(\ref{13probability;P^d})$, in its form. Nevertheless, the magnitude is much smaller, since the angular frequency
$\omega_e$ is much larger than $\omega_{\gamma}$.  

For the system of a photon density $n_{\gamma}$, the total probability
$P^{(d)}_{e}(total)$ that one photon makes a transition is
 \begin{eqnarray}
  P^{(d)}_{e}(total)= n_{\gamma} \times P^{(d)}_{e}.
 \end{eqnarray}

%%%%%%%%%%%%%%%%%%%%%%%%%%%%%%%%%%%%%%%%%%%%%%%%%%%%%%%%%%%%%%%
\subsubsection{electron acceleration }
%%%%%%%%%%%%%%%%%%%%%%%%%%%%%%%%%%%%%%%%%%%%%%%%%%%%%%%%%%%%%%
In the case 
$p_{e_1}=(E_{e_1},0,0,p_{e_1}),p_{\gamma_1}=(E_{\gamma_1},0,0,E_{\gamma_1})$,
the  electron's energy distribution is obtained by integrating the
photon momentum in the final state first. Accordingly, the phase space and the
angular velocity  are changed to 
\begin{eqnarray}
& & s-m_e^2-2p_{e_2}\cdot (p_{e_1}+k_{\gamma_1}) \geq 0, \\
& & \omega_e=E_e(p,e)-p_e. \nonumber
\end{eqnarray}
For $E_e \gg m_e$, $E_{\gamma_1} > 2p_{e_1}$, the frequency is
 \begin{eqnarray}
  \omega_{e_2}=\frac{m_e^2}{2p_{e_2}},
 \end{eqnarray}
and the probability is
\begin{align}
 P^{(d)}=(\frac {e^2}{2 m_e})^2
 \sigma_{e}\frac{8}{4 \pi^2}
 \frac{2k_{\gamma_1}^2 }{m_e^2}\int^{E_{\gamma_1}-\frac{p_{e_1}^2}{E_{\gamma_1}}} dE_{e_2} E_{e_2}^2\label{electron-accelation} 
 = (\frac {e^2}{2m_e})^2
 \sigma_{e}\frac{2}{3 \pi^2}
 \frac{2 k_{\gamma_1}^2}{m_e^2} (E_{\gamma_1}-\frac{p_{e_1}^2}{E_{\gamma_1}})^3.
\end{align}
The probability that the electrons get the energy from the higher energy photon is determined by this $P^{(d)}$.
The lower energy photon does not give the energy.    

The probability $P^{(d)}$ for the photon and the electron are
proportional to the square of the electron radius ${\frac{e^2}{m_e}}$,
and to the range in space covered by the photon wave function 
$\sigma_{\gamma}$ or the electron wave function $\sigma_{e}$.  By tuning the latter
parameter, the enhancement of the probability is possible. The asymptotic values for the photon is proportional to $\frac{1}{m_{\gamma}^2}$ and for the electron is proportional to $\frac{1}{m_{e}^2}$.    
Eq. $(\ref{electron-accelation})$ is inversely proportional to ${m_e^2}$, Since 
the electron mass is much larger than the photon's effective mass, 
$P^{(d)}$ for the electron is much smaller than that of the photon.  However, that is  still sizable for a large  $\sigma_e$, which may
be realized in dilute gas such as the atmosphere and the space.

\subsubsection{ The case :(2,2) }
In the case that one of the initial states and one of the final states are 
plane waves and others  have small sizes, $(2,2)$ case, the transition 
amplitude  is given by
\begin{eqnarray}
& & \mathcal{M}=- i{e^2 \over m_e} \int d^4x \langle p_{e_2}|
 J(x)|p_{e_1} \rangle \langle {\vec k}_{\gamma_2},{\vec
 X}_{\gamma_2},T_{\gamma_2}|A_{\mu}(x) A^{\mu}(x)| {\vec k}_{\gamma_1},{\vec X}_{\gamma_1},T_1
 \rangle. \nonumber 
\end{eqnarray}
The probability is written with the amplitude. $T_1\tilde g(\omega_{\gamma_2} T_1)$ is substituted instead of  $T\tilde g(\omega_{\gamma_2} T)$ in  Eq.$(\ref{13probability})$ and the others, where $T_1$ is determined by the initial wave packet. 
%%%%%%%%%%%%%%%%%%%%%%%%%%%%%%%%%%%%%%%%%%%%%%%%%%%%%%%%%
\subsubsection{$e+E(B) \rightarrow e+\gamma$}
%%%%%%%%%%%%%%%%%%%%%%%%%%%%%%%%%%%%%%%%%%%%%%%%%%%%%%%
One photon production in the electron  scattering with the electric or
magnetic field has the probability $\Gamma T$ and $P^{(d)}$. The cross
sections for the bremsstrahlung or the synchrotron radiation are known
well.  $P^{(d)}$ in the processes that the electron
interacts with  a macroscopic electric or magnetic field gives a
scattering probability in the extreme forward direction.   The range
in space covered by these photon's wave functions can be much larger than
the atomic size, and  $P^{(d)}$ can be enhanced over $\Gamma
T$.  

   %%%%%%%%%%%%%%%%%%%%%%%%%%%%%%%%%%%%%%%%%%%%%%%%%%%%%%%%%%%%%%%%%%%%%
\subsection{Other QED processes}
%%%%%%%%%%%%%%%%%%%%%%%%%%%%%%%%%%%%%%%%%%%%%%%%%%
Other QED processes, 
$e^{+}+e^{-} \rightarrow 2 \gamma $ and $\gamma+\gamma \rightarrow \gamma+\gamma $ and others are also subject of the corrections of $P^{(d)}$. $P^{(d)}$ of these processes are expressed in the same manner as that of the Thomson scattering.

%%%%%%%%%%%%%%%%%%%%%%%%%%%%%%%%%%%%%%%%%%%%%%%%%%
\subsection{ Magnitude of $P^{(d)}_{\gamma}$}
%%%%%%%%%%%%%%%%%%%%%%%%%%%%%%%%%%%%%%%%%%%%%%%%%%%%%
 $P(T)$ at a small $T$ is mainly given by $P^{(d)}_{\gamma}$, and at a 
large $T$ by $\Gamma T$ and $P^{(d)}$.    
%%%%%%%%%%%%%%%%%%%%%%%%%%%%%%%%%%%%%%%%%%%%%%%%%%%%%
\subsubsection{ $\omega_{\gamma} T \ll1$}
%%%%%%%%%%%%%%%%%%%%%%%%%%%%%%%%%%%%
From Eq.$(\ref{P^d vs Gamma})$,  
 $P^{(d)}$ is proportional to $T$ in the region $\omega_{\gamma}T \ll 1 $.The effective rate defined by $\Gamma_{eff}^{(d)}=\frac{P^{(d)}}{T}$  
is proportional to the  range in space covered by the  wave function of the object that the photon interacts with, $\sigma_{\gamma}$, which  is evaluated from its  mean free
 path $l_{\text {mfp}}$ as 
\begin{eqnarray}
\sigma_{\gamma}=\pi l_\text{mfp}^2,  
\end{eqnarray}
and   
\begin{eqnarray}
\sigma_{\gamma} E_{\gamma}^2=\pi (\frac{l_{mfp}}{\lambda_{\gamma}})^2, \lambda_{\gamma}=\frac{h}{p_{\gamma}}. 
\end{eqnarray} 
If the mean free path is longer than the wave length, $\Gamma^{(d)}_{eff}$ is larger than $\Gamma $. The mean free path is determined by 
\begin{eqnarray}
 l_\text{mfp}= \frac{1}{ n_\text{charge} \times \sigma_\text{cross section}}, 
\end{eqnarray}
where the cross section $\sigma_{cross~section}$  is the Rutherford cross section for the charged particles in plasma  
%\begin{eqnarray}
%\sigma_{charge}= \sigma_\text{Ru}=4 \pi \left(\frac{\alpha \hbar c}{ k_B \text T} \right)^2 \text {log} \Lambda
 %, ~~5 \leq   \text {log} \Lambda \leq 20,
%\end{eqnarray}
and for the neutral atoms is  $ \pi ( r_{atom})^2$,
where $r_{atom}$ is the atomic size. The mean free path  is as long as $10^5$ meter or longer in dilute gas, and $\Gamma^{(d)}_{eff}$ is more important than $\Gamma$. That is not the case in liquid and solid.

%%%%%%%%%%%%%%%%%%%%%%%%%%%%%%%%%%%%%%%%%%%%
\subsubsection{$\omega_{\gamma} T >1$ } 
%%%%%%%%%%%%%%%%%%%%%%%%%%%%%%%%%%%%%%%%%%%
In the asymptotic region of $\omega_{\gamma} T \gg 1$ and $\Gamma T <1$, the probability behaves as Eq.$( \ref{two-components})$. $P^{(d)}$  is proportional to   $r_d=c \hbar n_e \frac{\sigma_{\gamma}}{m_{\gamma}^2}$, which 
 depends on  the photon's effective mass, $m_{\gamma}$. That is determined by
the plasma frequency,
\begin{eqnarray}
m_{\gamma}=\hbar  \sqrt\frac{4 \pi \alpha N_e}{ m_e};\ N_e~ \text{the total
 electron density},
\end{eqnarray}
and is expressed as
\begin{eqnarray}
m_{\gamma} c^2= 30 \sqrt{ \frac{N_e}{N_e^0}} eV, {N_e}^0=10^{30} {{\text m}}^{-3}.
\end{eqnarray}
Combining this with the $\sigma_{\gamma}$, and substituting $c \hbar =2 \times 10^{-7} \text{ eV ~m}$,   
\begin{align}
&r_d= c \hbar n_e \frac{\sigma_{\gamma}}{m_{\gamma}^2}= c \hbar n_e  {\pi l_{mfp}^2 }  {\frac{N_e^0}{N_e}}\frac{1}{900} 
=\frac{n_e}{N_e} \pi l_{mfp}^2N_e^0 \frac{1}{900}  \frac{ 2\times 10^{-7} \text{eV m}}{\text{eV ~m }} 
\end{align}
is large in dilute systems such  as  the ionophere, the solar corona, and the inter space. In the Ionosphere, parameters are roughly
$l= 10^7 \text{m}$, $n_e=10^{10} /{\text{m}^3}$, and $N_e=10^{15} /{\text{m}^3}$, and 
$r_d >1$. 

At larger $T$ of $\Gamma T >1$, the average life-time determines  the exponential behavior.

In the normal scattering of $P^{(d)}\approx 0$ and $\Gamma \neq 0$,  the sum of the kinetic energies 
of final states agrees with the initial energy, but 
in the processes due to  $P^{(d)}$, the substantial portion of the 
final states  have less kinetic energy than the initial energy. The  rest 
of the energy is stored in the  interaction energy.   Thus the physical effect due to 
 $P^{(d)}$ remains.  
 
 %%%%%%%%%%%%%%%%%%%%%%%%%%%%%%%%%%%%%%%%%%%%%%%%%%
\subsection{ Magnitude of $P^{(d)}_{e}$}
%%%%%%%%%%%%%%%%%%%%%%%%%%%%%%%%%%%%%%%%%%%%%%%%%%%%%
%%%%%%%%%%%%%%%%%%%%%%%%%%%%%%%%%%%%%%%%%%%%%%%%%%%%%
 
The product $n_{\gamma} \frac{\sigma_{e}}{m_{e}^2}$ determines the magnitude of $P^{(d)}_{e}$ at $\omega_e T >1$. The electron mass is much heavier than the photon's effective mass,
and   $P^{(d)}_{e}$ is much smaller that   $P^{(d)}_{\gamma}$, if $\sigma_e \approx \sigma_{\gamma}$.

%%%%%%%%%%%%%%%%%%%%%%%%%%%%%%%%%%%%%%%%%%%%%%%%%%%%%
%%%%
\section{Unique features of QCS }
%%%%%%%%%%%%%%%%%%%%%%%%%%%%%%%%%%%%%%%%%%%%%%%%%

  In this section $c$ and $\hbar$ are written.  Unique features of QCS were obtained using the   wave packets  constructed from 
 the eigenstates of the free Hamiltonian $H_0$.   As these wave packets  form a complete set, the results are universal. The case that the initial normalized states are 
 defined from the eigenstates of the $H$ was  studied in 
Section 2 H, and other cases are studied in Appendix D.    

In QED,   $  \Gamma_{Thom}$ is smooth in the angle  and characterized  by 
the de Broglie wave length 
$ \frac{h}{p}$.  Now,    $P^{(d)}$  
peaks at $\theta =0$ and the value is determined  by 
$\sigma_{\gamma}$ and   ${ \hbar E \over c^3
m_{\gamma}^2}$, which in fact is of a macroscopic magnitude for the
light particles. This component is unique in the quantum
mechanics and has no classical counterpart.  
Even in the low energy limit, the additional 
component  $P^{(d)}_{Thom}$ does not vanish and appears in the
extreme forward direction. 
%%%%%%%%%%%%%%%%%%%%%%%%%%%%%%%%%%%%%%%%%%%%%%%%%%%%
\subsection{Lorentz transformation and  $P^{(d)}$}
%%%%%%%%%%%%%%%%%%%%%%%%%%%%%%%%%%%%%%%%%%%%%%%%%%%%
  ${ \hbar E \over c^3
m_{\gamma}^2}$ characterizing  $P^{(d)}$ is huge and may be longer than  the  distance between the positions of the source and the detector, $L=cT$. $P^{(d)}$  is tranformed differently under the Lorentz transformation. The length $L$ in the rest system is transformed  to $\frac{L}{\sqrt{1-(v/c)^2}}$ in the moving frame of
  the velocity $v$.  The space time positions $(t,0,0,z)$ is tranformed  to $(t',0,0,z')$ by the boost in z-direction, 
 \begin{eqnarray}
(t,z) \rightarrow (t',z') =(t \cosh \xi+z \sinh \xi, t\sinh \xi +z \cosh \xi),
 \end{eqnarray}
where  $\cosh \eta=\frac{1}{\sqrt{1-(v/c)^2}}$.  Two space-time positions $X_1$ and $X_2$ of the distance $L_0$ are tranformed as
 \begin{eqnarray}
& &X_1, (0,0,0,0) ; X_2,(0,0,0,L_0) \rightarrow X_1',(0,0,0,0) ;  X_2', (L_0\sinh \eta,0,0,L_0\cosh \eta ), \\
& &L_0 \rightarrow L=L_0\cosh \eta, \nonumber 
 \end{eqnarray}
where $L$ is the transformed distance.   The  energy-momentum  of a particle of the rest mass $m$ is tranformed as 
  \begin{eqnarray}
(m,0,0,0) \rightarrow (E,0,0, p), E=m\cosh \eta,  p= m\sinh \eta .\label{T_E_ratio2}
 \end{eqnarray}
 Thus the ratio $\frac{T}{E }$ satisfies    
 \begin{eqnarray}
\frac{T}{E}=\frac{T_0 \cosh \eta}{m \cosh \eta}=\frac{T_0 }{m }.\label{T_E_ratio}
 \end{eqnarray}

    In 
   the region $ \omega T_1 \approx 0$, $P^{(d)}$ of the decay and the Thomson scattering are expressed from Eq.$(\ref{p^d-distribution})$   
   and $(\ref{13probability})$ as  
\begin{eqnarray}
& &{P^{(d)}|_{scalar}} =  \frac{T_1}{E_2({\vec P})} {8g^2} 
 \sigma_1\tilde g(0)\int \frac{d^3 p_1}{(2\pi)^3 E({\vec p}_1)}  \theta(m_2^2-m_1^2-2P \cdot p_1),\label{p^d-distribution3} \\
& &P^{(d)}|_{Thom, \gamma}\times {E_{e_1}} = \frac{T_1}{E_{\gamma_1}}  {2e^4 }\sigma_{\gamma} \tilde g(0)\int \frac{d {\vec k}_2}{ (2\pi)^3 E_{\gamma_2}  }    \label{photon-diffraction3}  \theta(s-m_e^2-2k_{\gamma_2}\cdot(p_{e_1}+k_{\gamma_1}) ),  
\end{eqnarray}
where the phase space are defined by the step functions.  The first factors in the right-hand sides  are the same in the
 moving frame  from Eq.$(\ref{T_E_ratio})$ and so are the integrals.    Eq.$(\ref{p^d-distribution3} )$  and Eq.$(\ref{photon-diffraction3} )$  
are manifestly Lorentz invariant. 

In the region $\omega T \gg 1 $, 
\begin{eqnarray}
& &P^{(d)}= {8g^2}
 \sigma_1 \frac{2}{m_1^2} \int \frac{d^3 p_1}{(2\pi)^3 E({\vec p}_1)}  \frac{E({\vec p_1})}{E_2({\vec P})}\theta(m_2^2-m_1^2-2P \cdot p_1), \label{p^d-distribution4}\\
& &P^{(d)}|_{Thom, \gamma}\times {E_{e_1}} =   {2e^4 }\sigma_{\gamma} \frac{2}{m_1^2}\int \frac{d {\vec k}_2}{ (2\pi)^3 E_{\gamma_2}  }\frac{E_{\gamma_2}}{E_{\gamma_1}}    \label{photon-diffraction4}  \theta(s-m_e^2-2k_{\gamma_2}\cdot(p_{e_1}+k_{\gamma_1}) ). 
\end{eqnarray}
From Eq.$(\ref{T_E_ratio2})$,  the right-hand sides of Eqs.$(\ref{p^d-distribution4})$ and $(\ref{photon-diffraction4}  )$  are invariant, despite of   complicated form.  This is understandable because $P^{(d)}$  gets contributions from the long distance. In general situations, $P^{(d)}$ from  Eqs.$(\ref{p^d-distribution})$   
   and $(\ref{13probability})$ are more complicated functions. The variable $y=\frac{m_1^2 T_1}{2E}$ is the invariant combination, and $P^{(d)}$ in the decay and $P^{(d)}_{\gamma} \times E_e$   in the scattering are invariant.   $P^{(d)}_e$  is  the same.

%%%%%%%%%%%%%%%%%%%%%%%%%%%%%%%%%%%%%%%%%%%%%%%%%%%%%%%%%%%%%%%%%%%%%%%%%%%
\subsection{ QCS interaction with matter \label{ section-detection}}
%%%%%%%%%%%%%%%%%%%%%%%%%%%%%%%%%%%%%%%%%%%%%%%%%%%%%%%%%%%%%%%%%%%%%%%%%%
 Interactions of QCS with matter  are   governed
by   the local interaction of the fundamental fields at the same or
nearly the same positions, and  the transition amplitude and the probability are   
composed of $\Gamma T$ and  $P^{(d)}$. 
The former has the  typical scale of the de Broglie length
$\frac{h}{p}$, and  the latter 
has the scale of  $\frac{ \hbar E}{m^2 c^3}$. Two particles in an event in the former have  a short correlation length in time, whereas  those in the
 latter have a long length.

 %%%%%%%%%%%%%%%%%%%%%%%%%%%%%%%%%%%%%%%%%%%%%%%%%%%%%%%%%
   \subsubsection{Interaction of QCS with microscopic states  }
%%%%%%%%%%%%%%%%%%%%%%%%%%%%%%%%%%%%%%%%%%%%%%%%%%%%%%%%%%
A reaction of QCS with microscopic states is analyzed  for an 
  initial state $|\Psi \rangle$ that includes  QCS,
\begin{eqnarray}
|\Psi \rangle=|B \rangle \otimes [|C' \rangle +|\Psi_{QCS}(C, \gamma) \rangle ].
\end{eqnarray}
where   $|B \rangle$ and $|C'\rangle$  are the  particle state
 in a detector  and  in the beam, and  QCS composed of a particle $C$ and a photon is  expressed by
$|\Psi_{QCS}(C,\gamma)\rangle $. For simplicity, the interaction of $C$ with matters
is assumed local and negligibly weak, and  a photon  interacts with $B$  in the
    detector by the interaction,
\begin{eqnarray}
H_\text{int}(t)=\int d{\vec x} J^{\mu}(x) A_{\mu}(x), J^{\mu}= \bar \psi_{B'}(x) \Gamma^{\mu}\psi_{B}(x). 
\end{eqnarray}
  The probability  amplitude for the  state  $|\Psi \rangle$  to be  transformed to  a final state $|out \rangle $ is 
 \begin{eqnarray}
     \mathcal{M}= \langle out| \int dt H_\text{int}(t)| \Psi \rangle, \langle out |= \langle
  C| \otimes\langle B' |.
    \end{eqnarray}
 $|C' \rangle $ and $ |\Psi_{QCS}(C, \gamma) \rangle $ interact independently.     
Scattering of the former is  described by a normal amplitude,  
 \begin{eqnarray}
      \mathcal{M}= \langle out| \int dt H_\text{int}(t)| B ,C' \rangle, 
    \end{eqnarray}
which depends upon the energies of $B$, $B'$, $C$, and $C'$.  
That for  QCS is,  
\begin{eqnarray}
     \mathcal{M}_{QCS}= \langle out  | \int dt H_{int}(t)| \Psi_{QCS}(C,\gamma) ,B \rangle =\langle B'| \int dt H_{int}(t)|B , \gamma \rangle \langle C|\Psi_{QCS} \rangle , 
    \end{eqnarray}
   where  the   
 photon in QCS interacts with  $B$, but others do not.  
That is expressed in Fig. 4.
%%%%%%Fig%%%%%%%%%%%%%%%%%%%%%%%%%%%%%%%%
\begin{figure}[t]
\centering{ \includegraphics[scale=.8,angle=0]{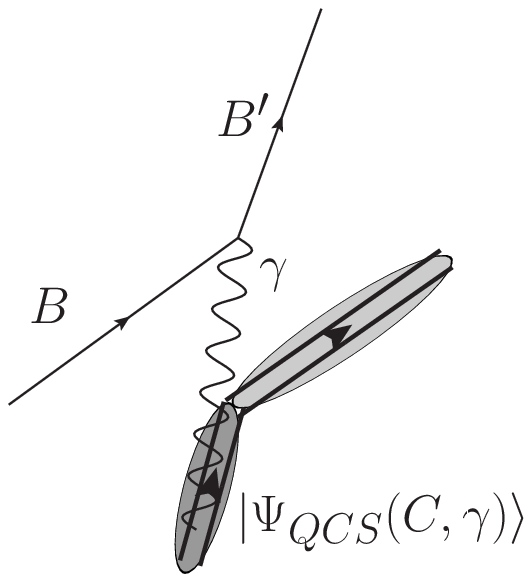}
\caption{Diagram that the photon in QCS interacts with the state $B$ in the detector and is detected. Other part of QCS does not interacts.}
}\label{fig:qcs-detection}
\end{figure}
%%%%%%%%%%%%%%%%%%%%%%%%%%%%%%%%%%%%%%%%%%%
Transition  of the photon is described by either  $\Gamma$  or $P^{(d)}$.  
 The   former amplitude   is
    proportional to  the following integral over the time that depends
    on the  kinetic energy,  
\begin{eqnarray}
& & \int dt e^{i(E_B+E_{\gamma}-E_{B'})t} F_{B,B'} \approx 2\pi
  \delta (E_B+E_{\gamma}-E_{B'})\epsilon_{\mu}F_{B,B'}^\mu, \\
& &F_{B,B'}^{\mu}= \int d{\vec x} \langle B'|J^{\mu} (0,{\vec x})|B \rangle \langle C\ | \Psi_{QCS} \rangle. \nonumber
\end{eqnarray} 
The kinetic energy of the photon  is expressed from $E_B$ and $E_{B'}$
    \begin{eqnarray}
     E_{\gamma}= E_{B'}-E_{B}. \label{kinetic-energy}
    \end{eqnarray} 
The transition amplitude depends upon this  energy, and so does the
probability. The  kinetic energy of $C$ and the interaction energy decouple from the amplitude,
 and are irrelevant to the transition. Thus  the energy $E_{\gamma}$
is detectable in the experiments, and is able to transfer to  other
states in the natural processes, but the rest is not. For the process governed by $P^{(d)}$, the energies satisfy 
 \begin{eqnarray}
     E_{\gamma} \leq E_{B'}-E_{B} \label{kinetic-energy2},
    \end{eqnarray} 
instead of   Eq.$(  \ref{kinetic-energy})$.   

If the
particle $C$ has  some charge 
that interacts with one of gauge fields, the kinetic energy of $C$ is also
detectable in experiments and is able to transfer to other states. In
general QCS composed of the constituent particles that interact with
an atom or a nucleus, their kinetic energy is detectable, but the 
interaction energy decouples and is not detectable.

The fact that   kinetic energy is extractable, but 
interaction energy is not,  is equivalent to  a transmission law of
 energy in classical  mechanics.
Kinetic energy of a massive body is determined by the velocity, and  its
transmission  to others is easily made by a contact with another of lower
 velocity.
Now  potential energy is determined by the position, and is
the same between two bodies in contact with each other, and does not
transmit directly.  To
 transmit  potential energy, that is transferred to  kinetic energy
 first. For instance, a potential
 energy of the water at a higher position is transferred first to its kinetic
 energy,  and  is transmitted next to a turbine or others. That can be 
 used for generating an electricity in the hydro-electric generation.  
Similarly,  interaction energy of QCS does
not appear in the transition amplitude, and is
 neither  detectable with a normal detector nor transmittable to other
 matter.

It is heuristic to make a comparison of  QCS with a bound state (BS). QCS is loosely bound, but  BS is tightly bound in a microscopic region of a
finite energy gap and
is a stationary state that is stable under the Poincare transformation.   The 
matrix element of an operator $J(x)$ is written by the four dimensional 
momentum,
\begin{eqnarray}
\langle BS; {\vec p}_1| J(x)| BS; {\vec p}_2 \rangle =e^{i(p_2-p_1)\cdot x} \langle
 BS;{\vec p}_1|J(0)| BS;{\vec p}_2 \rangle. 
\end{eqnarray}
This is valid for any operator and  bound state that are composed of 
fundamental fields. Furthermore, for a Poincare covariant state, the
matrix element is described by a boost
operator $U({\vec p}_1,{\vec 0})$ as
\begin{eqnarray}
 \langle
 BS;{\vec p}_1|J(0)| BS;{\vec p}_2 \rangle= \langle
 BS;{\vec 0}|U^{\dagger}({\vec p}_1,{\vec 0})J(0)U({\vec p}_2,{\vec
 0})| BS;{\vec 0} \rangle.
\end{eqnarray}
Kinetic
energy and  interaction energy are not separable, and the amplitude is written with  total energy.  Total energy is measured in BS, whereas     interaction 
energy is separated and  decouples from the transition   in QCS.  

For a weakly  correlated state, such as QCS,  kinetic energy is separable 
from  total energy, and is detected independently.  
Interaction energy  decouples and is not detectable by microscopic processes.  An example is 
  Feynman's parton
model for  a nucleon in deep inelastic scattering. That is expressed by 
three valence quarks and soft sea quarks and anti-quarks  and
gluons. Their kinetic energies can be probed by the photon or the weak
bosons, but the   interaction energy is undetectable, and so is
invisible.  That is like a missing energy.  Interaction energy  or an effect of 
boost  that includes
the interaction is non-detectable  by the scatterings through the
electroweak currents also. 

Weakly correlated states appear in wide area. They include halos 
in nuclei, atoms, molecules, and other larger physical systems such as
the star and galaxy. It is challenging to measure the interaction energy in
these systems
\cite{Ekstein-Siegert, Gaemers-Visser, particle-data}.

%%%%%%%%%%%%%%%%%%%%%%%%%%%%%%%%%%%%%%%%%%%%%%%%%%
 \subsubsection{Interaction with a large wave  in space}
%%%%%%%%%%%%%%%%%%%%%%%%%%%%%%%%%%%%%%%%%%%%%%%%%%%
If  $B$ and $B'$ are not in the normal matter, but are extended in wide area,  their interaction with QCS is different from those of the previous case.
    Large $P^{(d)}$ may appear in their processes, and  
 determines the transition. Due 
to the positive semi-definite interaction energy,    kinetic 
energy  shows 
  \begin{eqnarray}
     \omega =E_{\gamma}  -  E_{B'}+E_{B} \leq  0.
    \end{eqnarray}
$\sigma$  is large   in dilute systems, or  in
highly correlated quantum states such as superconductor, super-fluid,
and others. A magnetic field or an electric field
can be uniform in large area, and $\sigma$ of quantum states  become
macroscopic size in magnitude. That   may couple with
QCS. A large $P^{(d)}$ there may be detectable   in 
laboratory or observable in natural processes.  
%%%%%%%%%%%%%%%%%%%%%%%%%%%%%%%%%%%%%%%%%%%%%%%%   
\subsubsection{Coupling with gravity}
%%%%%%%%%%%%%%%%%%%%%%%%%%%%%%%%%%%%%%%%%%%%%%%%
Energy momentum tensor is composed of the kinetic part and the
interaction part. The latter  is ignorable for uncorrelated plane waves
\footnote{If the fields are normalized in a  volume $V$,
the integral of  n-th power form of the fields is proportional to
$V^{\frac{2-n}{2}}$, and
vanishes for $n \geq 3$ in $V \rightarrow \infty$.The interaction energy
of QCS is in the
forward angle does not follow this, and remains.} but
is sizable for  QCS.    Thus the energy momentum tensor is written as,
     \begin{eqnarray}
     T_{\mu\nu}=T^{(0)}_{\mu\nu}+T^{(d)}_{\mu\nu}, \label{energy-tensor}
     \end{eqnarray}
where the first term is from the kinetic energy  and the
 second term is from the  interaction energy, which  
is positive from a Virial theorem in Sec.2.3.  The
former is equivalent to that of the particle  energy
and the latter is proportional to
the metric $g_{\mu\nu}$.
\begin{eqnarray}
 T^{(0)}_{\mu\nu}=T_{\mu\nu}(matter), T^{(d)}_{\mu\nu}=g_{\mu\nu} \Lambda^{(d)},
 \end{eqnarray}
where $\Lambda^{(d)}$ is a constant.
The tensor Eq.$(\ref{energy-tensor})$ becomes a source of the gravitational field.  The  positive definite
energy-momentum tensor proportional to  the metric $g_{\mu\nu}$
from   QCS   are added constructively, and those of 
a macroscopic number can form  a macroscopic gravitational field.
Consequently they affect a motion of a  massive body there. Conversely
the  interaction energy of QCS affects the  macroscopic motion of   a
massive  object, but is not detected by the ordinary measurement
that uses the standard interaction. The ratio of the two components is
\begin{eqnarray}
r=\frac{T^{(0)}}{T^{(d)}}=\frac{2 r_{ke}}{P^{(d)}}.
\end{eqnarray}
Interaction energy  gives the equivalent effects as
 a dark matter or a dark energy \cite{WMAP-neutrino}.

%%%%%%%%%%%%%%%%%%%%%%%%%%%%%%%%%%%%%%%%%%%%%%%%%%%%%%%%%%%%%%%
\subsection{Fraction of the QCS and of  invisible energy}
%%%%%%%%%%%%%%%%%%%%%%%%%%%%%%%%%%%%%%%%%%%%%%%%%%%%%%%%%%%%%%
The average kinetic energy of QCS in the scalar decays Eq.$(\ref{p^d-distribution})$, and   in 
the Tomson scattering Eq.$(\ref{energy-fraction2})$ are studied  for the case 
of $P \ll 1 $ and of $ P^{(d)} \geq 1$. 
%%%%%%%%%%%%%%%%%%%%%%%%%%%%%%%%%%%%%%%%%%%%%%%%%%%%%%%%%%%%%%%%
\subsubsection{Fraction of QCS}
%%%%%%%%%%%%%%%%%%%%%%%%%%%%%%%%%%%%%%%%%%%%%%%%%%%%%%%%%%%%%
For $P^{(d)}_{\gamma}(total) \ll  1$, the total probability that one photon is scattered in a gas of the electron density $n_e$,  at $T$ is $P=\Gamma(total) T+P^{(d)}(total)$, where $\Gamma(total)=n_e \Gamma, P^{(d)}(total)=n_e P^{(d)}_{\gamma}$.
%    \begin{eqnarray}
%     P=\Gamma(total) T+P^{(d)}_{\gamma}(total),
%    \end{eqnarray}
 %   with $ \Gamma(total)=n_e \Gamma, P^{(d)}_{\gamma}(total)=n_e P^{(d)}_{\gamma} $.  
 Suppose that 
$P$ becomes the unity at  a time $T_{max}$, 
\begin{eqnarray}
    \Gamma(total) T_{max}=1-P^{(d)}_{\gamma}(total).     
    \end{eqnarray}
$T_{max}$ is the mean free time of the initial state.    By the time $T_{max}$, due to the scattering with the electrons all states are scattered, and transformed to the final states.  
  The number of the final states at $T_{max}$ is proportional to those that are transferred  in the time interval between $T=0$ and  $T=T_{max}$,  
\begin{eqnarray}
& &P_\text{conservig}=\frac{1}{2} T_{max} (1-P^{(d)}_{\gamma}(total)),  \label{transferrd-1}\\
& &P_\text{non-conserv.}=T_{max} P^{(d)}_{\gamma}(total). \nonumber
\end{eqnarray}

For $P^{(d)}_{\gamma}(total) \geq   1$, the total probability reduced to
    \begin{eqnarray}
     P=\frac{ \Gamma(total) T}{1+P^{(d)}_{\gamma}(total)}+\frac{P^{(d)}_{\gamma}(total)}{1+P^{(d)}_{\gamma}(total)},
    \end{eqnarray}
and becomes $P=1$ at  $ T_{max}=\frac{1}{\Gamma(total)}$. The probabilities Eq.$(\ref{transferrd-1})$  become 
\begin{eqnarray}
& &P_\text{conservig}=\frac{T_{max}}{2(1+P^{(d)}_{\gamma}(total))} \label{transferrd-2},\\
& &P_\text{non-conserv.}=\frac{T_{max} P^{(d)}_{\gamma}(total)}{1+ P^{(d)}_{\gamma}(total)}, \nonumber 
\end{eqnarray}
We denote    a fraction of the average kinetic energy of the QCS  as ${r_{ke}}$. That in  the scalar decays Eq.$(\ref{p^d-distribution})$ and the Thomson scattering Eq.$(\ref{energy-fraction2})$ is,
\begin{eqnarray}
r_{ke}=
\begin{cases}
    \frac{1}{2} ;\text{scalar decay}  \\
      \frac{1}{2}, ~\frac{2}{3}; \text{Thomson scattering}. 
\end{cases}
\end{eqnarray} 
The average fraction of the observed energy  for
$P^{(d)}_{\gamma}(total) <1 $ is
    \begin{eqnarray}
     \frac{E_\text{visible}}{E_\text{total}}= \frac{1}{2}(1-P^{(d)}_{\gamma}(total))+P^{(d)}_{\gamma}(total) {r_{ke}},
    \end{eqnarray}
and     for $P^{(d)}_{\gamma}(total) \geq   1$, 
    \begin{eqnarray}
     \frac{E_\text{visible}}{E_\text{total}}=
      \frac{1}{2(1+P^{(d)}_{\gamma}(total))}+\frac{P^{(d)}_{\gamma}(total)}{1+P^{(d)}_{\gamma}(total)}
      {r_{ke}}.
    \end{eqnarray}

    If $P^{(d)}_{\gamma}(total) \gg 1$, the majority of the energy is stored in the
     interaction energy, and is invisible with  the detectors that
    use microscopic processes. 

The statistical behaviors of the final states due to $\Gamma(total)$ is determined by
 Gibbs ensemble,
but those due to $P^{(d)}$ is not. As is shown in Appendix E, they show  behaviors of non-stationary states.  

%%%%%%%%%%%%%%%%%%%%%%%%%%%%%%%%%%%%%%%%%%%%%%%%%%
\section{Confirmation  of QCS}
%%%%%%%%%%%%%%%%%%%%%%%%%%%%%%%%%%%%%%%%%%%%%%%%%%

%%%%%%%%%%%%%%%%%%%%%%%%%%%%%%%%%%%%%%%%%%%%%%%%%%%%%%%%%%%%%%%  
%   \subsubsection{Space-time symmetry of QCS  }
%%%%%%%%%%%%%%%%%%%%%%%%%%%%%%%%%%%%%%%%%%%%%%%%%%%%%%%%%%%%%%%%

%%%%%%%%%%%%%%%%%%%%%%%%%%%%%%%%%%%%%%%%%%%%%%%%%%%%%%%%

%%%%%%%%%%%%%%%%%%%%%%%%%%%%%%%%%%%%%%%%%%%%%%%%%%%%%%%%%%%%%%%%%%%%
\subsection{Comparison with previous experiments}
%%%%%%%%%%%%%%%%%%%%%%%%%%%%%%%%%%%%%%%%%
QCS and the probability $P^{(d)}$ have been barely paid attentions by the researchers.  This is because  experimental signals are wide in  energy and narrow in  forward direction in the configuration space. Hence  identifications are hard, \cite{ TS}.These signals have been considered as backgrounds,  and serious studies have been barely made. This does not mean that they do not exit. When  $P^{(d)}$ becomes larger than $\Gamma T$, that is necessary  for correct understandings of the phenomena.

%%%%%%%%%%%%%%%%%%%%%%%%%%%%%%%%%%%%%%%%%%%%%%%%%%%%%%%%%%%%%%%%%%%%%

%%%%%%%%%%%%%%%%%%%%%%%%%%%%%%%%%%%%%%%%%%%%%%%%%%%%%%%%%%%%%%%%%%%%
\subsection{Comparison with the axiomatic field theory}
Scatterings of plane waves derived from the limit $\sigma \rightarrow \infty$ have  the 
  infinity large $P^{(d)}$.  $\Gamma T$ remains finite, hence its relative weight  vanishes.  All the final states are QCS, and the scattering process described  $P^{(d)}$,  which arises in  extreme forward angle, takes place.  The scattering expressed by the golden rule does not occur.     This is partly equivalent to the Haag theorem \cite{Haag}.  The theorem proved that the  Lorentz invariant S-matrix is the unity, whereas the  present paper showed  that normal scattering disappears in the   limit  $\sigma \rightarrow \infty$. They are equivalent.   The remaining scattering in the extreme forward angle is not invariant in the 3+1 dimension  but  in  the 1+1 dimension.

%%%%%%%%%%%%%%%%%%%%%%%%%%%%%%%%%%%%%%%%%%%%%%%%%%%%%%%%%%%%%%%%%%%%%
%%%%%%%%%%%%%%%%%%%%%%%%%%%%%%%%%%%%%%%%%%%%%%%%%%%%%
%%%%%%%%%%%%%%%%%%%%%%%%%%%%%%%%%%%%%%%%%%%%%%%%%%%%%
%%%%%%%%%%%%%%%%%%%%%%%%%%%%%%%%%
\subsection{Experimental confirmation}
%%%%%%%%%%%%%%%%%%%%%%%%%%%%%%%%%%
There are various ways to test QCS. 

(1) Direct  observation of the missing energy. 

The kinetic energy is visible by its interaction with atoms, but the interaction energy of QCS is invisible.  Hence  the energy measured with an ordinary detector is a partial portion, and deviates from the total energy. Observation of the missing energy  and the  kinetic energy transferred from the  interaction energy may be able to confirm  QCS.  

(2)Observation of the absolute probability. 

Another way to test  QCS  experimentally is to measure the absolute  transition probability $P(T)$ and $\Gamma T$.  It would be hard to distinguish the events due to $P^{(d)}$ from   backgrounds.  If that is made,  and  a clear difference between two is found, the transition  due to QCS will be proved.   

(3)Observation of the rapid transitions. 

The rapid change of $P^{(d)}$  in the region $\omega_{\gamma} T \ll1$,    is   independent of 
the thermal effects. Furthermore,
 in the dilute system  $P^{(d)}$ is enhanced  and governs the total transition probability. It would be possible to identify these    features  using  advanced laser technology. 

(4)Interaction with gravity. 

The interaction energy carried by the overlapping waves becomes a source of  the gravitational field.  Hence, macroscopic number of QCS is observable through their energy and momentum.

\section{ Summary }
%%%%%%%%%%%%%%%%%%%%%%%%%%%%%%%%%%%%%%%%%%%%%%%%%%%%
%%%%%%%%%%%%%%%%%%%%%%%%%%%%%%%%%%%%%%%%%%%%%%%%%%%%%
The scattering theory for the finite time intervals was  formulated with the wave packets and the finite probability  was derived. Based on FQM, the difficulty raised by 
 Stueckelberg for the plane waves that the probability diverges   was resolved by  using the normalized one-particle states. It was found that the transition 
 probability at $T$, $P(T)$,   is  convergent and composed of   $\Gamma T$  and $P^{(d)}$. $\Gamma T$   is derived from the bulk states  and computed easily at the
  asymptotic space-time region under ASI, where    
 the interaction Hamiltonian is $e^{-\epsilon|t|}H_{int} $, and  vanishes at $|t| \rightarrow \infty$.  The states there are composed of
  independent free waves.  They are  particle-like states,  of $O_{qcs}=0$,  in the initial 
and final states, and the transitions among them are expressed by the average rate  $\Gamma T$ from the ratios of the fluxes. 
 $P^{(d)}$ is derived from the boundary states,  which  includes  the  wave-like solutions of the 
many-body Schr\"odinger equation,  QCS. QCS  is  the  state of  the finite-interaction energy, and appears in the final states of  the scattering 
or decaying states, and contributes to the  transition probability at later times. 
Thus the transition probability  $\Gamma T+ P^{(d)}$ governs the physical phenomena.

Due to intriguing properties of   $P^{(d)}$, new phenomena are expected to be observed.  They reveal the rapid transition   in the extreme forward angle. Although 
these experiments   have  been  difficult so far, they are   getting feasible.       

In the Thomson scattering,  QCS in the extreme forward direction and  $P^{(d)}$ were computed.   
 Because $P^{(d)}$  is proportional to  the new scale ${\hbar E \over  m^2 c^3} \sigma$,  where $m$ is the effective mass of the photon, that is   large. Moreover,  $P^{(d)}$ behaves differently from $\Gamma$ in  energy and in  time.  That is widely spread in   kinetic energy and changes rapidly at  short time and remains constant later.  From these  behaviors, $P^{(d)}$ has not been paid attentions.    However, $P^{(d)}$     plays important roles in natural phenomena, and new  experiments  dedicated to $P^{(d)}$   may supply valuable information. 
The transition probability of the Thomson scattering is modified in  intermediate energy but the low-energy theorem \cite{gell-mann-goldberger}  is valid in the low energy limit.   The critical energy depends on the mean free path of the charged matter and becomes extremely low in the dilute system. 
     
 Interaction energy of QCS is not observed like kinetic energy of particles.  Normally   kinetic energy of the wave  proportional to the frequency by the
Einstein relation $E=h \nu$
transmits from the initial state to another state in the microscopic transition processes. 
This energy   is
detectable  with the normal detectors, and transmittable to other systems 
in natural phenomena, and    visible.
 Contrary to  kinetic energy,   interaction energy is stored in overlapping waves
 and neither  transmittable nor detectable unless that is converted to  kinetic energy. Hence this is
 invisible with microscopic processes. Nevertheless,
 this  non-detectable energy is one part of the energy momentum tensor
 and becomes the source of  the gravitational field and affects the  
macroscopic motion of a massive body.   For $P^{(d)}
 \geq 1$, the dominant part of the energy of the final states is
 invisible.

  $P^{(d)}$ depends upon the fundamental constants of the underlining theory,    and   has  intriguing properties.  
 This could  have  many implications, especially in  dilute systems, and   changes also basic understandings of the natural
 phenomena \cite{corona-heating, 1987A-neutrino}.  $P^{(d)}$   gives effects  
similar to  a new interaction.

  %%%%%%%%%%%%%%%%%%%%%%%%%%%%%%%%%%%%%%%%%%%%%%%%%%%%%%%%%%%%%%%%%%%%%%%%%%%
%\newpage
{\bf Acknowledgements.} 
The present work is  partially supported by a
Grant-in-Aid for Scientific Research (Grant No.24340043), and JSPS KAKENHI  Grant Number  15H05885(J-Physics). 
The authors  thank Dr. Takashi Kobayashi,
Dr. Takasumi Maruyama, Dr. Tsuyoshi Nakaya, and Dr.Koichiro Nishikawa for useful discussions on 
the neutrino experiments, Dr. Asao Arai, Dr. Shoji Asai, Dr. Tomio Kobayashi, Dr. Makoto Minowa, Dr. Toshinori Mori, 
Dr. Sakue Yamada, Dr.Terry Sloan, Dr.Kin-ya Oda, Dr. Izumi Ojima,Mr. Hiromasa Nakatsuka, Dr. Toshiki Tajima, Dr. Kiyoshi Kuramoto, Dr. Mitsuteru Sato, Dr. Masaki Takesada, Dr. Tomobumi Mishina, 
Dr. Shigeto Watanabe, and Dr. Junji Watanabe for useful discussions on
quantum  interferences. 

%{ \bf Note  added}:

%After we submitted the manuscript for the publication, the work of Stueckelberg  \cite{Stuckelberg} came to our attention. 
%In this paper, he studied the transition 
%occuring during a finite time interval. It was found (1) that divergences occur for processes which are localized in space-time by a sharply defined boundary in 
%the transition probability of the plane waves, (2) that  the convergent results were obtained by the diffuse boundaries, and (3) that a prescription for  the unitary S-matrices for the diffuse boundaries was presented. The convergent result in (2)  was independent of time and  photon emission by a free electron was presented.  The states were 
% plane and not normalized. 

 % Now in the present paper, the initial and final states are normalized and the transition probability is defined by a sharply defined boundary following the von Neumann's fundamental principle of the quantum mechanics. Accordingly, the unitarity is manifest and the probability $P$ is finite and less than or equal to the unity and agrees with that measured in experiments. That
 %  is linear in the time interval
  % of the form $P=\Gamma T +P^{(d)}$ for $P \ll 1$, where $\Gamma$  agrees with the standard formula.  $P^{(d)}$ is the correction term and vanishes for a free electron and those     for the Thomson scattering and the decays are presented.       Present paper studies a similar problem but the formalism and the results are different from those  of  Stueckelberg.

{}
\appendix
%%%%%%%%%%%%%%%%%%%%%%%%%%%%%%%%%%%%%%%%%%%%%%%%%%%%%%%%%%%%%%%%%%%%%%
\section{Integration over space-time coordinates  }
%%%%%%%%%%%%%%%%%%%%%%%%%%%%%%%%%%%%%%%%%%%%%%%%%%%%%%%%%%%%%%%%%%%%%The vacuum 

\subsection{Integration in the bulk and  in the boundary}

The  amplitude  that  a state is transformed to other state   in the lowest order of the coupling strength is the integral over the space-time coordinates of the product 
of the initial and final wavefunctions in the coordinates space.   For a transition  that the parent of an average momentum ${\vec
p}_{2}$ at $X_{2}=(T_{2},\vec{\text{X}}_{2})$  and the daughters of  average momenta ${\vec
k}_{1_i}$ at ${ X}_{1_i }=(T_{1_i},{\vec X}_{1_i}),i=1,2$
%. The   amplitude   in the lowest order of the coupling strength for the parent of an average momentum ${\vec
%p}_{2}$ at $X_{2}=(T_{2},\vec{\text{X}}_{2})$  
\begin{eqnarray}
 \mathcal{M}=\int_{T_{2}}^{T_{1}} dt\int d^3x  \langle {\vec k}_{1},{\vec k}_{1'};{\vec X}_{1},{\vec X}_{1'}
 |H_{int}(x)| {\vec p}_{2} , {\vec X}_{2}   \rangle. \label{decay-amplitude}
\end{eqnarray}
Substituting 
\begin{eqnarray}
& &\langle  0 |\varphi_2(x)|{\vec p_{2}},{ X}_{2} \rangle=N_{2} \rho_{2} (\frac{2\pi}{\sigma_{2}})^{3/2} 
e^{-\frac{1}{2 \sigma_{2}}  ({\vec x}-{\vec X}_{2}-{\vec v}_{2}(t-T_{2}))^2} 
 e^{-i(E({\vec p}_{2}) (t-T_{2})-{\vec p}_{2} ({\vec x}-{\vec X}_{2}))},   \nonumber \\
 & &\langle {\vec k_{1_i}},{X}_{1_i}|\varphi_1(x) |0 \rangle=N_{1_i} \rho_{1_i} (\frac{2\pi}{\sigma_{1_i}})^{3/2} 
e^{-\frac{1}{2 \sigma_{1_i}}  ({\vec x}-{\vec X}_{1_i}-{\vec v}_{1_i}(t-T_{1_i}))^2} 
 e^{i(E({\vec p}_{1_i}) (t-T_{1_i})-{\vec p}_{1_i} ({\vec x}-{\vec X}_{1_i}))} , \nonumber \\
& &N_{2(1_i)}=(\frac{\sigma_{2(1_i)}}{\pi})^{3/4}, \rho_{2(1_i)}=(\frac{1}{2E_{2(1_i)}(2\pi)^3})^{1/2};(x-X_{2(1)})^2 \geq 0 ,  
\end{eqnarray}   
that is  reduced to  
\begin{align}
\label{amplitude2}
&\mathcal{M} = 
g N_1 \int_{T_{2}}^{T_{1}} d t \int d^3x   e^{ -{\frac{1}{2 \sigma_s}{({\vec x}-{\vec x}_0(t))^2-\frac{1}{2 \sigma_t }( t-T_0)^2}}+i \delta \omega (t-T_0)-i {\delta {\vec p}}\cdot({\vec x}-{\vec x}_o) +R+i\Phi},\\
& \frac{1}{ \sigma_s}=\frac{1}{\sigma_{2}}+\frac{1}{\sigma_{1_1}} +\frac{1}{\sigma_{1_2}}, \frac{1}{\sigma_t}=
\sum_j\frac{{\vec v}_j^2}{\sigma_{j}}-\frac{{\vec v}_0^2}{\sigma_s}  , {\vec v}_0=\sigma_s \sum \frac{{\vec v}_j}{\sigma_j},\\
& \delta {\vec p}={\vec p}_{2}-{\vec k}_{1_1} -{\vec k_{1_2}}, \delta E= E_{2}-E_{1_1} -E_{1_2} , \delta \omega= \delta E- {\delta {\vec p}}\cdot {{ \vec v}_0}, 
\end{align}
where  $\Phi$  depends on the momenta and positions of the initial and final states \cite{Ishikawa-Shimomura}, and 
\begin{eqnarray}
& &T_0=\sigma_t( \frac{1}{\sigma_s}{\vec v}_0 \cdot {\vec x}_0 -\sum{\frac{1}{\sigma_j} {{\vec v}_j }{ \vec {\tilde X}_j}}) ,\vec x_0(t)= {\vec v}_0 t+\vec x_0, \vec x_0=\sigma_s ( \sum_l{\frac{1}{\sigma_j}{ { {\vec {\tilde X}}_j }}}),  \\
& &R=-\sum_l\frac{1}{2\sigma_j} {\vec {\tilde X}_j}^2+2\sigma_s(\sum_j\frac{1 }{2\sigma_j} {\vec {\tilde X}_j})^2 +2\sigma_t (\sum_j(\frac{{\vec v}_0}{\sigma_s}-\frac{{\vec v}_j}{\sigma_j}) \cdot {\vec {\tilde X}_j})^2,    \\
& &N_1=i \left( (4 \pi)^3 {\sigma_{1_1} } {\sigma_{{1_2} }}  {\sigma_{{2}} }   \right)^{-\frac{3}{4}}\left( {
E_{1_1} E_{1_2} E_{2}}\right)^{-\frac{1}{2}}, \vec {\tilde X}_j =\vec { X}_j -{\vec v}_j T_j. 
\end{eqnarray}  

In Eq.$(\ref{amplitude2})$,  the variable ${\vec x}$ is unlimited 
and the integration  over the space is constant irrespective of the center position. On the other hand,   $t$ has a upper and lower bounds, and the integral varies depending on the central position $T_0$. For  $T_0 < 0-\sqrt{2 \sigma_t}$ and $ T+\sqrt{2 \sigma_t}<T_0$ that is negligibly small, but   for   $0 +\sqrt{ 2 \sigma_t }\ll T_0 \ll  T-\sqrt{2 \sigma_t}$, and  $0-\sqrt{2 \sigma_t} < T_0 < 0+ \sqrt{2 \sigma_t}, T-\sqrt{2 \sigma_t} < T_0 < T+ \sqrt{2 \sigma_t}$ that is sizable. In the  former, the central position is inside the bulk region, and the integral over $t$ is equivalent to that of the region $-\infty \leq t \leq \infty$, 
\begin{eqnarray}
& &I_{bulk}=\int_0^T dt \int d{\vec x} e^{-\frac{1}{2\sigma_t}(t-T_0)^2-\frac{1}{2\sigma_s}({\vec x}-{\vec x}_0)^2+i \delta \omega (t-T_0)-i{\delta {\vec p}}\cdot ({\vec x}-{\vec x}_0)} = ({2\pi})^2 {(\sigma_t  \sigma_s^3)^{1/2}} e^{-\frac{\sigma_t}{2} (\delta \omega)^2-\frac{\sigma_s}{2} (\delta {\vec p})^2}, \label{Gaussian-integral1} \nonumber \\
& & 
\end{eqnarray}
 $I_{bulk}$  decreases exponentially with $|\delta \omega|$ and $|\delta {\vec p}|$. 
In the latter, the central position is at the boundaries,  and the integral is approximately 
\begin{eqnarray}
& &I_{boundary}=\int_{T_1}^{T_2} dt \int d{\vec x} e^{-\frac{1}{2\sigma_t}(t-T_0)^2-\frac{1}{2\sigma_s}({\vec x}-{\vec x}_0)^2+i \delta \omega (t-T_0)-i{\delta {\vec p}}\cdot ({\vec x}-{\vec x}_0)}  \label{Gaussian-integral2} \\
& &=\int_{0}^{\infty} dt \int d{\vec x} e^{-\frac{1}{2\sigma_t} t^2-\frac{1}{2\sigma_s}({\vec x}-{\vec x}_0)^2+i \delta \omega t-i{\delta {\vec p}}\cdot {\vec x}} \\
& &=
\begin{cases}
({2\pi})^{3/2} {( \sigma_s^3)^{1/2}}  \frac {1}{i \delta \omega}e^{-\frac{\sigma_s}{2} (\delta {\vec p})^2} ; \delta \omega \rightarrow \infty. \\ 
({2\pi})^{3/2} {( \sigma_s^3)^{1/2}}  \frac{\sqrt{ 2 \pi \sigma_t}}{2} e^{-\frac{\sigma_s}{2} (\delta {\vec p})^2} ; \delta \omega \rightarrow 0. \nonumber 
\end{cases}
\end{eqnarray}
  $I_{boundary}$  decreases inversely proportional to $|\delta \omega |$, which gives a large contribution  from the large $|\delta \omega|$ region.  A number of states of different positions ${\vec X}$ in the former is proportional to the length in temporal direction $T$, whereas that of the latter  is independent of $T$.  The former corresponds to $\Gamma T$ and the latter corresponds to $P^{(d)}$. 
 
 $I_{bulk}(\delta \omega )$ decreases  fast with $\delta \omega $, but $I_{boundary}(\delta \omega )$  decreases slowly, and their square are given by 
 \begin{eqnarray}
& & |I_{bulk}( \delta \omega)|^2 =(2 \pi)^4 ( \sigma_t \sigma_s^3) e^{-\sigma_s (\delta p)^2} e^{-\sigma_t(\delta \omega )^2}  \label{momentum_factor}\\
& &  |I_{boundary}(\delta \omega ;+)|^2 +|I_{boundary}( \delta \omega ;-)|^2= 2(2 \pi)^3 ( \sigma_s^3) e^{-\sigma_s (\delta p)^2} \frac{1}{ (\delta \omega)^2+  \epsilon^2 } ;\frac{1}{ \epsilon^2}= \pi \sigma_t,  \nonumber
\end{eqnarray}
in which the phase factor $\Phi$ drops.

In higher order diagrams, the integration is made over  more than one variables. That in one variable has  the boundary  terms which  decrease 
slowly as $\frac{1}{\delta \omega} $ or $\frac{1}{|\delta {\vec p}|}$ of the corresponding momentum. The integral of the amplitude over
 the intermediate states  is affected 
 by the phase factor $\Phi$,  and results to  the common  divergence in the bulk and the boundary. These   divergences are subtracted by the counter terms.    

The position dependences are expressed by $R$ in Eq.$(\ref{amplitude2})$.   $R\geq 0$ in general and for the scattering of the wave packets in which the trajectories do not intersect,    $R >R_{min}$ with $R_{min}  >0$, where $R_{min}$ is determined by the distance of trajectories. Then  the probability is suppressed by 
a factor $e^{-R_{min}}$. For the scattering of wave packets of intersecting   trajectories,  $R_{min}=0$, and the probability is not suppressed.

%%%%%%%%%%%%%%%%%%%%%%%%%%%%%%%%%%%%%%%%%%%%%%%%%%%%%%%%%%%%%%%%%%%%%%
\section{Amplitudes and probabilities  of wave packets  }
%%%%%%%%%%%%%%%%%%%%%%%%%%%%%%%%%%%%%%%%%%%%%%%%%%%%%%%%%%%%%%%%%%%%%The vacuum 
\subsection{ Vacuum amplitude}
The amplitude that  the vacuum  state is transformed to   three particle state, Eq.$(\ref{vacuum-amplitude})$, is
\begin{eqnarray} 
   & &N_3(\chi_1,\chi_2,\chi_3,T)=g\int_0^T dt \int d{\vec x} \langle 0|\varphi_1 (x) \varphi_1 (x) \varphi_2 (x) A_1(\chi_1)^{\dagger} A_1(\chi_2)^{\dagger} A_2(\chi_3)^{\dagger} |0\rangle  \\
  & & =g\int_0^T dt \int d{\vec x} C_1({\vec x};\chi_1)C_1({\vec x};\chi_2)C_2({\vec x};\chi_3) \nonumber \\
& &=g N_0  \int_0^T dt \int d{\vec x}    e^{ - \frac{1}{2 \sigma_s}({\vec x}-{\vec x}_0)^2 -\frac{1}{2 \sigma_t}(t-t_0)^2} e^{R+i\phi},  \nonumber \\
 & &  N_0=i \left( (4 \pi)^3 {\sigma_{1_1} } {\sigma_{{1_2} }}  {\sigma_{{2}} }   \right)^{-\frac{3}{4}}\left( {
E_{1_1} E_{1_2} E_{2}}\right)^{-\frac{1}{2}}. \nonumber 
\end{eqnarray}  
where ${\vec x}_0$ and $t_0$ are  functions of coordinates $\chi_i$ \cite{Ishikawa-Shimomura}, and $\phi$ is a phase.  $0 \ll t_0 \ll T$, and 
 for symmetric cases  $\sigma_l=\sigma$, $\sigma_s=\frac{\sigma}{3}$,
\begin{eqnarray}
& &\frac{1}{\sigma_t}=\sum_l \frac{v_l^2}{\sigma_l}-\sigma_s \sum_l \frac{{\vec v}_l}{\sigma_l}=\frac{1}{3\sigma}\sum_{l \neq m}({\vec v}_l-{\vec v}_m)^2, \label{wave-packets-parameters}\\
& &R_{momentum}=-\frac{\sigma_t}{2}(\delta E-{\vec v}_0 \cdot \delta {\vec p})^2-\frac{\sigma_s}{2}(\delta {\vec p})^2, \delta E=\sum_l E_l , \delta {\vec p}=\sum_l {\vec p}_l. \nonumber
\end{eqnarray}
Now,
\begin{eqnarray}
|R_{momentum} |> \frac{\sigma_t}{2}( E_g )^2,
\end{eqnarray}
 and        
\begin{eqnarray}
|  N_3(\chi_1,\chi_2,\chi_3,T )| <  N_0 g (2\pi \sigma_s)^{3/2} (2 \pi \sigma_t)^{1/2} e^{-\frac{\sigma_t}{2} E_g^2}.
\end{eqnarray}

 The integral over the momenta and positions of final states converges.   The last factor  in Eq.$(A5)$ satisfies 
 $e^{-\frac{\sigma_t}{2} E_g^2}  < e^{- 10^{10}}$ for parameters of our interests $\sigma=(10^{-8})^2$ meter$^2$ and $E_g=1$ MeV.  Thus the probability from the bulk  $P=0$.  
 
   The   contributions from   states of finite energy gap vanish for $E_g \geq 1$ MeV and $\sigma \geq  (10^{-8})^2$ meter$^2$, or for $E_g \geq 10^{-1}$ eV and $\sigma \geq  (10^{-1})^2$ meter$^2$. 
   
   A contribution from the boundary in time is inversely proportional to  $\frac{1}{ \omega^2}$ Eq.($\ref{momentum_factor})$, 
   which  is not exponentially suppressed.  The classical trajectories stating from the  positions  
   ${\vec X}_i$ should be one space-time point, and  these positions are restricted to a  region of a length $\sqrt \sigma_s$. The total probability 
   integrated over these  ${\vec X}_i$  is less than that of QCS  by a factor $\frac{{\sigma_s}^{3/2}}{V_D}$  and is negligible, where $V_D$ is the volume of the detector.
  The probability from the boundary can be considered $P=0$. 
  
  The probability for the vacuum vanishes.

%%%%%%%%%%%%%%%%%%%%%%%%%%%%%%%%%%%%%%%%%%%%%%%%%%
\subsection{ Particle decay probability }
%%%%%%%%%%%%%%%%%%%%%%%%%%%%%%%%%%%%%%%%%%%%%%%%
The decay  $2 \rightarrow 1+1'$ is described by Eq.$(\ref{decay-amplitude})$ and  occurs for $m_2 > 2 m_1$, and $E_g=0$.  A probability   for a particle 1 is  found for a large $\sigma_2 $, from the light-cone singularity
  due to $p_1' \rightarrow \infty$   was presented  in \cite{Ishikawa-Tobita-PTEP, Ishikawa-Tobita-ANA}.
 Here the probability is studied without using the light-cone singularity. 

Substituting Eq.$(\ref{momentum_factor})$ , we have the squares 
\begin{eqnarray}
& &| \mathcal{M}|^2=| \mathcal M_{bulk}|^2+|\mathcal M_{boundary}|^2\\
 & &|\mathcal{M}_j|^2= g^2 N_1^2 |I_j( \delta \omega )|^2 e^{2R}; {j=bulk, boundary}.
\end{eqnarray}

\subsubsection{ integral  over ${\vec X_i}$ }

The  process $ 1 \rightarrow 2+1'$ has the gap  $E_g=m_2$, and  $\Gamma=0$.   $P^{(d)}=0$,   as  a causality condition, 
\begin{eqnarray}
(p_1-p_2)^2 -m_1^2=0 
\end{eqnarray} 
is not satisfied.  
%%%%%%%%%%%%%%%%%%%%%%%%%%%%%%%%%%%%%%%%
\subsubsection{  Integral over the momenta  }
%%%%%%%%%%%%%%%%%%%%%%%%%%%%%%%%%%%%%%%%%%
It is shown that (1) the total probability for  a  scalar decay for wave packets is convergent in $\varphi_1^2 \varphi_2$ interaction but that is   divergent for $(\partial \varphi_1)^2 \varphi_2$ interaction, and (2) the probability for $\cos \Theta +1 \leq \epsilon$ is convergent, where $\Theta$ is the angle between two particles in the final states and $\epsilon$ is a constant.    Because the bulk term  
converges and is  in agreement with the golden rule, we focus on the 
 boundary term.   That    is given by  
\begin{eqnarray}
\int \prod_i \frac{d^3 {\vec k}_i}{ 2E(k_i) (2 \pi)^3} (k_i)^q e^{-\sigma_s {(\delta {\vec p})^2}} \frac{1}{(\delta \omega)^2+a^2}, \label{wave-packet-probability}
\end{eqnarray}
  where $a$ is a constant, and $q=0$ for  $\varphi_1^2 \varphi_2$ interaction or $q=1$ for   $(\partial \varphi_1)^2 \varphi_2$ interaction. 
   We study     the initial state at rest, $|{\vec p}_2|=0$,  and mass-less final state  $m_1=0$. 
  
% {\bf  Formula for the integration over the angle} 
  
   Two particle final states are written as
   \begin{eqnarray}
& &   {\vec k}_i=k_i( \sin \theta_i \cos \phi_i, \sin \theta_i \sin \phi_i, \cos \theta_) ;i=1,2\\
%& &   {\vec k}_2=k_2( \sin \theta_2 \cos \phi_2, \sin \theta_2 \sin \phi_2, \cos \theta_2)  \nonumber \\
& &d {\vec k_i}= k_i^2 dk_i d\Omega_i, d \Omega_i=d \cos \theta_i d \phi_i  \nonumber \\ 
  & &({\vec k}_1, {\vec k}_2)=k_1 k_2 \cos \Theta, \cos \Theta=\cos \theta_1\cos \theta_2+\sin \theta_1\sin \theta_2 \cos (\phi_1-\phi_2). \nonumber 
  \end{eqnarray} 
with these variables 
\begin{eqnarray}
& &(\delta {\vec p})^2=k_1^2+k_2^2+2 k_1k_2 \cos \Theta \\
& &(\delta \omega)^2= ((k_1+k_2)( 1  - \frac{\sigma_s}{\sigma_{1}}( 1+\cos \Theta)) -m_{2}  )^2.\nonumber
\end{eqnarray}
  Using  spherical functions  $P_l(\cos \Theta)$ 
% \begin{eqnarray}
 %& & P_0=1,P_1(z)=z,P_2(z)=\frac{1}{2}(3z^2-1), P_3(z)= \frac{1}{2}(5z^3-3z),  \cdots \\
 %& &P_{\nu}^0(z)=P_{\nu}(z) , \cdots \nonumber 
 %\end{eqnarray}
% \begin{eqnarray}
%& &\int_{-1}^{1} d \cos \theta P_{l_1}( \cos \theta )P_{l_2}( \cos \theta) =\frac{2}{2l_1+1} %\delta_{l_1 l_2} \\
%  & &P_{\nu}( \cos \Theta)=P_{\nu}( \cos \theta_1) P_{\nu}( \cos \theta_2)+ 2 \sum_{m=1}^{\infty}  (-1)^m  P_{\nu}^{m}( \cos \theta_1) P_{\nu}^{-m}( \cos \theta_2) \cos m (\phi_1-\phi_2), \nonumber
% \end{eqnarray}
a function  of $\Theta$ is expanded, 
% and its integral over the angle, 
 \begin{eqnarray}
& & f(\cos \Theta)=\sum_l a_l P_l(\cos \Theta) , a_l= \frac{2l+1}{2} \int_{-1}^{1} d \cos\theta  f(\cos \Theta) P_l(\cos \Theta)
\end{eqnarray}
and its integral over the angles $ d \Omega_i=d \cos \theta_i d \phi_i, -1 \leq \cos \theta_i  \leq 1, 0 \leq \phi_i \leq 2\pi $, is
\begin{eqnarray}
& & \int d\Omega_1  d\Omega_2 f(\cos \Theta)=\sum_l a_l  \int d\Omega_1  d\Omega_2  P_l(\cos \Theta)  \\
& &= \sum_l a_l  \int d\Omega_1  d\Omega_2  [ P_{l}( \cos \theta_1) P_{l}( \cos \theta_2)+ 2 \sum_{m=1}^{\infty}  (-1)^m  P_{\nu}^{m}( \cos \theta_1) P_{l}^{-m}( \cos \theta_2) \cos m (\phi_1-\phi_2) ] \nonumber  \\
& &=\sum_l a_l  (2 \pi)^2 \int d\cos \theta_1  d\cos \theta_2 [ P_{l}( \cos \theta_1) P_{l}( \cos \theta_2)] \nonumber \\
& &=\sum_l a_l  (2 \pi)^2  2^2 \delta_{ l,0} \delta_{ l ,0} =(4 \pi)^2 a_0. \nonumber
\end{eqnarray}

Because now  ${ \vec p}_2=0 $,
\begin{eqnarray}
g(\cos \Theta)=\frac{1}{((k_1+k_2)( 1  - \frac{\sigma_s}{\sigma_{1}}( 1+\cos \Theta)) -m_{2}  )^2+a^2} e^{x \cos \Theta},  x=2 \sigma_s k_1k_2
\end{eqnarray}
\begin{eqnarray}
& &I_g= \int d \Omega_1 d \Omega_2g(\cos \Theta)=(4 \pi)^2 b_0\\
& & b_0=\frac{1}{2} \int_{-1}^{1} dt g(t) .
\end{eqnarray}
For large $k_1+k_2$, 
\begin{eqnarray}
& & ((k_1+k_2)( 1  - \frac{\sigma_s}{\sigma_{1}} 2 ) -m_{2}  )^2\leq ((k_1+k_2)( 1  - \frac{\sigma_s}{\sigma_{1}}( 1+\cos \Theta)) -m_{2}  )^2 \leq ((k_1+k_2) -m_{2}  )^2 \nonumber \\
& &\frac{1}{((k_1+k_2) -m_{2}  )^2+a^2} e^{x \cos \Theta}\leq g(t) \leq \frac{1}{((k_1+k_2)( 1  - \frac{\sigma_s}{\sigma_{1}}2) -m_{2}  )^2+a^2} e^{x \cos \Theta}, \nonumber
\end{eqnarray}
and 
\begin{eqnarray}
& &\frac{1}{((k_1+k_2) -m_{2}  )^2+a^2} a_0 \leq b_0  \leq \frac{1}{((k_1+k_2)( 1  - \frac{\sigma_s}{\sigma_{1}}2) -m_{2}  )^2+a^2} a_0, \\
& &\frac{e^x-e^{-x}}{2x (((k_1+k_2) -m_{2}  )^2+a^2) } \leq b_0  \leq \frac{e^x-e^{-x}}{2x(((k_1+k_2)( 1  - \frac{\sigma_s}{\sigma_{1}}2) -m_{2}  )^2+a^2)}. 
\end{eqnarray}
Thus 
\begin{eqnarray}
\frac{e^{2 \sigma_s k_1k_2}}{2\sigma_sk_1k_2 (((k_1+k_2) -m_{2}  )^2+a^2) } \leq b_0  \leq \frac{e^{2 \sigma_s k_1k_2} }{2\sigma_sk_1k_2(((k_1+k_2)( 1  - \frac{\sigma_s}{\sigma_{1}}2) -m_{2}  )^2+a^2)}.
\end{eqnarray}
The total probability for $q=0$
\begin{eqnarray}
& &P=\int \prod_i \frac{d^3 {\vec k}_i}{ 2E(k_i) (2 \pi)^3} (k_i)^q e^{-\sigma_s {(\delta {\vec p})^2}} \frac{1}{(\delta \omega)^2+a^2},\\
& &=(4 \pi)^2 \int \prod_i \frac{k_i^2 dk_i} { 2E(k_i) (2 \pi)^3}  e^{-\sigma_s (k_1^2+k_2^2)} b_0, \nonumber
\end{eqnarray}
is bounded as  $P_{min} \leq P \leq P_{max}$, where 
\begin{eqnarray}
& &P_{min}=   (4 \pi)^2 \int  \frac{ dk_{+}  dk_{-} } { 2^2 (2 \pi)^6}  e^{-\sigma_s  k_{-}^2} \frac{1}{2\sigma_s (( 2k_{+}  -m_{2}  )^2+a^2) }      \\
& &=     \frac{  (4 \pi)^2}{ 2^2 (2 \pi)^6 2\sigma_s} \sqrt{ \frac{\pi}{ \sigma_s}}  \int_0^{\infty} dk_{+} \frac{1}{ (2 k_{+}  -m_{2}  )^2+a^2 } \nonumber \\
& &P_{max} =  (4 \pi)^2 \int  \frac{ dk_{+} dk_{-} } { 2^2 (2 \pi)^6}  e^{-\sigma_s  k_{-}^2} \frac{1}{2\sigma_s ( 2k_{+} (1-2\frac{\sigma_s}{\sigma_1}) -m_{2}  )^2+a^2) }                                              \\
& &=     \frac{  (4 \pi)^2}{ 2^2 (2 \pi)^6 2\sigma_s} \sqrt{ \frac{\pi}{ \sigma_s}}  \int_0^{\infty} dk_{+} \frac{1}{ (2 (1-2\frac{\sigma_s}{\sigma_1})k_{+}  -m_{2}  )^2+a^2 },  \nonumber 
\end{eqnarray}
where $k_{+}=\frac{k_1+k_2}{2},k_{-}=k_1-k_2$.
$P_{min} , P_{max}$ are finite, and $P$ is finite. 

For $q=1$,  $k_{+}^2$ is multiplied to the above  integrands,  
%\begin{eqnarray}
%& &P_{min}=     \frac{  (4 \pi)^2}{ 2^2 (2 \pi)^6 2\sigma_s} \sqrt{ \frac{\pi}{ \sigma_s}}  \int_0^{\infty} dk_{+} \frac{k_{+}^2 }{ (2 k_{+}  -m_{2}  )^2+a^2 }\\
%& &P_{max} =     \frac{  (4 \pi)^2}{ 2^2 (2 \pi)^6 2\sigma_s} \sqrt{ \frac{\pi}{ \sigma_s}}  \int_0^{\infty} dk_{+} \frac{k_{+}^2 }{ (2 (1-2\frac{\sigma_s}{\sigma_1})k_{+}  -m_{2}  )^2+a^2 },  
%\end{eqnarray}
and the integrals  are divergent. $P_{min} , P_{max}$ are infinite, and $P$ is infinite. 

For $\cos \Theta +1 \leq \epsilon$, the exponential suppression factor $e^{-2 k_1k_2 \epsilon }$ is multiplied to the above  integrands, and the integrals  converge for arbitrary $q$. 
This is the case  for large ${| \vec p_2 |  }$, in which  it is worthwhile to estimate the integrals using the behaviors at small angles $\theta_i \leq \epsilon$.  For ${\vec p}_2$ in the z-direction, the integrand is given by
\begin{eqnarray}
g(\cos \Theta;\cos \theta_1,\cos \theta_2)=e^{2 \sigma_s p_2 (k_1 +k_2) }  \frac{1}{((k_1+k_2)( 1  - 2 \frac{\sigma_s}{\sigma_{1}}) -m_{2}  )^2+a^2} e^{-2 \sigma_s k_1k_2 },  
\end{eqnarray}
The probability is bounded as  $P_{min} \leq P \leq P_{max}$, where 
\begin{eqnarray}
& &P_{min}=   \frac{(4 \pi)^2 }{ { 2^2 (2 \pi)^6}2\sigma_s} \int_0^{\infty}  dk_{+} \int_{-k_{+}}^{k_{+}}   dk_{-}  ( k_{+}^2-\frac{1}{4} k_{-}^2)^{q}  \frac{ e^{-\sigma_s ( p_2-2 k_{+})^2 }}{ ( 2k_{+}  -m_{2}  )^2+a^2 }      \\
& &P_{max} = \frac{(4 \pi)^2 }{ { 2^2 (2 \pi)^6}2\sigma_s}   \int_0^{\infty}   dk_{+} \int_{-k_{+}}^{k_{+}}   dk_{-}  ( k_{+}^2-\frac{1}{4} k_{-}^2)^{q}     \frac{e^{-\sigma_s ( p_2-2k_{+})^2}}{   { ( 2k_{+} (1-2\frac{\sigma_s}{\sigma_1}) -m_{2}  )^2+a^2 }}          \nonumber                  
 \nonumber 
\end{eqnarray}
$P_{min} , P_{max}$ are finite due to the exponential factor, and $P$ is finite  even for arbitrary $q$.

%%%%%%%%%%%%%%%%%%%%%%%%%%%%%%%%%%%%%%%%
\subsection{Divergences  due to the intermediate sates in higher order corrections}
%%%%%%%%%%%%%%%%%%%%%%%%%%%%%%%%%%%%%%
The summation over  the intermediate states of the central values of $( {\vec X}_i, {\vec P}_i)$ for the wave packets in higher order corrections   is equivalent to that of the plane waves  and lead divergences of the universal properties from Section (2-I). The divergences are the same in the bulk and the boundary, and are subtracted by the counter terms in the Lagrangian. 
Accordingly, the ultraviolet divergences in the intermediate states are subtracted with the counter terms and  $P^{(d)}$ in the renormalized theory 
are computed in the standard manners of  the renormalization procedure.     

      The contributions form the bulk in  intermediate states of the higher order corrections  have $e^{-\sigma (\omega)^2}$, and $\delta \omega \geq E_g$ . These vanish practically as is given in Appendix, and the contribution from the boundary is inversely proportional to $\delta \omega$ and  finite.  is  and  intermediate states
ultraviolet divergences in the corrections of the order $e^2$  from  the intermediate states   are due to  the boundary terms, and are  cancelled by the counter terms.   Now, the boundary effects   arise in the lowest order of $e$  and in the higher orders, and are long-distance effects. They are convergent for  the normalized initial and final states.   Accordingly, that   in the lowest order is studied in detail.
   %%%%%%%%%%%%%%%%%%%%%%%%%%%%%%%%%%%%%%%%%%%%%%%%%%%
%\subsection{Higher order effects}
%%%%%%%%%%%%%%%%%%%%%%%%%%%%%%%%%%%%%%%%%%%%%%%%%
 The divergences due to the intermediate states in the  higher order corrections are common in  the bulk terms and the boundary terms 
  for  the plane waves and for  the wave packets. These divergences     caused by the fluctuations of the fields in the extremely  short-distance region 
  are subtracted by the local counter terms in the Lagrangian Eq.$(\ref{renormalized-QED})$.  $P^{(d)}$  is expressed with the  physical masses and the  coupling
   constants.     
%%%%%%%%%%%%%%%%%%%%%%%%%%%%%%%%%%%%%%%%%%%%%%%%%%%%%%%%%%%%%%%%%%%%%%%%
 \section{Interaction Lagrangian of the total derivative}
%%%%%%%%%%%%%%%%%%%%%%%%%%%%%%%%%%%%%%%%%%%%%%%%%%%%%%%%%%%%%%%%%%%%%%%%
 An example  of    showing $\Gamma=0, P^{(d)} \neq 0$ was given in  \cite{Ishikawa-Tajima-Tobita-PTEP}, and another  is a system of 
a scalar field $\varphi(x)$ of the mass $m_1$ and a vector field
$\varphi_{\mu}(x)$  of the mass $m_2$ described by a standard kinetic action of fields of masses  $m_2 > 2m_1$ and $L_\text{int}= -g \partial_{\mu}( \varphi^{\mu}(x)\varphi(x)^2)$.   
The  action integral becomes a surface term
\begin{eqnarray}
S_\text{int}=-g \int d^3 S_{\mu} ( \varphi^{\mu}(x)\varphi(x)^2), \label{total-deivative} 
\end{eqnarray}  
and the interaction does neither modify the equation of motion in the bulk nor  the canonical structure of the variables. If ASI used in  the Tomonaga-Feynman-Schwinger formalism \cite{Tomonaga, Feynman, Schwinger} were apllied,   
the vertex part of the interaction Eq. $( \ref{total-deivative} )$ in the momentum representation 
\begin{eqnarray}
\mathcal M = -ig N (2\pi)^4 (p_{V}-p_1-p_1')_{\mu} \delta^{4} (p_V-p_1-p_1') \epsilon^{\mu}(p_V),
\end{eqnarray}  
vanishes  from the property of the Dirac's delta function.   Thus $\Gamma=0$, but  the amplitude  in configuration space, 
\begin{eqnarray}
\mathcal{M}=-ig N \int_{\lambda \geq 0} d^4x \frac{\partial}{\partial x_{\mu}}[ e^{i(-p_V+p_1') \cdot x +ip_1\cdot( x-X_{\mu}) -\xi(x)}] \epsilon^{\mu}(p_V) , \lambda =(x-X)^2
\end{eqnarray}
does not vanish. $P^{(d)}$ for a large $\sigma_1$ is  given in Eq.$(\ref{t_derivative_P})$.
The  probability computed with the Fermi's golden rule vanishes, but   $P^{(d)} \neq 0$.   
  $P^{(d)}$   is not computed   with ASI, but by the FQM with the normalized state. 
%%%%%%%%%%%%%%%%%%%%%%%%%%%%%%%%%%%%%%%%%%%%%%%%%%%%
\section{Changing  the initital states}
%%%%%%%%%%%%%%%%%%%%%%%%%%%%%%%%%%%%%%%%%%%%%%%%%%%%
The initial states  satisfy  the free wave equations   in many processes. This is partly  because   fundamental interactions 
are local, and the range of the binding force is short. The atom, nucleus, and hadrons are formed almost instantaneously.  Furthermore,   lighter particles produced in the decays   separate quickly due to  higher speed than the parent.  The subsequent transition of these products are studied by the wave functions of this  initial condition 
and     by the interaction that switches on  at this instant of time $t =0$.    Now, for a case of strong  $H_{int}$ which acts rapidly,  the initial state may transfer  instantaneously to the eigenstate of $H$. The amplitude then is expressed with the wave packets defined from eigenstates of $H$. From Section 2H,  that  gives the equivalent transition amplitude with  those  of $H_0$.  General cases  are studied  here.

%%%%%%%%%%%%%%%%%%%%%%%%%%%%%%%%%%%%%%%%%%%%%%%%%%%%%%%%%%%%%%%%
\subsection{Wavefunction}
%%%%%%%%%%%%%%%%%%%%%%%%%%%%%%%%%%%%%%%%%%%%%%%%%%%%%%%%%%%%%%%%%%%

 For a  normalized state of a linear combinations  Eq.$(\ref{complete-set})$,     
\begin{eqnarray}
& &|\Psi_{\alpha}(0)  \rangle =|\Psi_{\alpha}(0), {\vec P},{\vec X},T_0 \rangle+\sum_{M}  C_{\alpha}^{M} | M, \chi_i ,i=1,M \rangle , C_{\alpha}^M=\langle \chi_i,  M|  \Psi_{\alpha}(0) \rangle,\label{wavefunction} \\
& &\sum_{\alpha} {C_{\alpha}}^{M} ({{C_{\alpha}}^{M'}})^{*} =\langle \chi_i,i=1-M|\chi_j,j=1- M'\rangle,  \nonumber
\end{eqnarray}
  where  $| | M, \chi_i ,i=1,M \rangle $ are states constructed from $H_0$, which are  denoted as $\chi$ states.   The wave function at $T$ for the initial state  $|\Psi_{\alpha}(0) \rangle$ is expanded with the  $\chi$ states 
\begin{eqnarray}
& &|\Psi(T) \rangle=U(T) |\Psi_{\alpha} (0) \rangle =U(T) |  \chi_1 \rangle \langle  \chi_1 |\Psi_{\alpha}(0) \rangle \nonumber \\
& &=\sum_{\chi_1}(|\Psi_{(p)},  \chi_1 \rangle \langle  \chi_1 |\Psi_{\alpha}(0) \rangle + |\Psi_{(w)},   \chi_1 \rangle \langle  {\chi_1} |\Psi_{\alpha}(0) \rangle) ,
\end{eqnarray}
where the decomposition in the $\chi$ space ,
\begin{eqnarray}
U(T)|\chi \rangle =  |\Psi_{(p)}(T), \chi \rangle+ |\Psi_{(w)}(T), \chi \rangle
\end{eqnarray}
of  two components was substituted.   
\subsection{S-matrix :S[T] }
%%%%%%%%%%%%%%%%%%%%%%%%%%%%%%%%%%%%%%%%%%%%%%%%%%%%%%%%%%%%%%%%%%
The S-matrix $S[T,\alpha]$ for the initial states $|\Psi_{\alpha} \rangle$ is written as
\begin{eqnarray}
& &\langle \alpha, \chi_1|S[T]| \alpha, \chi_2 \rangle = \sum_{\chi_1',\chi_2'}\langle \alpha, \chi_1|\chi_1' \rangle \langle \chi_1'|S[T]|\chi_2' \rangle \langle \chi_2'| \alpha,\chi_2 \rangle  \nonumber\\
& &= \sum_{\chi_1',\chi_2'}\langle \alpha, \chi_1|\chi' \rangle [ \langle \chi_1'|S[T]|\chi_2' \rangle^{(p)} +\langle \chi_1'|S[T]|\chi_2' \rangle^{(w)}] \langle \chi_2'| \alpha,\chi_2 \rangle. 
\end{eqnarray}
The probability is
\begin{eqnarray}
P=|\langle \alpha, \chi_1|S[T]| \alpha, \chi_2 \rangle |^2 ,
  \end{eqnarray}
and the total probability to certain states is
\begin{eqnarray}
& &\sum_{final~states}  P_{f} \nonumber\\
& &=\sum_{\chi_1} |\langle \alpha, \chi_1|S[T]| \alpha, \chi_2 \rangle|^2  \nonumber  \\
& &=\sum_{\chi_1, \chi_1'} \langle \alpha, \chi_1|\chi_1' \rangle \langle \chi_1'|S[T]|\alpha, \chi_2 \rangle \langle  \alpha,\chi_2   |S[T]| \chi_1'' \rangle \langle \chi_1''|\alpha,\chi_1 \rangle \nonumber \\
& &=\sum_{\chi_1'}  | \langle \chi_1'|S[T]|\alpha, \chi_2 \rangle |^2    
 \end{eqnarray}
and the probability averaged over the initial states is
\begin{eqnarray}
 \bar P_f  =\sum _{\chi_1',\chi_2'}|\langle \chi_1'|S[T]|\chi_2' \rangle |^2 . 
 \end{eqnarray}

%%%%%%%%%%%%%%%%%%%%%%%%%%%%%%%%%%%%%%%%%%%%%%%%%%%%%%%%%%%%%%%%%%%%%%%%%
   \section{Gibbs ensemble, Boltzmann equation, and Ergodic
   hypothesis}
%%%%%%%%%%%%%%%%%%%%%%%%%%%%%%%%%%%%%%%%%%%%%%%%%%%%%%%%%%%%%%%%%%%%%%%%
Probability due to $\Gamma T$  comes from the microscopic region of the order of de Broglie wave length, and satisfies the kinetic-energy conservation.  This  term leads the Boltzmann equation consistent with the Gibbs ensemble.  
 The states in the space of the wave
  functions  $|\Psi_{(p)}(t)\rangle $ reveal these  particle properties  and the probabilities Eqs. $(\ref{exponential1})$, and $(\ref{exponential2})$.  Transitions occur  independently,
   the distribution   follows the principle of equal a priori probability and the Boltzmann distribution,
\begin{eqnarray}
P_G = e^{-\beta H},
\end{eqnarray}
using an average energy per each freedom  $\frac{1}{\beta}=k_B \text {T}$, here $k_B$ is the Boltzmann constant and T is the temperature.
The states  follow the normal  thermodynamics expressed by the Gibbs
 ensemble \cite{Tolman}. Many phenomena in fact follow this
 distribution. 

Now $P^{(d)}$ due to QCS, $|\Psi_{(w)}(t) \rangle$, revealing  the wave's overlap is rapid, and remains the same long period.  That does neither satisfy
 the conservation law of the kinetic energy, nor the Markov nature.  In the
decays, they follow the
time-independent power law  Eq. ($\ref{power-law})$ instead of the 
exponential behavior.  The probability $P^{(d)}$ determines the initial probability for the Boltzmann equation, and   follows   the power
 law in the kinetic energy, and is independent of  the temperature.
This kind of behavior has been observed in many area in dilute systems, and  will be studied in a forthcoming work.    The  total statistical
ensemble of these waves are the product of this term and that of Gibbs ensemble 
\begin{equation}
P=P_{G} \times P_{D},
\end{equation}
where $P_{D}$ is the distribution corresponding to $P^{(d)}$.    

The probability in the scattering or the decay from $\Gamma T $   
covers the wide angle in each reaction. Accordingly the whole phase 
space is covered ultimately, and leads us to the Ergodic
hypothesis. Those from $P^{(d)}$, however,  is restricted to the
extreme forward angle in the time-independent manner, and does not cover 
the whole phase space. Hence this component does not follow the Ergodic
hypothesis.

 %The light-cone singularity  \cite{Wilson-OPE} is connected with the $P^{(d)}$  \cite{Ishikawa-Tobita-PTEP}. 
%%%%%%%%%%%%%%%%%%%%%%%%%%%%%%%%%%%%%%%%%%%%%%%%%%%%%%%%%%%%%%%%%%%%%%%%%%%%%%%%%%%%%%%% 


\begin{thebibliography}{}
\bibitem{Dirac-t}   P. A. M. Dirac,  \textit{Fundamental principle of quantum menchanics}  Oxford Press,  (1927).   As was emphasized in this book, the cross section or the decay rate defined from  fluxes are  the average values, which are useful for many practical purposes. They 
 differ from the rigorous probability derived from FQM, which are studied here.
\end{thebibliography}
\end{document}